\documentclass[a4paper,11pt]{article}
\pdfoutput=1 

\usepackage{jheppub} 
\usepackage{xcolor}
\usepackage{pagecolor,lipsum}
\usepackage{mdframed}
\usepackage{verbatim}
\usepackage{subcaption,graphicx}
\usepackage[T1]{fontenc} 
\usepackage{mathrsfs}
\usepackage{arydshln}
\usepackage{color}
\usepackage{slashed}
\usepackage{epsfig}
\usepackage{enumerate}
\usepackage{color}
\usepackage{wrapfig}
\usepackage{amsfonts}
\usepackage{multirow}
\usepackage{amsthm}
\usepackage{amsmath}
\usepackage{tikz}
\usepackage{float}
\usepackage[numbers]{natbib}
\usepackage{amssymb,amsmath,amsbsy,bbold}
\usepackage{mathrsfs}
\usepackage{axodraw}
\usepackage{dsfont}

 \usepackage[normalem]{ulem}

\newcommand{\ba}{\begin{array}}
\newcommand{\ea}{\end{array}}
\newcommand{\bd}{\begin{displaymath}}
\newcommand{\ed}{\end{displaymath}}
\def\be{\begin{equation}}
\def\ee{\end{equation}}
\def\bsube{\begin{subequation}}
\def\esube{\end{subequation}}
\def\bea{\begin{eqnarray}}
\def\eea{\end{eqnarray}}
\def\bal{\begin{align}}
\def\ealign{\end{align}}
\def\eal{\end{align}}
\def\beq{\begin{equation}}
\def\eeq{\end{equation}}
\newenvironment{Eqnarray}%
     {\arraycolsep 0.14em\begin{eqnarray}}{\end{eqnarray}}
\def\beqa{\begin{Eqnarray}}
\def\eeqa{\end{Eqnarray}}
\def\bea{\begin{Eqnarray*}}
\def\eea{\end{Eqnarray*}}
\def\ben{\begin{enumerate}}
\def\een{\end{enumerate}}

\def\beq{\begin{equation}\ba{rcl}}
\def\eeq{\ea\end{equation}}

\def\lsup#1{^{\lower 6pt\hbox{$\scriptstyle#1$}}}

\def\ddel{\!\!\mathrel{\raise1.5ex\hbox{$\leftrightarrow$\kern-.85em
\lower1.7ex\hbox{$\partial$}}}}

\title{\boldmath Lepto-philic 2-HDM + singlet scalar portal induced fermionic dark matter}
\author[a,\,\dagger ]{Sukanta Dutta,}
\author[b,\,\#]{Ashok Goyal,}
\author[a,b,\,\$]{Manvinder Pal Singh,}

\affiliation[a]{SGTB Khalsa College, University of Delhi, Delhi, India.}
\affiliation[b]{Department of Physics $\&$ Astrophysics, University of Delhi, Delhi, India.}

\emailAdd{${}^\dagger$Sukanta.Dutta@gmail.com}
\emailAdd{${}^\#$agoyal45@yahoo.com}
\emailAdd{${}^{\$}$ Corresponding~Author: manvinderpal666@yahoo.com}

\abstract{We explore the possibility that the discrepancy in  the observed  anomalous magnetic moment of the muon $\Delta a _\mu$ and the predicted relic abundance of Dark Matter by Planck data, can be explained in a lepto-philic 2-HDM  augmented by a real SM singlet scalar of mass $\sim$ 10-80 GeV. We constrain the model from the observed Higgs Decay width at LHC, LEP searches for low mass exotic scalars and anomalous magnetic moment of an electron $\Delta a_e$. This constrained light singlet scalar serves as a portal for the fermionic Dark Matter, which contributes to the required relic density of the universe. A large region of model parameter space is found to be consistent with the  present observations from  the Direct and Indirect DM detection experiments.}
\begin{document} 

\maketitle
\flushbottom
\section{Introduction}
 Investigations into the nature of dark matter (DM) particles and their interactions is an important field of research in Astro-particle physics. The Atlas and CMS collaborations at the Large Hadron Collider (LHC) are searching for the signature of DM particles involving missing energy ($\not \!\!\!E_T$) \cite{Garny:2013ama,GooDMan:2010yf} accompanied by a single or two jet events. Direct detection experiments  measure the nuclear-recoil energy and its spectrum in DM-Nucleon elastic scattering \cite{Hisano:2011um,Garny:2012eb}. In addition, there are Indirect detection experiment  \cite{Ackermann:2013yva} searching for the DM annihilation into photons and neutrinos in cosmic rays. These experiments have now reached a level of sensitivity where a significant part of parameter space required for the observed relic density, if contributed by the dark matter composed of Weakly interacting massive particles (WIMPs) that survive as thermal relics, has been excluded. The null results of these direct and indirect experiments have given rise to the consideration of ideas where the dark matter is restricted to couple exclusively to either Standard Model (SM) leptons (lepto-philic) or only to top quarks (top-philic). In these scenarios the DM-Nucleon scattering occurs only at the loop level and the constraints from direct detection are weaker.

\par Extended Higgs sector have been studied in literature \cite{Dedes:2001nx,Abe:2015oca,Chun:2015hsa} to explain discrepancy in anomalous magnetic moment of muon.  Recently a simplified  {\it Dark Higgs portal model}  of the order of $\lesssim$ few GeV, that couples predominantly to leptons with the coupling constant  $\sim \, m_l/v_o$  where $m_l$ is the lepton mass and $v_o $ is the Higgs  VEV, has been considered in the literature \cite{Chen:2015vqy}. This model induces large contribution to the anomalous magnetic moment of muon and can explain the existing discrepancy between the experimental observation  $a^{\rm exp}_{\mu}=$ 11 659 209.1(5.4)(3.3)$\times10^{-10}$ \cite{pdg2018-muon} and theoretical prediction $a^{\rm SM}_{\mu}=$ 116 591 823(1)(34)(26) $\times10^{-11}$  \cite{Hagiwara:2011af,pdg2018-muon} of the muon anomalous magnetic moment $\Delta a_{\mu}\equiv a_{\mu^-}^{\rm exp}-a_{\mu^-}^\text{SM}\,=\,268(63)\,\times10^{-11}$ \cite{pdg2018-muon} without compromising the experimental measurement of electron anomalous magnetic moment $a^{\rm exp}_e=(1159.65218091\pm 0.00000026)\times 10^{-6}$ \cite{pdg2018-electron}. It has been shown in the literature  \cite{Batell:2016ove}, that with the inclusion of an additional singlet scalar below the electro-weak scale to the lepto-philic 2-HDM makes the model UV complete. This UV complete model with an extra singlet scalar $\sim\,\, <$ 10 GeV successfully explains the existing 3 $\sigma$ discrepancy of muon anomalous magnetic moment and is consistent with the   constraints on  the model parameters from muon and meson decays  \cite{{Bird:2004ts},OConnell:2006rsp,Batell:2009jf,Krnjaic:2015mbs}. These results have been analysed for 0.01 GeV < $m_{S^0}$ < 10 GeV when  compared with those for the singlet neutral vector $Z^\prime$ searches at  $B$ factories such as BaBar \cite{TheBABAR:2016rlg}, from electron beam dump experiments \cite{Battaglieri:2014hga} and electroweak precision experiments \cite{Lebedev:2000ix} etc. 
\par In reference \cite{Agrawal:2014ufa}, the authors have explored the possibility of explaining the anomalous magnetic moment of muon with an  additional lepto-philic light scalar mediator  assuming the {\it universal coupling} of the scalar with all leptons constrained from the LEP \cite{Schael:2013ita} resonant production and the BaBar experiments \cite{TheBABAR:2016rlg}.  These constraints were found to exclude   all of the scalar mediator mass range except between 10 MeV and 300 MeV.

\par In the current paper we consider fermionic dark matter that couples predominately with SM leptons through the  {\it non-universal} couplings with the   {\it  scalar portal}  in the UV complete lepto-philic 2-HDM model. We relax the requirement of the very light scalar considered in \cite{Batell:2016ove} and investigate parameter space for a comparatively heavier scalar 10 GeV $\lesssim \, m_{S^0}\,\lesssim $ 80 GeV. In section \ref{section_TheFramework}, we give a brief review of this simplified model, using the full Lagrangian and  couplings of the Singlet scalar $S^0$ with all model particles. In section \ref{section_ElectroWeak_Constraints}, we calculate the contribution from  scalars\,($S^0,H^0,A^0,H^\pm$) to the anomalous magnetic moment of the muon and discuss bounds on the model parameters from  LEP-II, $\Delta a_e$ and upper bound on the observed total Higgs decay width. Implications of the model contributions to the lepton couplings non-universality and the oblique corrections are briefly discussed along-with  available constraints on them.

In the present study we are motivated to explore the possibility of simultaneously explaining the discrepancy in the observed anomalous magnetic moment of the muon on the one hand and the expected relic density contribution from DM on the other. Accordingly, in section \ref{section_ Dark_ matter}, we  introduce the DM contributing to the relic density through dark matter -  SM particles interactions induced by the additional scalar in the  model and scan for the allowed parameter space which is consistent with   direct and indirect experimental data  as well as with the observed value of the $\Delta a_\mu$. Section \ref{section_summary} is devoted to discussion and summary of results. 
\section{The Model}
\label{section_TheFramework}
 We consider a UV complete  lepton specific 2-HDM with a singlet  {\it  scalar portal} interacting with the fermionic DM. In this model the two Higgs doublets $\Phi_1$ and $\Phi_2$ are so arranged that $\Phi_1$ couples exclusively to leptons while $\Phi_2$ couples exclusively to quarks. The ratio of their VEV's $\left\langle \Phi_1\right\rangle/\,\left\langle \Phi_2\right\rangle \equiv \,v_2/\, v_1$ = $\tan\beta$ is assumed to be large. In this model  the scalars (other than that identified with the CP even $h^0\,\sim$ 125 GeV) couple to leptons and quarks with coupling enhanced and suppressed by $\tan\beta$ respectively. A mixing term in the potential $A_{12}\, \left[\Phi_1^\dagger\Phi_2+\Phi_2^\dagger\Phi_1\right]\,\varphi^0$ results in the physical scalar $S^0$ coupling to leptons with strength proportional to $m_l/v_o$ where
 \begin{equation}
  v_o\equiv \sqrt{v_1 ^2 + v_2 ^2}\, = \, 246 \,{\rm GeV}.
  \end{equation}
 \noindent The full scalar potential is given by
 \begin{equation}
V(\Phi_1,\Phi_2,\varphi^0) = V_{2- HDM} + V_{\varphi^0} + V_{\rm portal}     
 \end{equation}
where CP conserving $V_{2-HDM}$ is  given as
\begin{eqnarray}
V_{2-HDM}(\Phi_1,\Phi_2) &=& m^2_{11}\, \Phi_1^\dagger\Phi_1 +
m^2_{22}\,\Phi_2^\dagger\Phi_2 -\left(m^2_{12}\, \Phi_1^\dagger\Phi_2
+{\rm h.c.} \right) +\frac{\lambda_1}{2}\, \left(\Phi_1^\dagger\Phi_1
\right)^2 +\frac{\lambda_2}{2}\, \left(\Phi_2^\dagger\Phi_2 \right)^2
\nonumber \\ && +\lambda_3\, \left(\Phi_1^\dagger\Phi_1 \right)
\left(\Phi_2^\dagger\Phi_2 \right) +\lambda_4\, \left(\Phi_1^\dagger\Phi_2
\right) \left(\Phi_2^\dagger\Phi_1 \right) +\left\{\frac{\lambda_5}{2}
\,\left(\Phi_1^\dagger\Phi_2 \right)^2 +{\rm h.c.} \right\} \,
\label{notation1}
\end{eqnarray}
and $V_{\varphi^0}$ and $V_{\rm portal}$ is assumed to be
\begin{eqnarray}
V_{\varphi^0}&=&B\varphi^0 +\frac{1}{2} m_0^2 (\varphi^0)^2+\frac{A_{\varphi^0}}{2} (\varphi^0)^3+\frac{\lambda_{\varphi^0} }{4} (\varphi^0)^4.
\label{eq:VS}\\
V_{\rm portal} &=& A_{11}\, \left(\Phi_1^\dagger\Phi_1\right)\,\varphi^0   +    A_{12}\,\left(\Phi_1^\dagger\Phi_2 + \Phi_2^\dagger \Phi_1\right)\,\varphi^0   +    A_{22}\,\left(\Phi_2^\dagger\Phi_2 \right)\,\varphi^0.\label{eq:VPortal}    
\end{eqnarray}
where the scalar doublets
 \begin{eqnarray}
\Phi_1 =\frac{1}{\sqrt{2}} \begin{pmatrix} \sqrt{2} \omega_1^+ \\ \rho_1+v_o\,\cos\beta
  +i\,z_1 \end{pmatrix}; \,\,\,\,
\label{phi1}\,\Phi_2 =\frac{1}{\sqrt{2}} \begin{pmatrix} \sqrt{2} \omega_2^+ \\ \rho_2 +
 v_o\,\sin\beta +i\,z_2 \end{pmatrix} \,.
\label{phi2}
\end{eqnarray}
\noindent are written  in terms of the mass eigenstates $G^0$, $A^0$ , $G^\pm$ and $H^\mp$ as
\begin{eqnarray}
\begin{pmatrix} z_1 \\ z_2 \end{pmatrix} = \left( \begin{array}{ccc}\cos\beta& &
  -\sin\beta \\ 
\sin\beta & &\cos\beta \end{array} \right)\begin{pmatrix} G^0 \\ A^0 \end{pmatrix}; && \,\,\,
\begin{pmatrix} \omega_1 \\ \omega_2 \end{pmatrix} = \left(\begin{array}{ccc} \cos\beta 
  &&-\sin\beta \\ 
\sin\beta  &&\cos\beta \end{array}\right) \begin{pmatrix} G^\pm \\ H^{\pm} \end{pmatrix}.\nonumber\\
\label{mixing}
\end{eqnarray}
Here $G^0$ \& $G^\pm$ are Nambu-Goldstone Bosons absorbed by the $Z^0$ and W$^\pm$ vector Bosons, $A^0$ is the pseudo-scalar and $H^\pm$ are the charged Higgs.
The three CP even neutral scalar mass eigen-states mix among themselves  under small mixing angle approximations  
\begin{eqnarray}
\sin \delta_{13} \sim \delta_{13}\simeq-\frac{v_0\,A_{12}}{m_{H^0}^2}, \,\, {\rm and} \,\, \sin \delta_{23} \sim  \delta_{23}\simeq-\frac{v_0\,A_{12}}{m_{h^0}^2}\left[1+\xi^{h^0}_{\ell}\left(1-\frac{m_{h^0}^2}{m_{H^0}^2}\right)\right]\cot\beta
\label{eq:deltas}
\end{eqnarray}
to give three  CP even neutral weak eigen-states as 
\begin{eqnarray}
\left(
\begin{array}{c}
\rho_1 \\
\rho_2 \\
 \varphi^0 
\end{array}
\right)\simeq\left( 
\begin{array}{ccccc}
-\sin\alpha & &\cos\alpha & &\delta_{13} \\
\,\,\,\cos\alpha && \sin\alpha & &\delta_{23} \\
\delta_{13}\sin\alpha -\delta_{23} \cos\alpha && \,\, -\delta_{13}\cos\alpha -\delta_{23} \sin\alpha && 1
\end{array}
\right)
\left( 
\begin{array}{c}
h^0 \\
H^0 \\
S^0
\end{array}
\right);
\label{mixingmatrix}\end{eqnarray}
The mixing matrix given in equation \eqref{mixingmatrix} validates the orthogonality condition up to an order  $\lesssim\,\mathcal{O}\left(\delta_{13}^2,\delta_{23}^2,\delta_{13}\delta_{23} \right)$. $\xi_l^{h^0}$ is chosen to be  $\sim$ 1 in the alignment {\it i.e.} ($\beta - \alpha$) $\simeq \pi/2$. 
\par The spectrum of the model at the electro-weak scale is dominated by $V_{2- HDM}$.  The $V_{\varphi^0}$ and $V_{\rm portal}$ interactions are treated as perturbations.  After diagonalization of the scalar mass matrix, the masses of the physical  neutral scalars  are given by 
\begin{eqnarray}
m_{S^0}^2&\simeq& m_0^2+2\delta_{13}\, M_{13}^2 +2\delta_{23} \, M_{23}^2\\
m_{h^0,\,H^0}^2 &\simeq& \frac{1}{2} \left[ M_{11}^2 + M_{22}^2 \mp \sqrt{\left(M_{11}^2-M_{22}^2\right)^2 +4 \, M_{12}^4}\right]
\end{eqnarray}
where 
\begin{eqnarray}
M_{11}^2= m_{12}^2 \tan\beta + \lambda_1 v_0^2\cos^2\beta; &&
M_{12}^2= -\,m_{12}^2  + \left(\lambda_{3}+\lambda_{4}+\lambda_{5}\right)  v_0^2\cos\beta\,\sin\beta;\nonumber\\
M_{22}^2= m_{22}^2 \cot\beta + \lambda_2 v_0^2\sin^2\beta;&&
 M_{13}^2= v_0\,A_{12}\,\sin\beta; \,\,\, M_{23}^2 = v_0\,A_{12}\,\cos\beta;  
\end{eqnarray}
In the alignment limit,  one of the neutral CP even scalar $h^0$ $\approx$ 125 GeV is identified with the SM Higgs. 
\par The coefficients $m_0^2$, $m_{11}^2,\, m_{22}^2$ and $\lambda_i$ for $i=1,\cdot\cdot\cdot,5$ are explicitly defined in terms of the physical scalar masses, mixing angles $\alpha$ and $\beta$ and the free parameter $m_{12}^2$ and are given in  the Appendix \ref{model_pram}.   Terms associated with $A_{11}$ are proportional to $\cot\beta$ and therefore can be neglected as they are highly suppressed in the large $\tan\beta$ limit. Terms proportional to   $A_{22}$ are tightly constrained from the existing data at LHC on decay of a heavy exotic scalar to di-higgs  channel and therefore they are  dropped.
The coefficient $B$ is fixed by redefinition of the field $\varphi^0$  to avoid a non-zero VEV for itself.
\begin{table}
\begin{center}
\begin{tabular}{|c|c|c|c|c|c|} \hline
$\xi^\phi_{\psi}$/$\xi^\phi_V$ & $S^0$ & $h^0$ & $H^0$ &$A^0$ &$H^{\pm}$ \\ \hline
$\ell$ & $\delta_{13}/c_\beta$ & $-s_\alpha/c_\beta$ & $c_\alpha/c_\beta$ &$-s_\beta/c_\beta$ & $-s_\beta/c_\beta$ \\
$u_q$ & $\delta_{23}/s_\beta$ & $c_\alpha/s_\beta$ & $s_\alpha/s_\beta$ & $c_\beta/s_\beta$ & $c_\beta/s_\beta$
\\ 
$d_q$ & $\delta_{23}/s_\beta$ & $c_\alpha/s_\beta$ & $s_\alpha/s_\beta$ & $- c_\beta/s_\beta$ & $c_\beta/s_\beta$
\\
 $Z^0$/$W^{\pm}$ & $\delta_{13}c_\beta+\delta_{23}s_\beta$ & $s_{\left(\beta-\alpha\right)}$ & $c_{\left(\beta-\alpha\right)}$&- &- 
\\
 \hline
\end{tabular}
\caption{\small \em{Values of $\xi^\phi_{\psi}$ and $\xi^\phi_V$ for $\phi=S^0$, $h^0$, $H^0$, $A^0$ and $H^{\pm}$; $\psi=\ell$, $u_q$ and $d_q$; $V$= $W^\pm$ and $Z^0$ in the lepto-philic 2-HDM+$S^0$ model. These values coincide with couplings given in reference \cite{Batell:2016ove} in the  alignment limit {\it i.e.} ($\beta - \alpha$) $\simeq \pi/2$. In the table $s$ and $c$ stands for $\sin$ and $\cos$ respectively.}}
\label{tab:couplings}
\end{center}
\end{table}

\par The Yukawa couplings arising due to Higgs Doublets $\Phi_1$ and $\Phi_2$ in type-X 2-HDM is given by 
\begin{equation}
-{\cal L}_Y=\bar L Y_e\Phi_1 e_R+\bar Q Y_d\Phi_2 d_R+\bar Q Y_u\tilde\Phi_2 u_R+{\rm h.c.},
\label{eq:doubletyukawas}
\end{equation}
\be
m_e = \cos\beta \times \frac{Y_e v_o}{\sqrt{2}},~~ m_{u(d)} = \sin\beta \times \frac{Y_{u(d)} v_o}{\sqrt{2}}.
\ee
We re-write the Yukawa interactions of the physical neutral states  as
\begin{align}
\label{xi_psi}
-{\cal L}_Y&\supset\sum_{\phi\equiv S^0,h^0,H^0}\,\, \sum_{\psi=\ell,\,q}\xi^\phi_{\psi}\,\frac{m_\psi}{v_o}\,\,\phi\,\,\bar\psi\psi
\end{align}
\noindent The couplings $\xi_\psi ^\phi$  are given in the first three rows of table \ref{tab:couplings}. It is important to mention here that the Yukawa couplings are proportional to the fermion mass {\it i.e. non-universal} unlike the consideration in reference \cite{Agrawal:2014ufa}. 

\par The interaction of the neutral scalar mass eigenstates   with the weak gauge Bosons are given by
\begin{align}
\label{xi_V}
{\cal L}&\supset\sum_{\phi\equiv S^0, h^0, H^0}\,\frac{\phi}{v_o}\,\,
\left(2\,\,\xi^\phi_{W^\pm}\,\,m^2_{W^{\pm}}\,W^+_\mu {W^-}^\mu+\xi^\phi_{Z^0}\,\,\,\,m_{Z^0}^2\,{Z^0}_\mu {Z^0}^\mu\right).
\end{align}
The couplings $\xi_V ^\phi$   are given in the last row of table \ref{tab:couplings}.  It is to be noted that for $m_{H^0}>>m_{h^0}$,  the singlet scalar coupling with $Z^0$ Boson can be fairly approximated as 
$\simeq \delta_{23}\sin\beta$.
\par The recent precision measurements at LHC constrains $\left\vert \kappa_V\right\vert$ = 1.06$^{+0.10}_{-0.10}$ \cite{Aad:2015ona,pdg2018Higgs} (where $\kappa_V$ is the scale factor for the SM Higgs Boson coupling) to the vector Bosons  restricts  the generic 2HDM Models and its extension like the one in discussion  to comply with the alignment limit.  In this model, the Higgs Vector Boson coupling to gauge Bosons is identical  to that of  generic 2HDM model at tree level as the additional singlet scalar contributes to  $h^0 V V$ couplings only at the one loop level which is suppressed by  $\delta^2_{23}/\left(16\,\pi^2\right)$. 

\par The triple scalar couplings of the mass eigen states  are given in the Appendix \ref{TripleScalarCoupling}, some of which can be constrained from the observed Higgs decay width and exotic scalar Boson searches at LEP, TeVatron and LHC.
\section{Electro-Weak Constraints}
\label{section_ElectroWeak_Constraints}
\subsection{Anomalous Magnetic Moment of Muon}
\label{section Anomalous Magnetic Moment of muon}
 We begin our analysis by evaluating the parameter space allowed from 3 $\sigma$ discrepancy $a_{\mu^-}^{\rm exp}-a_{\mu^-}^\text{SM}\,\equiv \Delta a_{\mu}  =\, 268(63) \,\times 10^{-11}$ \cite{pdg2018-muon}.
 In lepto-philic 2-HDM + singlet {\it scalar portal} model  all five additional scalars $S^0,\,H^0,\,A^0,\, H^{\pm}$ couple to leptons with the coupling strengths  given in table \ref{tab:couplings} and  thus give contributions to $\Delta a_\mu$ at the one-loop level  and are expressed as:
 \begin{eqnarray}
    \Delta a_{\mu} &=& \Delta a_{\mu}\arrowvert_{S^0} +\Delta a_{\mu}\arrowvert_{H^0} +\Delta a_{\mu}\arrowvert_{A^0} +\Delta a_{\mu}\arrowvert_{H^{\pm}} \label{amu}
	\nonumber\\
   &=& \frac{1}{8 \pi^2} \frac{m^2_\mu}{v_o^2} \,\tan^2\beta\left[\delta_{13}^2\,\, I_{S^0} +  I_{H^0}+   I_{A^0} +    \,I_{H^\pm}\right].
\end{eqnarray}
Here $I_{i}$ are the integrals given as 
\begin{eqnarray}	 
        I_{S^0,\, H^0}  = \int^1_0 dz\frac{\left( 1+z \right)\left( 1-z \right)^2}
	{\left( 1-z \right)^2+z\,r_{{S^0},\, H^0}^{-2}};\,\,\,&&I_{A^0} =
	-\,\int^1_0dz\frac{z^3}
	{r_{A^0}^{-2}\left( 1-z \right)+z^2}; \,\, \,{\rm and}\nonumber \\
		I_{H^\pm} =
	\int^1_0 dz\frac{z( 1-z) }
{\left( 1-z \right)-r_{H^\pm}^{-2}} \,\,\, &&{\rm with} \,\,\, r_{i}\equiv\frac{m_l}{m_i} \,\, {\rm for } \,\, i\equiv S^0, \,H^0, \,A^0, \,H^\pm .
\end{eqnarray}

\begin{figure}[h!]
\centering 
\begin{subfigure}{.49\textwidth}\centering
  \includegraphics[width=\columnwidth]{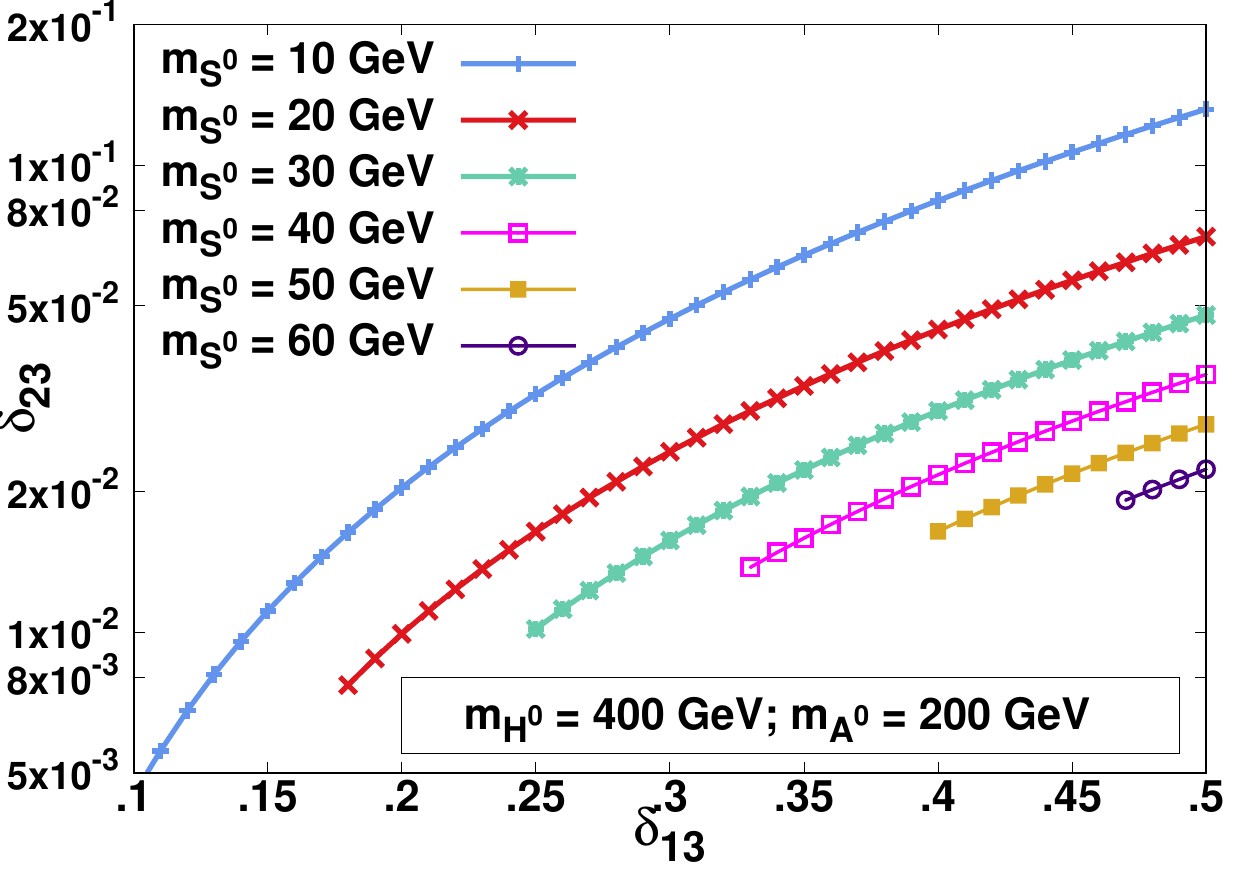}
  \caption{}
  \label{fig:d13_Vs_d23a}
\end{subfigure}%
\begin{subfigure}{.49\textwidth}\centering
  \includegraphics[width=\columnwidth]{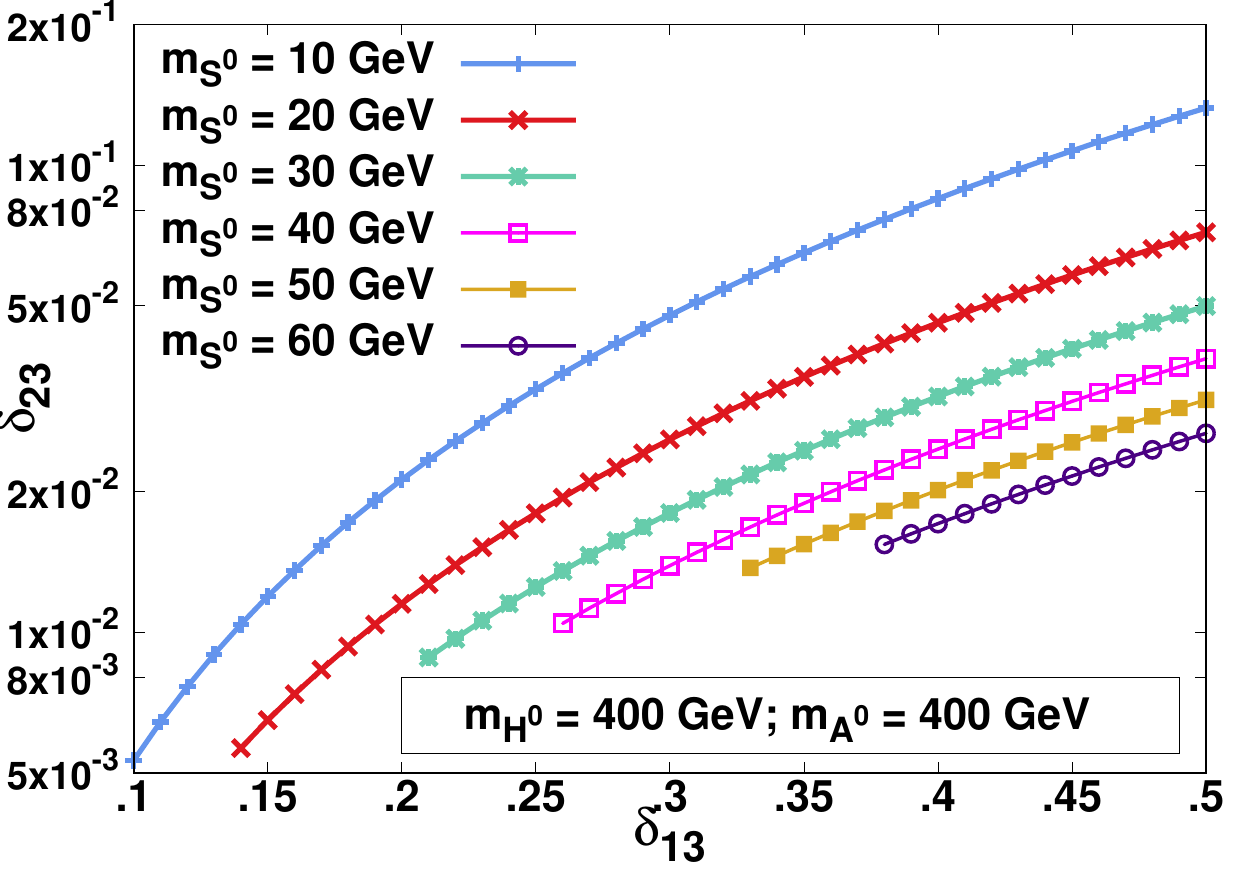}
  \caption{}
  \label{fig:d13_Vs_d23b}
\end{subfigure}%
\\
\begin{subfigure}{.49\textwidth}\centering
  \includegraphics[width=\columnwidth]{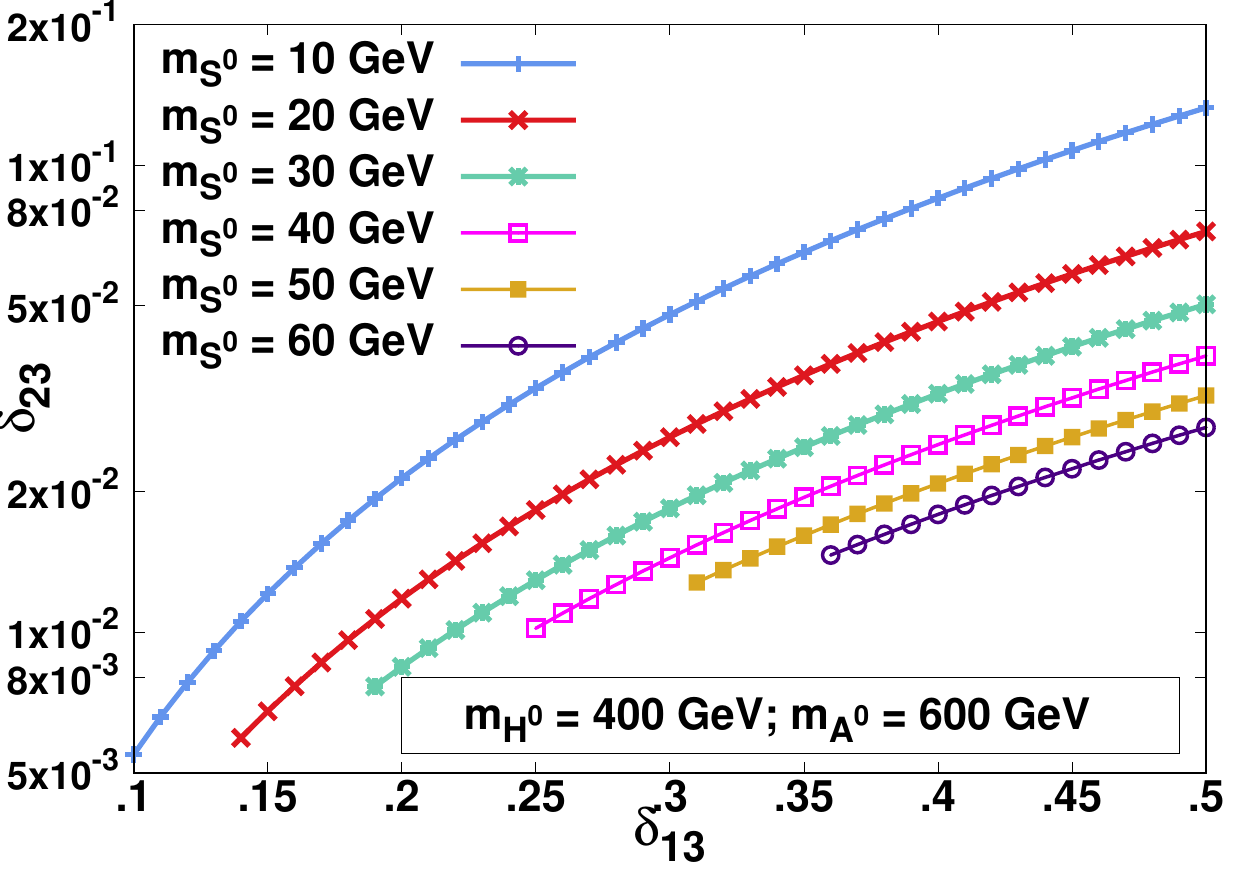}
  \caption{}
  \label{fig:d13_Vs_d23c}
\end{subfigure}%
\begin{subfigure}{.49\textwidth}\centering
  \includegraphics[width=\columnwidth]{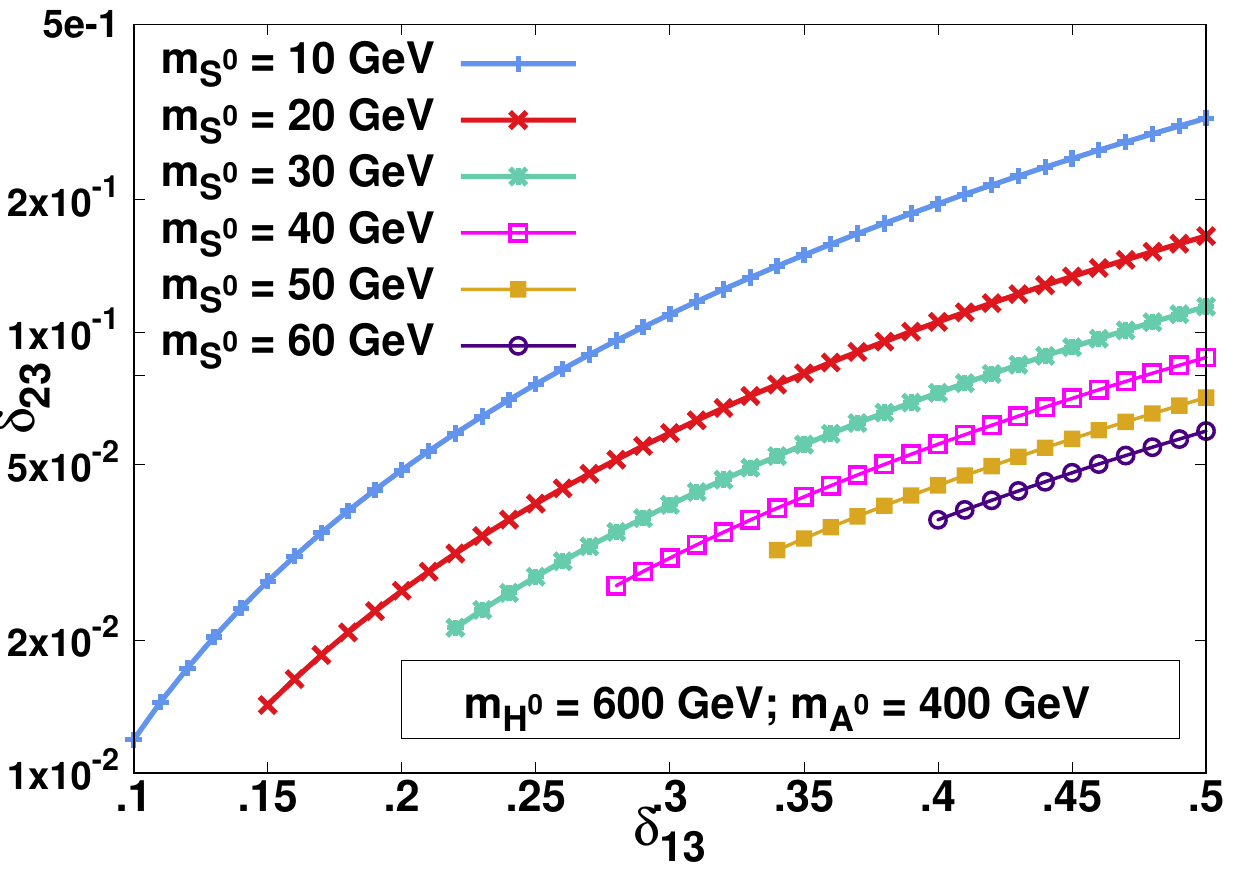}
  \caption{}
  \label{fig:d13_Vs_d23d}
\end{subfigure}%
\caption{\small \em{ Figures \ref{fig:d13_Vs_d23a}, \ref{fig:d13_Vs_d23b}, \ref{fig:d13_Vs_d23c} and \ref{fig:d13_Vs_d23d} show contours on the $\delta_{13}$ - $\delta_{23}$ plane satisfying $\Delta a_{\mu}=\,268(63)\,\times10^{-11}$ corresponding to four different combinations of $m_{H^0}$ and $m_{A^0}$ respectively. In each panel six contours   are depicted  corresponding to six choices of singlet masses 10, 20, 30, 40, 50 and 60 GeV respectively. }}
\label{fig:d13_Vs_d23}
\end{figure}

\par We observe  that in the limit $r_i<< 1$ the  charged scalar integral $I_{H^\pm}$ is suppressed by 2-3 orders of magnitude in comparison to the other integrals for the masses of the scalars varying between 150 GeV $\sim$ 1.6 TeV. Since the present lower bound on the charged Higgs mass from its searches at LHC is 600 GeV \cite{Aad:2015typ}, we can   neglect its contribution to the $\Delta a_{\mu}$ in our calculations.  

It is also important to note that the one loop  contribution from the pseudo-scalar integral $I_{A^0}$ is opposite in sign to that of  the other neutral scalars $I_{H^0,\,h^0,\,S^0}$, while  at the level of  two loops the  Barr-Zee diagrams \cite{Ilisie:2015tra}, it gives positive contribution to $\Delta a_\mu$ which may be sizable for low pseudo-scalar mass $m_{A^0}$ because of large value of the coupling $\xi^{A^0}_l$. However, for heavy $A^0$ and $H^0$ considered here, we can safely neglect the two loop contributions.

\par In the alignment limit  the mixing angle  $\delta_{23}\simeq\frac{\delta_{13}m^2_{H^0}}{m_{h^0}^2}\left[2-\frac{m_{h^0}^2}{m_{H^0}^2}\right]\cot\beta$ is fixed by constrains from $\Delta a_{\mu}$ and choice of $\delta_{13}$ and neutral CP-even scalar masses. To understand the model  we study the correlation of the two mixing parameters $\delta_{13}$ and $\delta_{23}$ satisfying the $\Delta a_\mu$ for a given set of input masses of the physical scalars and show four  correlation plots in figures \ref{fig:d13_Vs_d23a}, \ref{fig:d13_Vs_d23b},  \ref{fig:d13_Vs_d23c} and \ref{fig:d13_Vs_d23d}  for varying $\delta_{13}$.   We find that $\delta_{23}$ remains small enough for all the parameter space in order to fulfill the small angle approximation. We have chosen six   singlet scalar masses 10, 20, 30, 40, 50, and 60 GeV.   In each panel $m_{H^0}$ and $m_{A^0}$ are kept  fixed at values, namely (a) $m_{H^0}$ = 400 GeV, $m_{A^0}$ = 200 GeV, (b) $m_{H^0}$ = 400 GeV, $m_{A^0}$ = 400 GeV, (c) $m_{H^0}$ = 400 GeV, $m_{A^0}$ = 600 GeV, and (d) $m_{H^0}$ = 600 GeV, $m_{A^0}$ = 400 GeV. We find that relatively larger values of $\delta_{13}$ are required with the increase in scalar mass $m_{S^0}$. Increase in the pseudo-scalar mass $m_{A^0}$  for fixed $m_{H^0}$ results in the lower value of $\delta_{13}$ required to obtain the observed $\Delta a_{\mu}$.

\par On imposing the  perturbativity constraints  on the Yukawa coupling $\xi^{H^0}_\tau\equiv \tan\beta \, m_\tau /v_0$  involving the $\tau^\pm$ and $H^0$, we compute the upper bound on  the model parameter $\tan\beta\lesssim$ 485. As a consequence, we observe that the  values of $\delta_{23}$ also gets restricted for each variation curve  exhibited in figures \ref{fig:d13_Vs_d23a}, \ref{fig:d13_Vs_d23b}, \ref{fig:d13_Vs_d23c} and \ref{fig:d13_Vs_d23d}.

\begin{figure}[h!]
\centering 
\begin{subfigure}{.5\textwidth}\centering
  \includegraphics[width=\columnwidth]{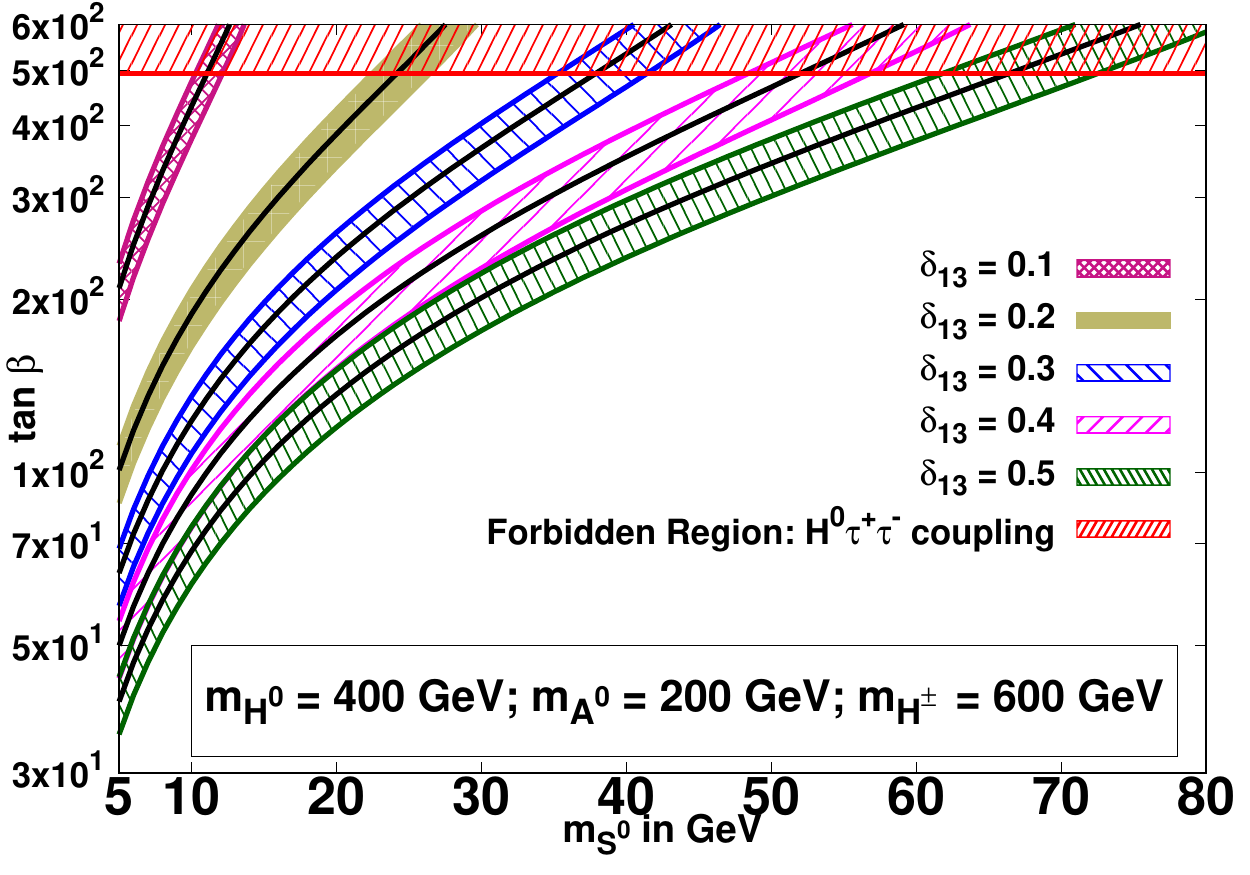}
  \caption{}
  \label{fig:tanbeta_Vs_ mSa}
\end{subfigure}%
\begin{subfigure}{.5\textwidth}\centering
  \includegraphics[width=\columnwidth]{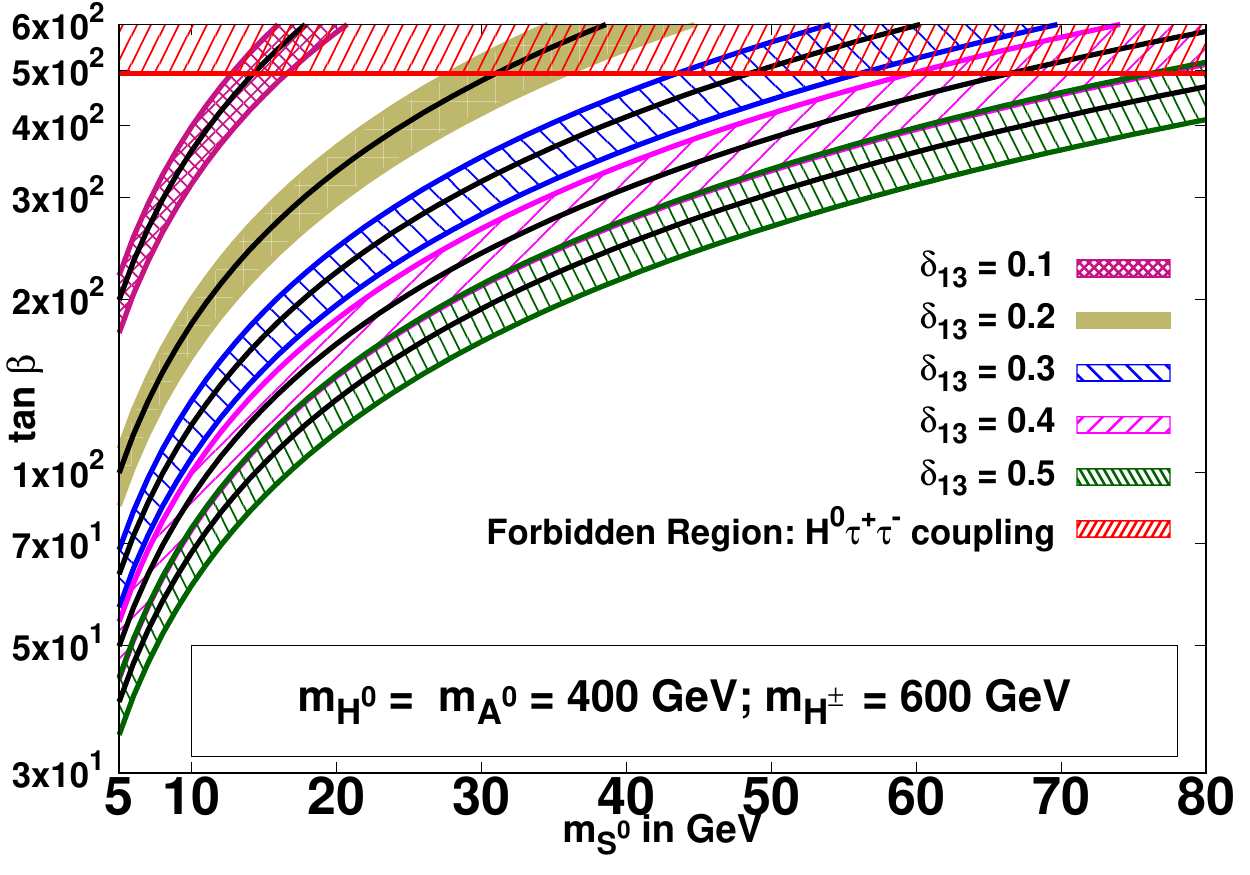}
  \caption{}
  \label{fig:tanbeta_Vs_ mSb}
\end{subfigure}%
\\
\begin{subfigure}{.5\textwidth}\centering
  \includegraphics[width=\columnwidth]{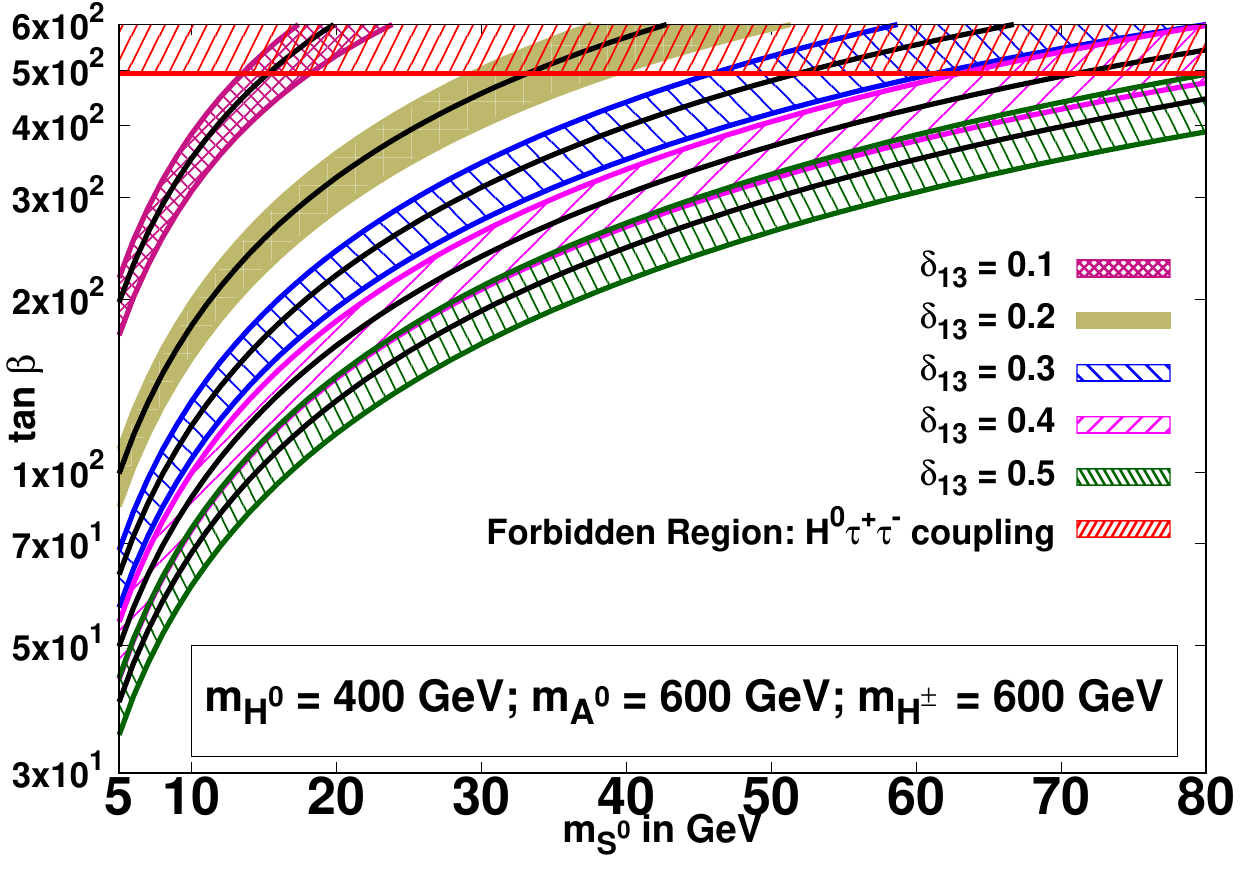}
  \caption{}
  \label{fig:tanbeta_Vs_ mSc}
  \end{subfigure}%
  \begin{subfigure}{.5\textwidth}\centering
  \includegraphics[width=\columnwidth]{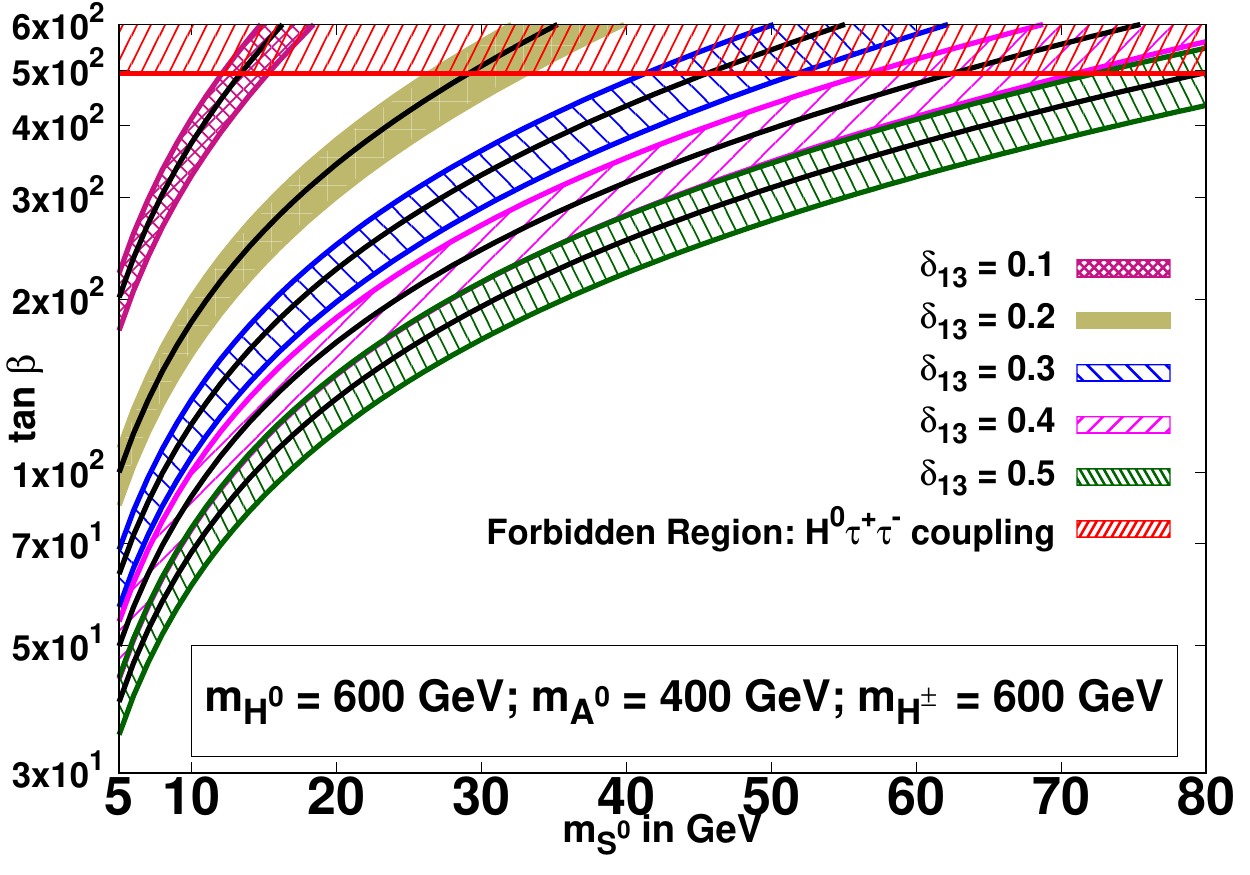}
  \caption{}
  \label{fig:tanbeta_Vs_mSd}
\end{subfigure}%
\caption{\small \em{ Figures \ref{fig:tanbeta_Vs_ mSa}, \ref{fig:tanbeta_Vs_ mSb}, \ref{fig:tanbeta_Vs_ mSc} and \ref{fig:tanbeta_Vs_mSd} show contours on  $m_{S^0}$  - $\tan\beta$ plane satisfying $\Delta a_{\mu}=\,268(63)\,\times10^{-11}$ for fixed $m_{H^\pm}$ = 600 GeV and four different combinations of $m_{H^0}$ and $m_{A^0}$ as shown. In each panel, five contours along-with shaded   one $\sigma$ bands of $\Delta a_{\mu}$  are depicted  corresponding to five choices of $\delta_{13}$ 0.1, 0.2, 0.3, 0.4 and  0.5 respectively. The top horizontal  band (shaded in red) in each panel shows the forbidden region on $\tan\beta$ due  to the perturbativity constraint on   the upper limit of $H^0 \tau^+ \tau^-$ coupling.}}
\label{fig:tanbeta_Vs_ mS}
\end{figure}

The contours satisfying $\Delta a_{\mu}=\,268(63)\,\times10^{-11}$ on 
$m_{S^0}$  - $\tan\beta$ plane for fixed charged Higgs mass $m_{H^\pm}$ = 600 GeV  are shown for four different combinations of heavy neutral Higgs mass and pseudo-scalar Higgs mass namely (a) $m_{H^0}$ = 400 GeV, $m_{A^0}$ = 200 GeV, (b) $m_{H^0}$ = 400 GeV, $m_{A^0}$ = 400 GeV, (c) $m_{H^0}$ = 400 GeV, $m_{A^0}$ = 600 GeV, and (d) $m_{H^0}$ = 600 GeV, $m_{A^0}$ = 400 GeV respectively in figures \ref{fig:tanbeta_Vs_ mSa}, \ref{fig:tanbeta_Vs_ mSb}, \ref{fig:tanbeta_Vs_ mSc} and \ref{fig:tanbeta_Vs_mSd}. In each panel the five  shaded regions, correspond to five choices of mixing angle $\delta_{13}$ = 0.1, 02, 0.3, 0.4 and 0.5 respectively depict the  3 $\sigma$ allowed regions  for  the discrepancy in $\Delta a_{\mu}$ around  its central value shown by the black lines. The horizontal band appearing at the top in all these panels shows the forbidden region on account of the perturbativity constraint on the upper limit of $H^0\tau^+\tau^-$ coupling as discussed above. 
\par  As expected the allowed value of $\tan\beta$ increases with the increasing singlet scalar  mass $m_{S^0}$ and  decreasing mixing angle $\delta_{13}$. We find that a very narrow region of the  singlet scalar mass is allowed by $\Delta a_{\mu}$ corresponding to  $\delta_{13}\leq 0.1 $. 
\subsection{LEP and $\Delta a_e$ Constraints}
Searches for the light neutral Bosons were explored in the Higgs associated vector Boson production channels  at LEP \cite{Abdallah:2004wy}. We consider the {\it s-channel} bremsstrahlung  process $e^+ e^- \rightarrow  Z^0/\gamma^0 + h^0 \rightarrow\tau^+ \tau^- \tau^+ \tau^-$ whose production cross-section  can be expressed in terms of the SM $h^0Z^0$ production cross-section and given as   
\begin{eqnarray}
\sigma_{e^+e^-\to S^0Z^0\to \tau^+\tau^- Z^0}&=&\sigma_{e^+e^-\to h^0Z^0}^{\rm SM} \times \left\vert \frac{\xi^{S^0}_{Z^0}}{\xi^{h^0}_{Z^0}} \right\vert^2\times {\rm BR}\left(S^0\to\tau^+\tau^-\right)\nonumber\\
&\equiv& \sigma_{e^+e^-\to h^0Z^0}^{\rm SM} \times \left\vert\frac{\delta_{13}c_\beta+\delta_{23}s_\beta}{\sin\left(\beta-\alpha\right)}\right\vert^2 \times {\rm BR}\left(S^0\to\tau^+\tau^-\right).
\end{eqnarray}
Since the BR$\left(S^0\to\tau^+\tau^-\right)\simeq$ 1, we can compute the exclusion limit on the upper bound on $\left\vert\xi^{S^0}_{Z^0}\right\vert\equiv \left\vert\delta_{13}c_\beta+\delta_{23}s_\beta\right\vert$  from the LEP experimental data \cite{Abdallah:2004wy}, which   are shown in table  \ref{table2} for some chosen values of singlet scalar masses in the alignment limit.

\par 
A light neutral Vector mediator ${Z^\prime}^0$ has also been extensively searched at LEP \cite{Schael:2013ita}. Vector mediator ${Z^\prime}^0$ of mass $\leq$ 209 GeV is ruled out for coupling to muons $\gtrsim$ 0.01 \cite{Agrawal:2014ufa}. Assuming the same production cross-section corresponding to a light scalar mediator, the constraint on vector coupling can be translated to scalar coupling by multiplying a factor of $\sqrt{2}$.  For the case of {\it non-universal} couplings where the  scalar couples to the leptons with the strength proportional to its mass as is the case in our model, a further  factor of $\sqrt{\frac{m_\mu}{m_e}}$ is multiplied.  We therefore find the upper limit on the Yukawa coupling for leptons to be $\xi^{S^0}_{l} \frac{m_l}{v_0}\lesssim$ 0.2.

\begin{table}
\begin{center}
\begin{tabular}{|c|c|c|c|c|c|c|c|c|c|c|c|c|} \hline
{$m_{S^0}$(GeV)} & 12 & 15 & 20 & 25 & 30 & 35 & 40 & 45 & 50 & 55 & 60 & 65
\\
 \hline
$\left\vert\xi^{S^0}_{Z^0}\right\vert\lesssim$ & .285 & .316 & .398 & .530 & .751 & 1.132 & 1.028 & .457 & .260 & .199 & .169 & .093
\\
\hline
\end{tabular}
\caption{\small \em{Upper limits on $\left\vert\xi^{S^0}_{Z^0}\right\vert$ from bremsstrahlung process $e^+e^-\to S^0Z^0\to \tau^+\tau^-\tau^+\tau^-$ LEP data \cite{Abdallah:2004wy}}}
\label{table2}
\end{center}
\end{table}

From the constrained parameter space of the model explaining the muon $\Delta a_\mu$, we find that the total contribution to anomalous magnetic moment of the electron comes out  $\sim 10^{-15}$. This is  two order smaller in the   magnitude than the  error in the measurement of $ a_e \simeq \pm 2.6\times 10^{-13}$ \cite{pdg2018-electron}. The present model is thus capable of accounting for the observed experimental discrepancy in the $\Delta a_\mu$ without  transgressing the allowed $\Delta a_e$.
 \begin{figure}[h!]
\centering 
\begin{subfigure}{.5\textwidth}
\centering
  \includegraphics[width=\columnwidth]{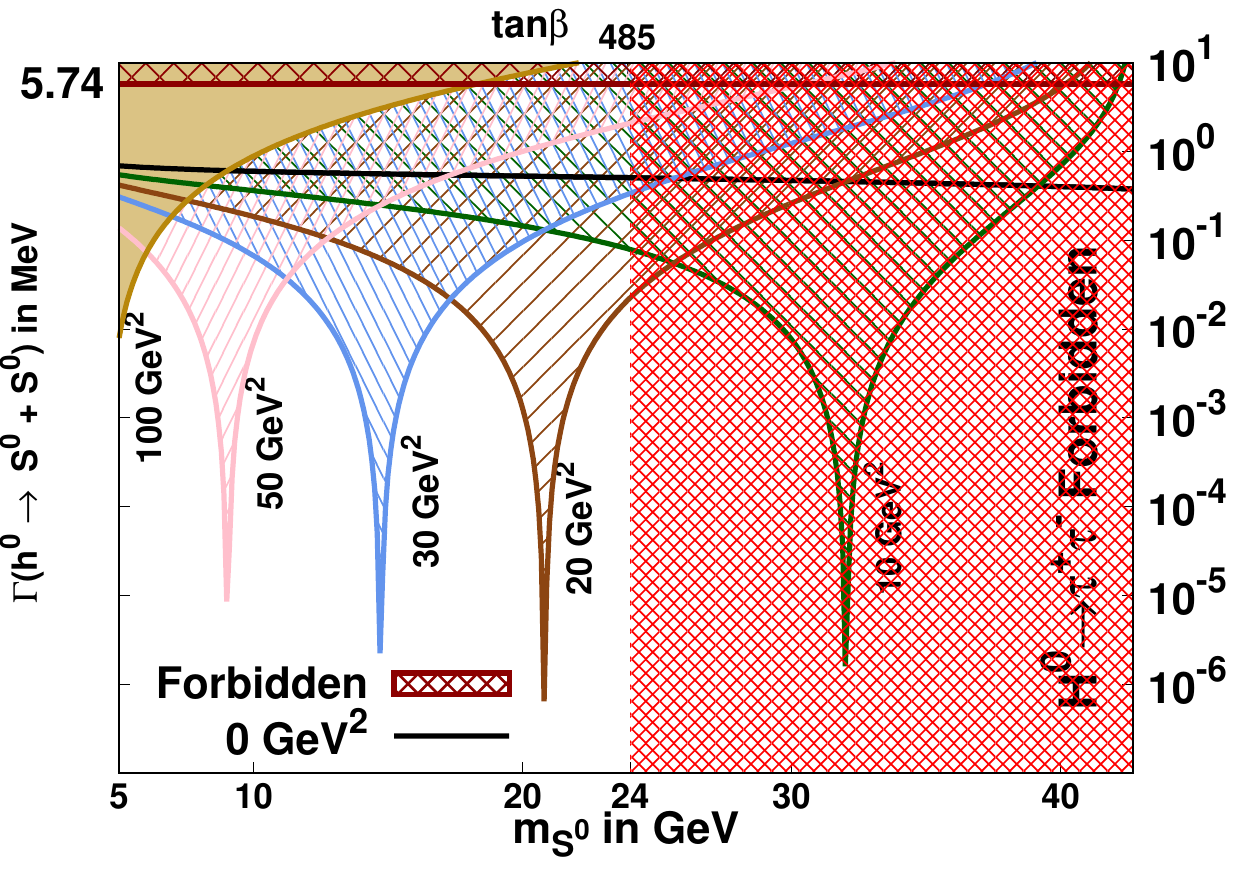}
  \caption{$m_{H^0}$ = 400 GeV; $m_{A^0}$ = 200 GeV}
  \label{fig:Higgs_Decay_Widtha_d13_p2}
\end{subfigure}%
\begin{subfigure}{.5\textwidth}\centering
  \includegraphics[width=\columnwidth]{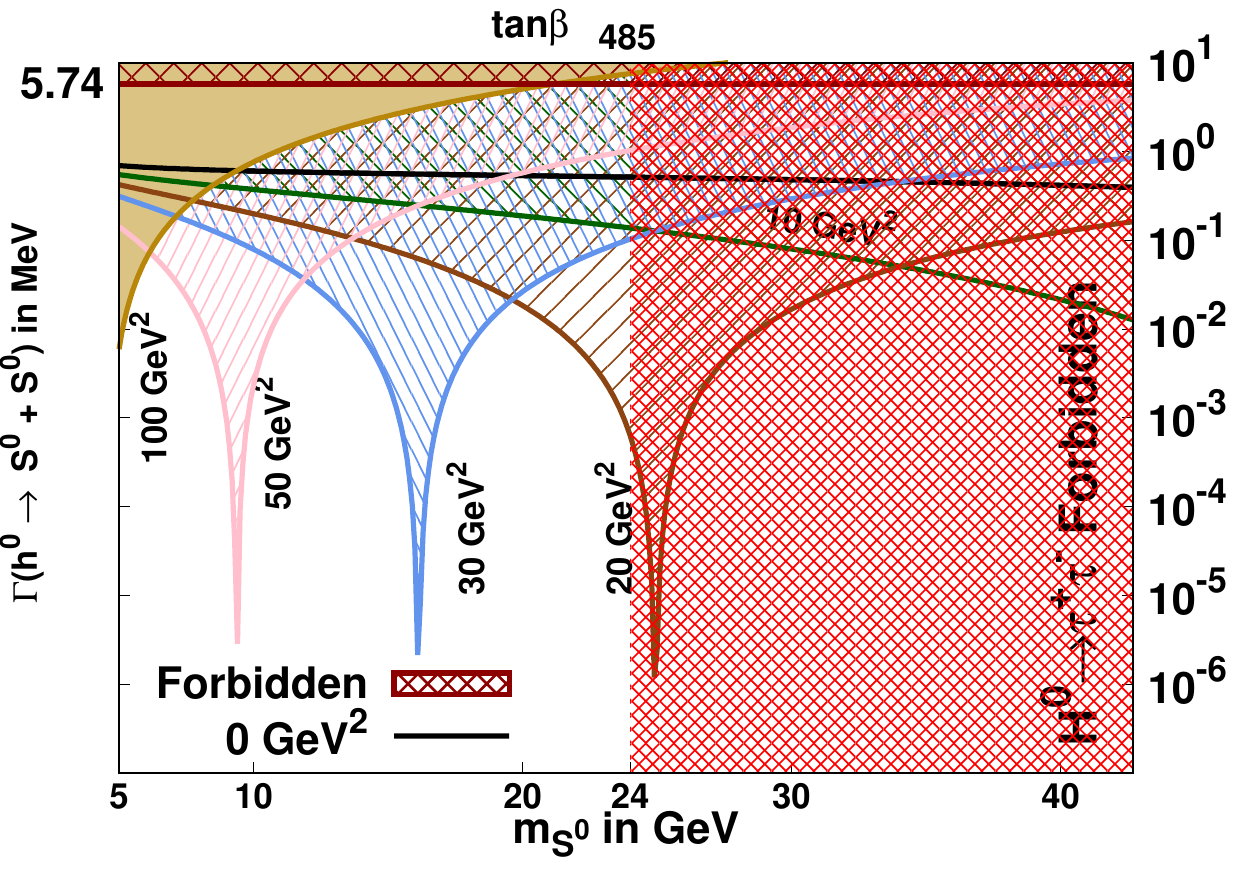}
   \caption{$m_{H^0}$ = 400 GeV; $m_{A^0}$ = 400 GeV}
  \label{fig:Higgs_Decay_Widthb_d13_p2}
\end{subfigure}%
\\
\begin{subfigure}{.5\textwidth}\centering
  \includegraphics[width=\columnwidth]{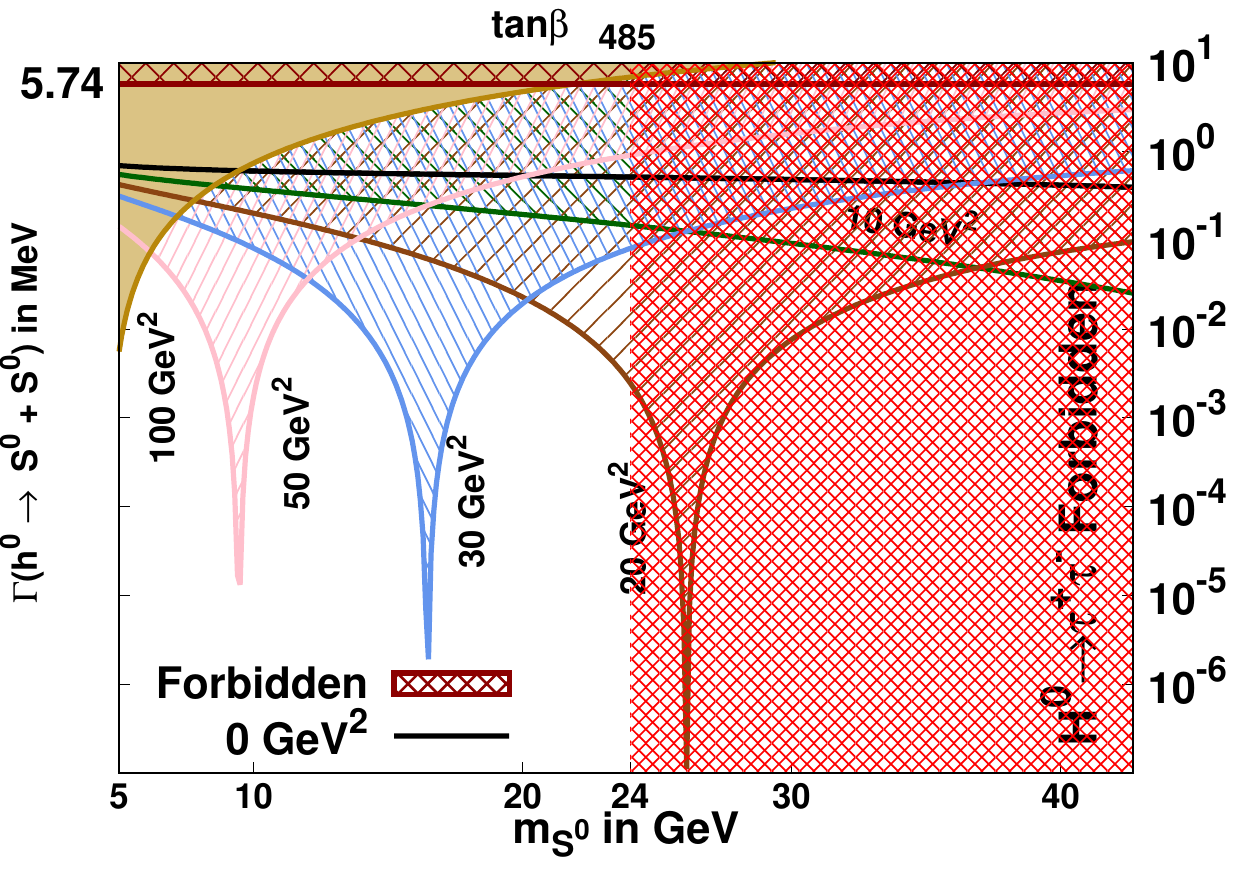}
  \caption{$m_{H^0}$ = 400 GeV; $m_{A^0}$ = 600 GeV}
  \label{fig:Higgs_Decay_Widthc_d13_p2}
  \end{subfigure}%
  \begin{subfigure}{.5\textwidth}\centering
  \includegraphics[width=\columnwidth]{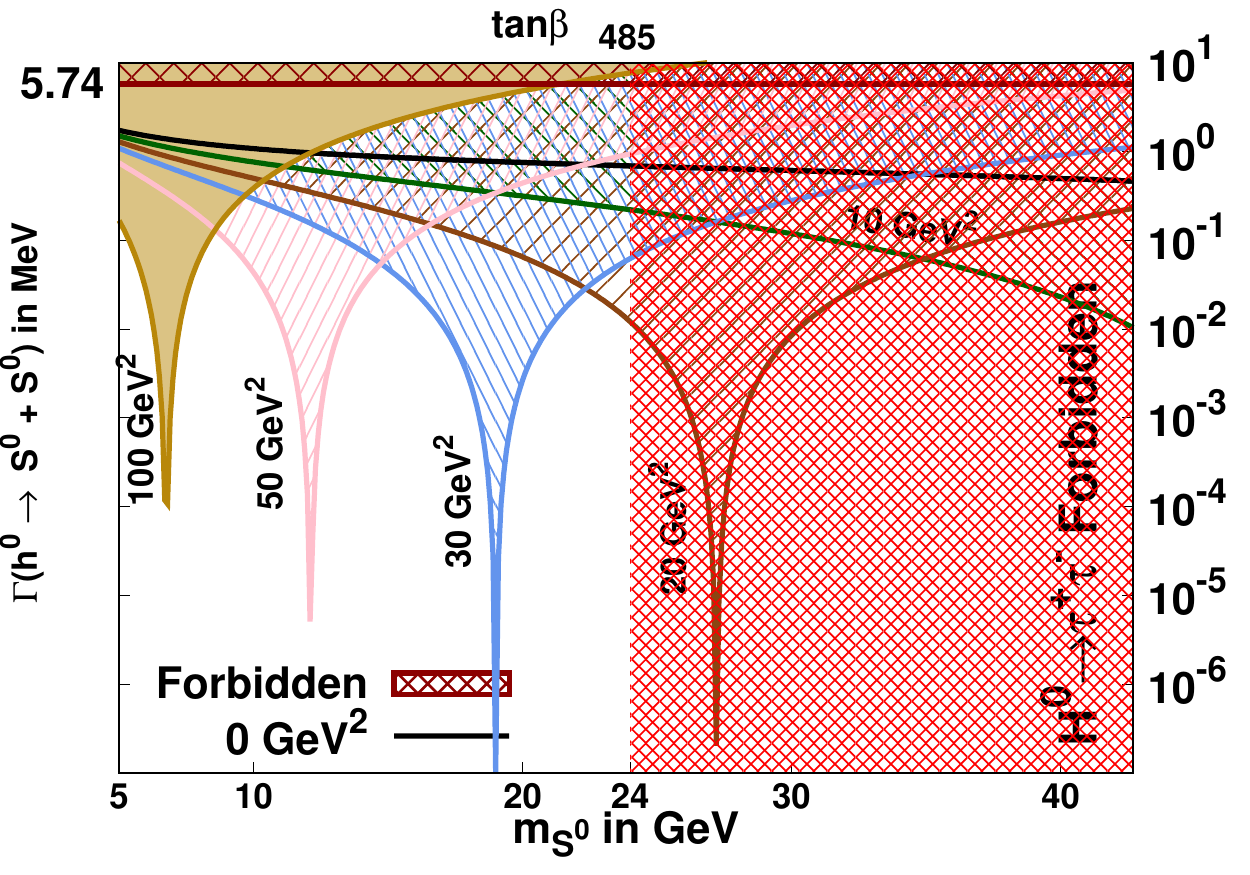}
   \caption{$m_{H^0}$ = 600 GeV; $m_{A^0}$ = 400 GeV}
  \label{fig:Higgs_Decay_Widthd_d13_p2}
\end{subfigure}%
 \caption{\small \em{ Figures \ref{fig:Higgs_Decay_Widtha_d13_p2}, \ref{fig:Higgs_Decay_Widthb_d13_p2}, \ref{fig:Higgs_Decay_Widthc_d13_p2} and \ref{fig:Higgs_Decay_Widthd_d13_p2} show $\Gamma_{h^0\to S^0S^0}$ variation with the $m_{S^0}$ for $m_{H^\pm}$ = 600 GeV, $\delta_{13}$ = 0.2 and four different combinations of $m_{H^0}$ and $m_{A^0}$ respectively. In each panel we shade five   regions   corresponding to  $m_{12}^2$ = 10, 20, 30, 50, 100 GeV$^2$ respectively. All  points on the solid curves  satisfy the discrepancy $\Delta a_{\mu}=\,268(63)\,\times10^{-11}$  and their corresponding values of $\tan\beta$ are shown in the upper x-axis of all the panels. We plot the contour corresponding to $m_{12}^2$ = 0 GeV$^2$ in black. The top horizontal band is   forbidden from the measurement of the total Higgs decay width at LHC. The red shaded region at the right in each  panel is forbidden due to non-perturbativity of $H^0\tau^+\tau^-$ coupling.}}

 \label{fig:Higgs_Decay_Widthp2}
\end{figure}
 \begin{figure}[h!]
\centering 
\begin{subfigure}{.5\textwidth}
\centering
  \includegraphics[width=\columnwidth]{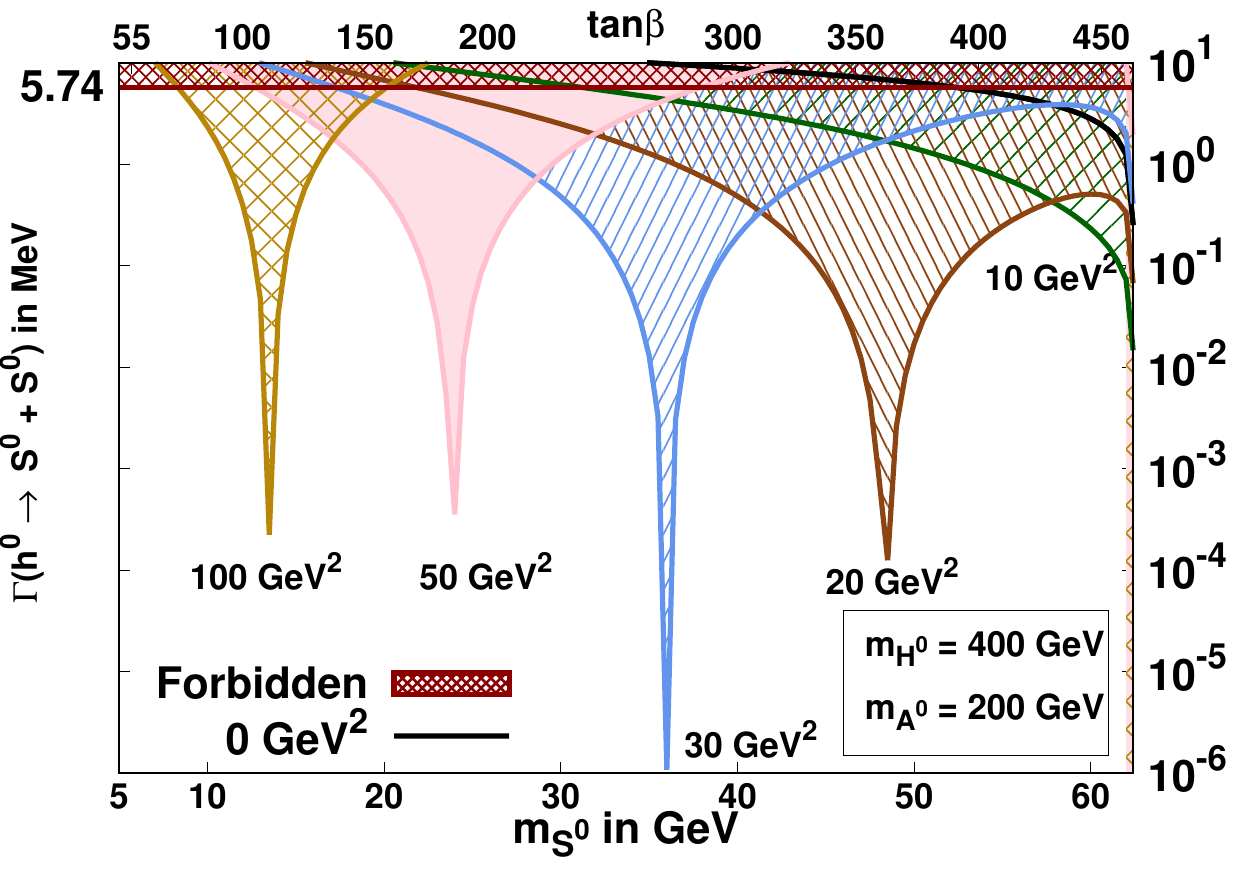}
  \caption{}
  \label{fig:Higgs_Decay_Widtha}
\end{subfigure}%
\begin{subfigure}{.5\textwidth}\centering
  \includegraphics[width=\columnwidth]{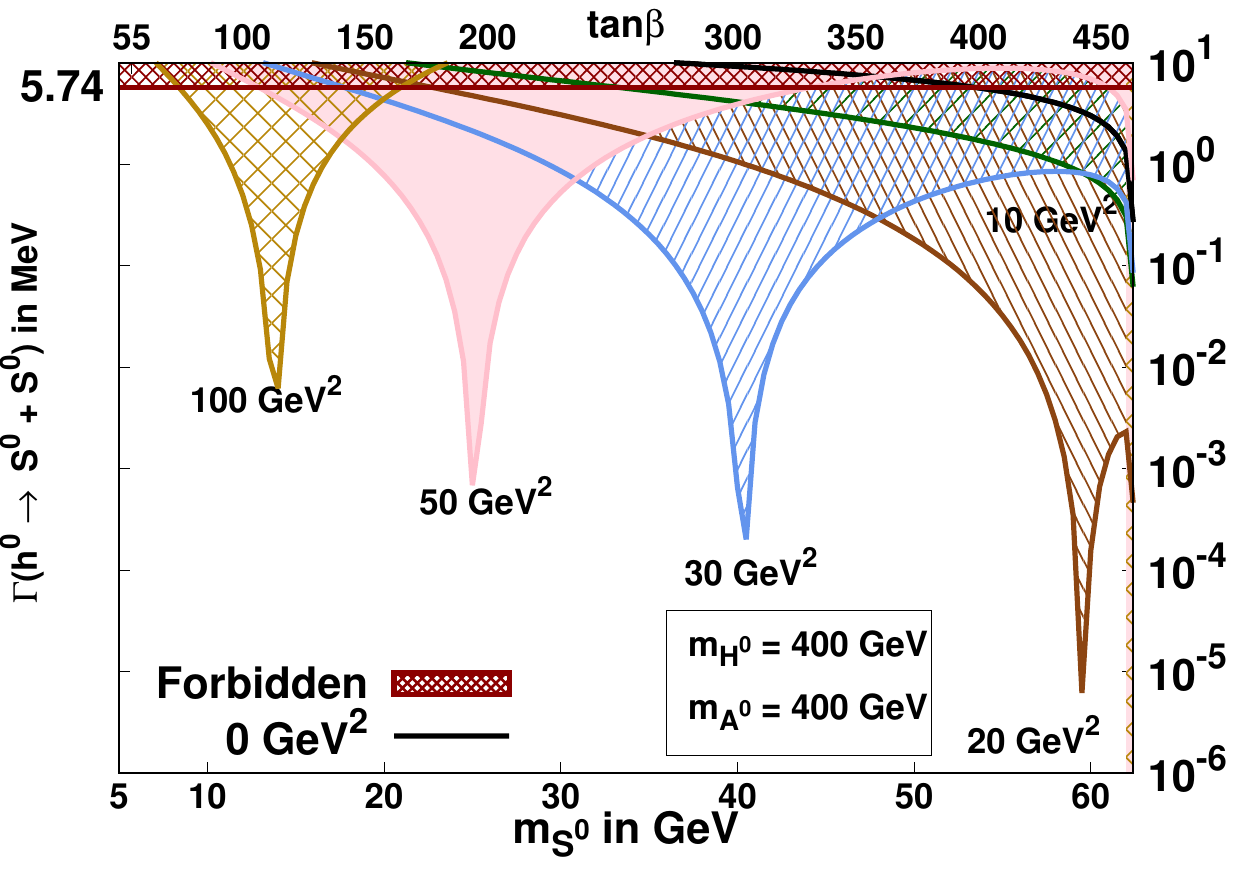}
  \caption{}
  \label{fig:Higgs_Decay_Widthb}
\end{subfigure}%
\\
\begin{subfigure}{.5\textwidth}\centering
  \includegraphics[width=\columnwidth]{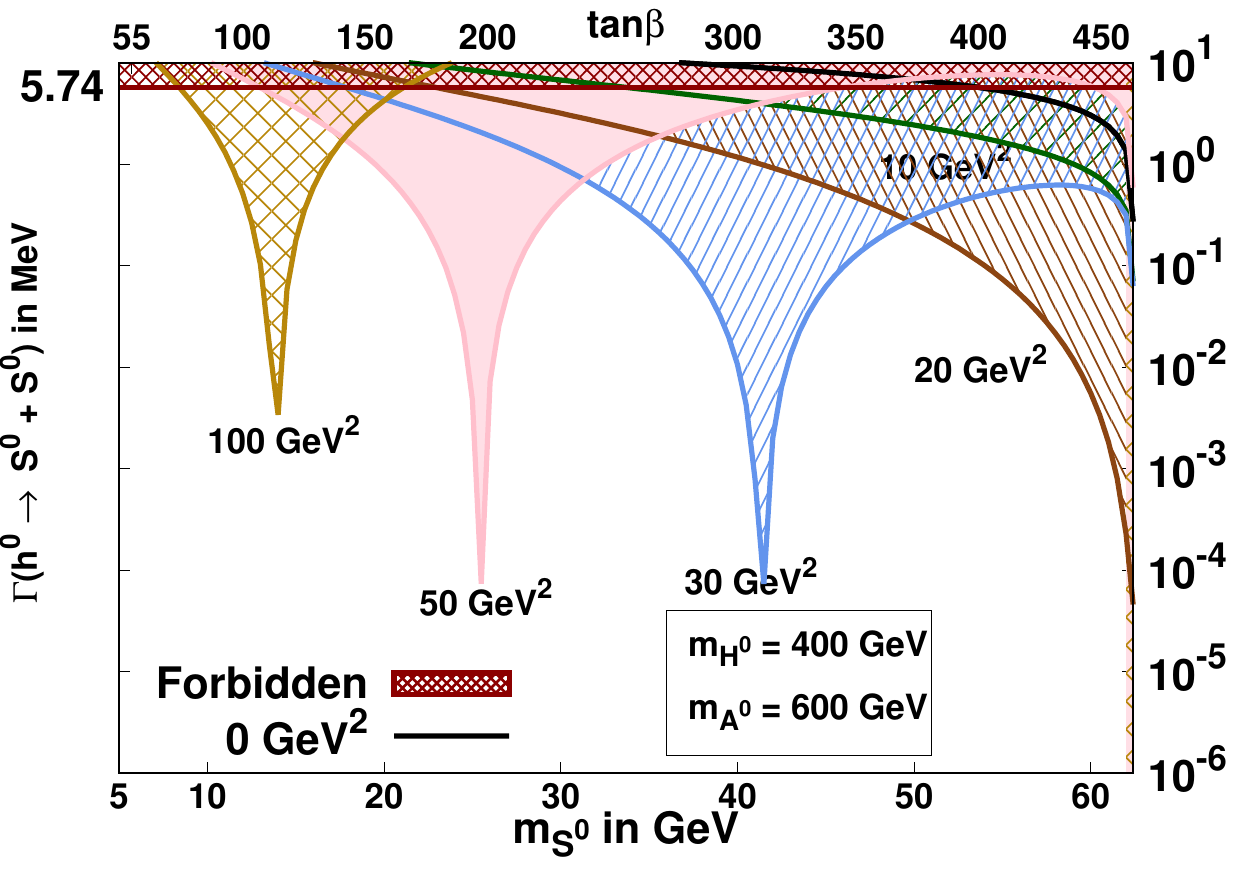}
  \caption{}
  \label{fig:Higgs_Decay_Widthc}
  \end{subfigure}%
  \begin{subfigure}{.5\textwidth}\centering
  \includegraphics[width=\columnwidth]{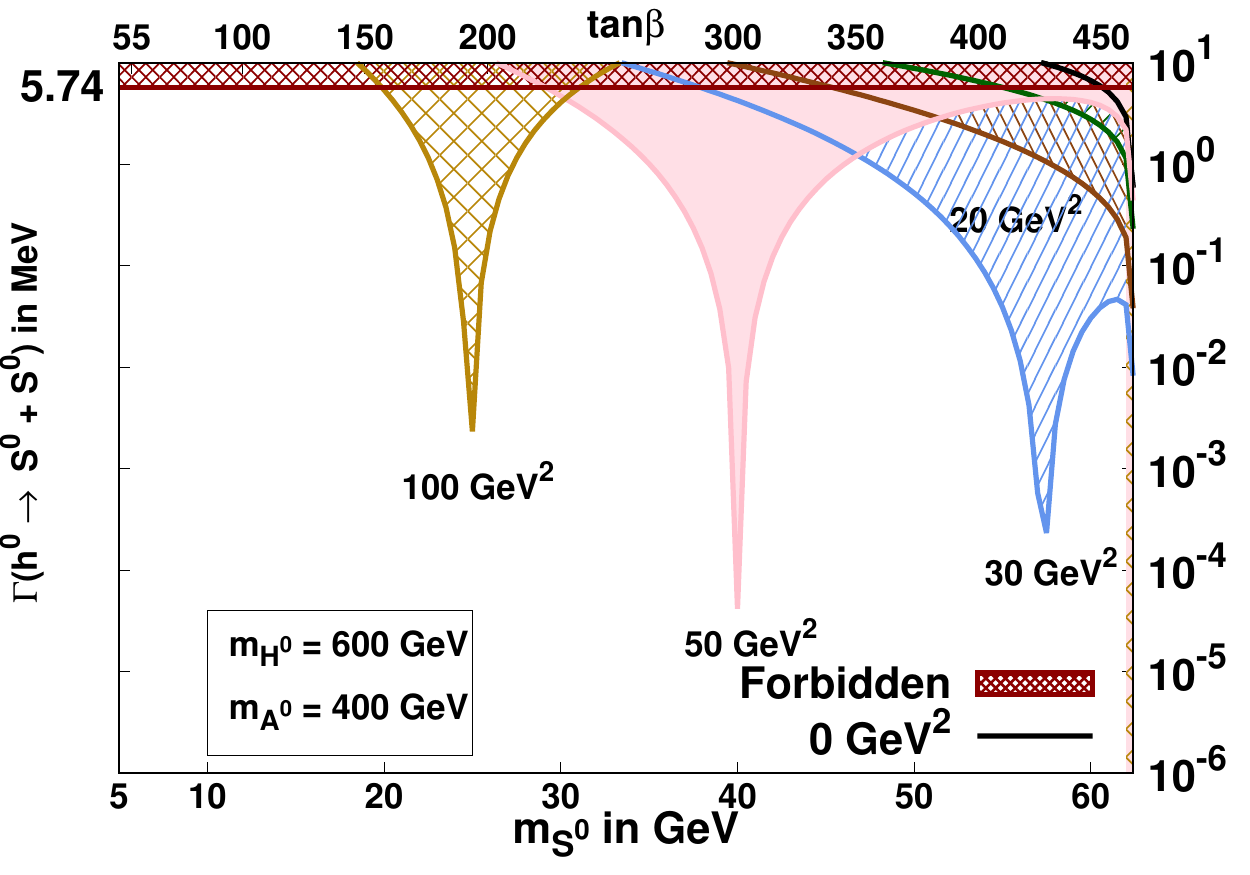}
  \caption{}
  \label{fig:Higgs_Decay_Widthd}
\end{subfigure}%
 \caption{\small \em{ Figures \ref{fig:Higgs_Decay_Widtha}, \ref{fig:Higgs_Decay_Widthb}, \ref{fig:Higgs_Decay_Widthc} and \ref{fig:Higgs_Decay_Widthd} show $\Gamma_{h^0\to S^0S^0}$ variation with the $m_{S^0}$ for $m_{H^\pm}$ = 600 GeV, $\delta_{13}$ = 0.4 and four different combinations of $m_{H^0}$ and $m_{A^0}$ respectively. In each panel we shade five   regions   corresponding to  $m_{12}^2$ = 10, 20, 30, 50, 100 GeV$^2$ respectively. All  points on the solid curves  satisfy the discrepancy $\Delta a_{\mu}=\,268(63)\,\times10^{-11}$  and their corresponding values of $\tan\beta$ are shown in the upper x-axis of all the panels. We plot the contour corresponding to $m_{12}^2$ = 0 GeV$^2$ in black. The top horizontal band is   forbidden from the measurement of the total Higgs decay width at LHC.}}
 \label{fig:Higgs_Decay_Width}
\end{figure}

 \begin{figure}[h!]
\centering 
\begin{subfigure}{.5\textwidth}\centering
  \includegraphics[width=\columnwidth]{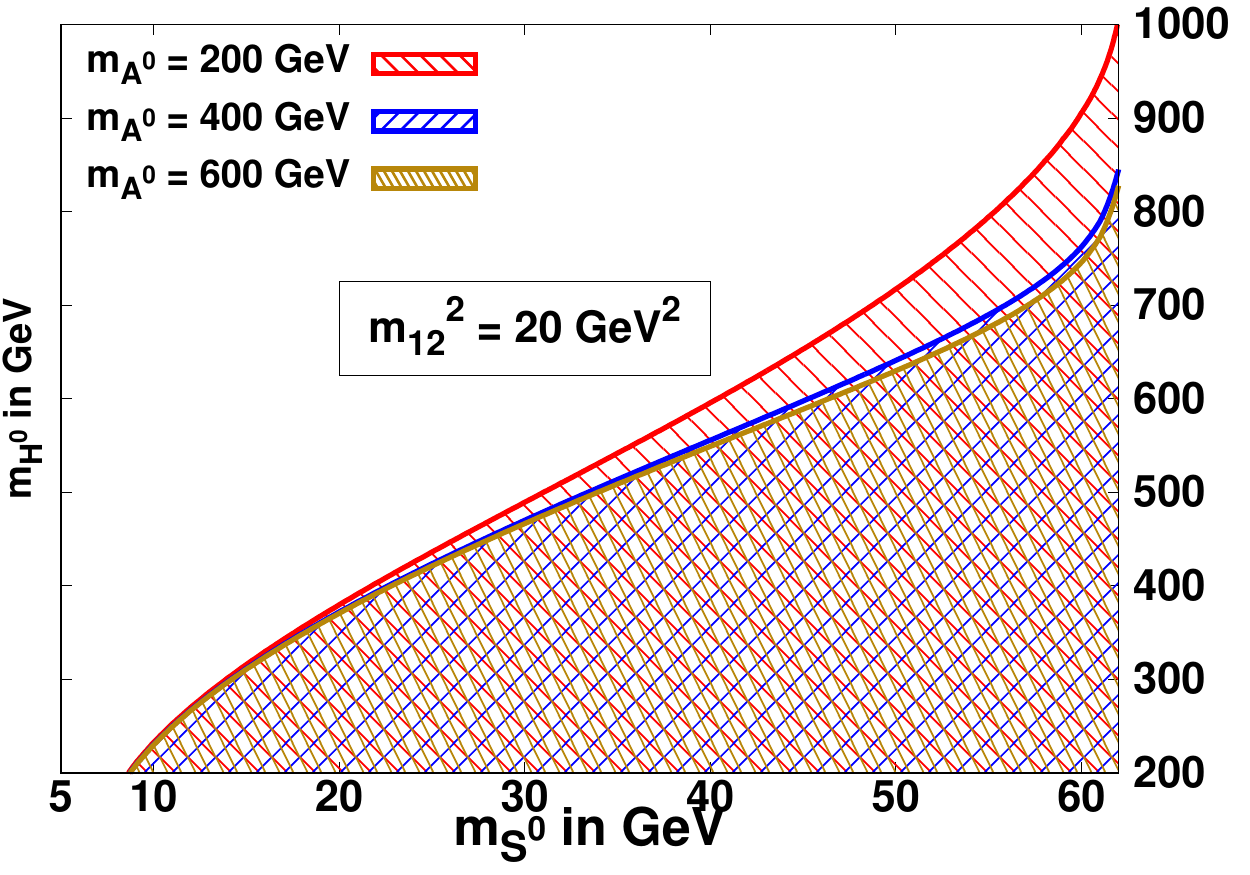}
  \caption{}
  \label{fig:contmh2vsmh3a} 
\end{subfigure}%
\begin{subfigure}{.5\textwidth}\centering
  \includegraphics[width=\columnwidth]{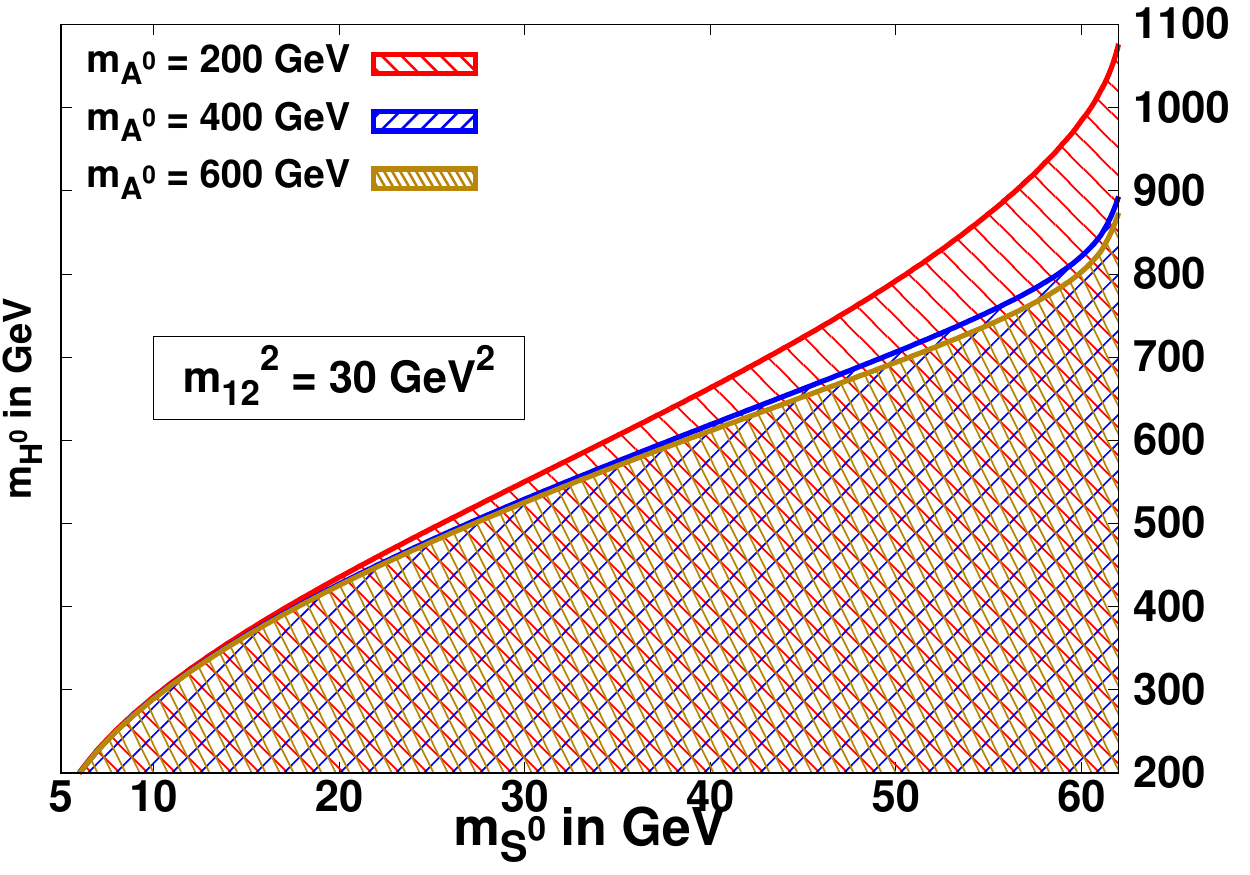}
  \caption{}
  \label{fig:contmh2vsmh3b} 
\end{subfigure}%
\\
\begin{subfigure}{.5\textwidth}\centering
  \includegraphics[width=\columnwidth]{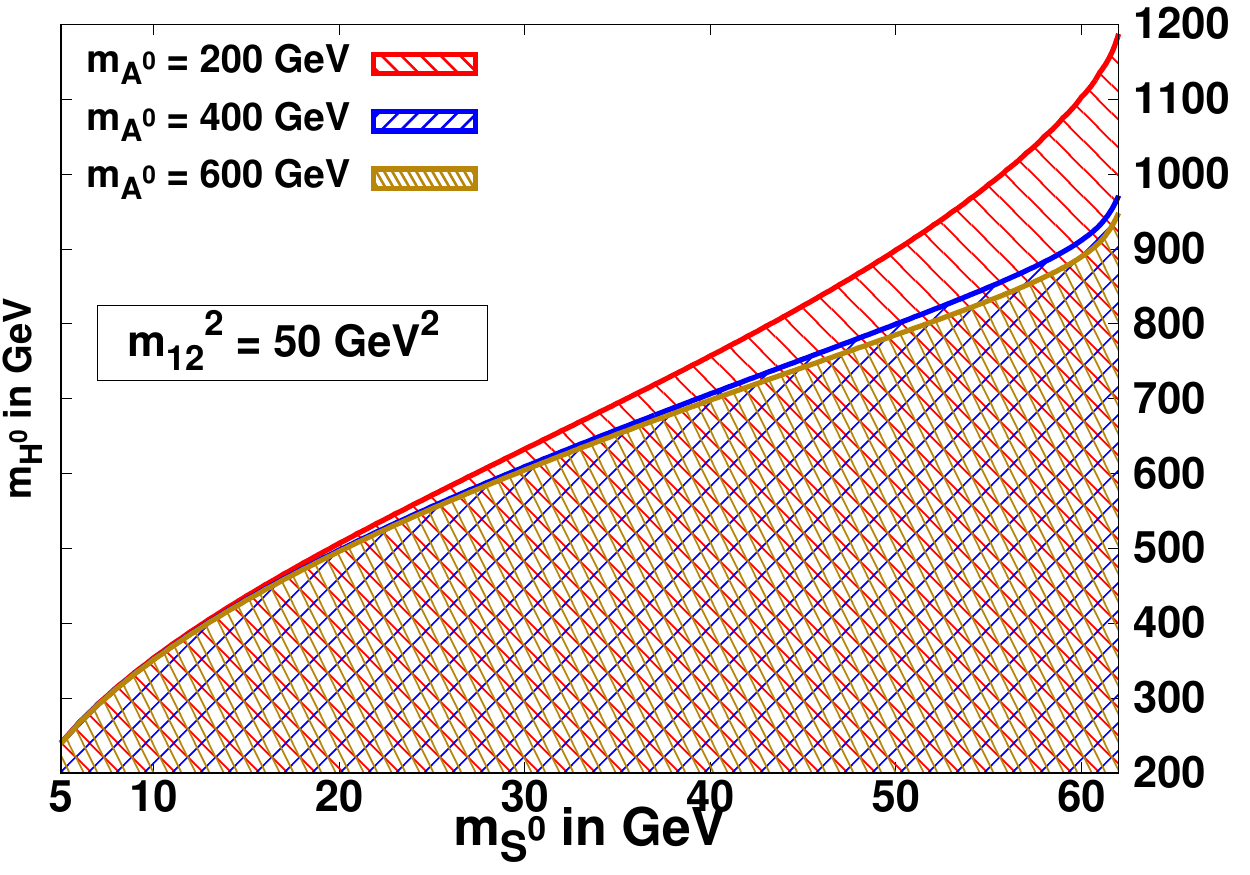}
  \caption{}
  \label{fig:contmh2vsmh3c} 
  \end{subfigure}%
\begin{subfigure}{.5\textwidth}\centering
  \includegraphics[width=\columnwidth]{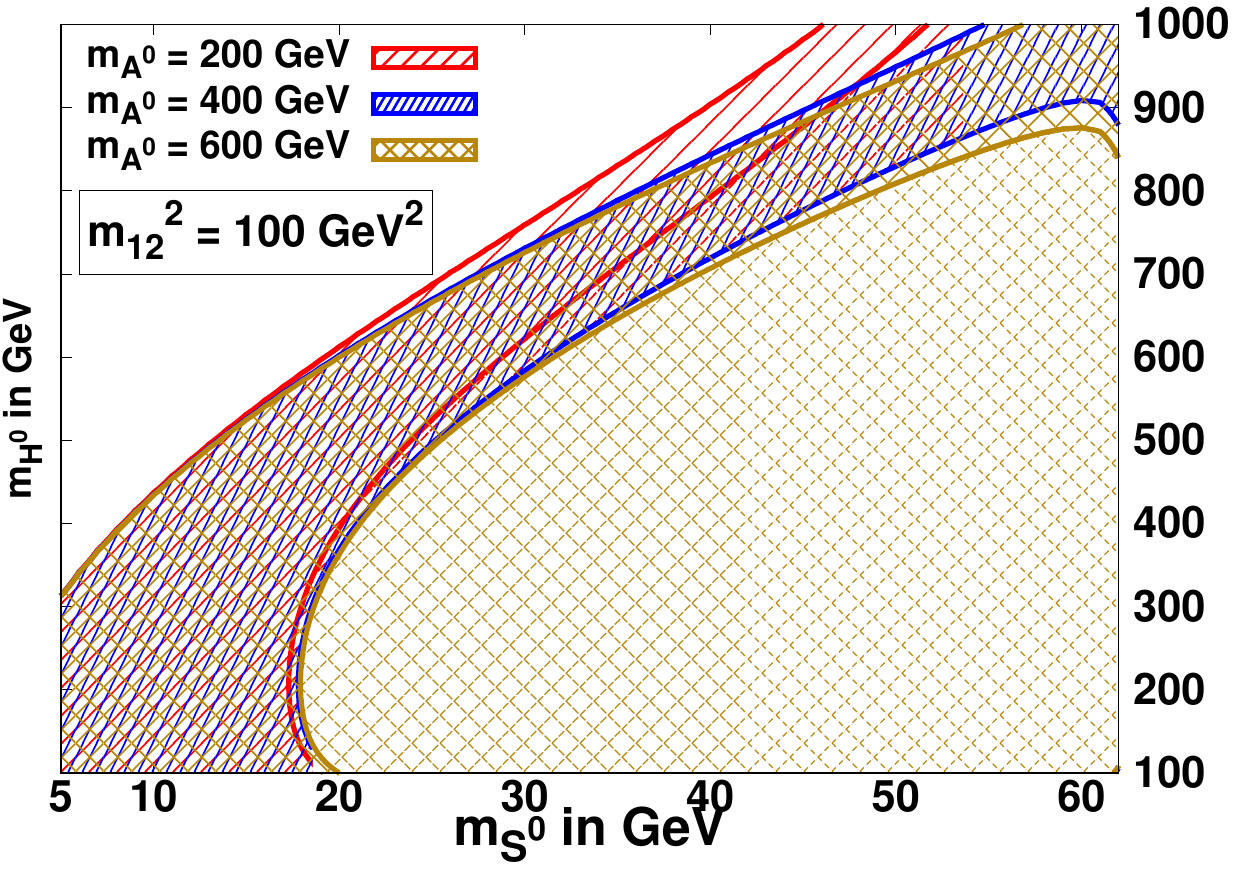}
  \caption{}
  \label{fig:contmh2vsmh3d} 
  \end{subfigure}%
 \caption{\small \em{ Figures \ref{fig:contmh2vsmh3a}, \ref{fig:contmh2vsmh3b}, \ref{fig:contmh2vsmh3c} and \ref{fig:contmh2vsmh3d} show  contours on the $m_{S^0}$  - $m_{H^0}$ plane  satisfying  the 95 \% C.L. upper limit on  the total observed Higgs decay width  $\Gamma_{\rm obs.}^{h^0} \le 2.4\times \Gamma_{\rm SM}^{h^0}$  for fixed $m_{H^\pm}$ = 600 GeV, $\delta_{13}$ = 0.4  and four different choices of $m_{12}^2$ = 20, 30, 50 and 100 GeV$^2$.  All points on the contours satisfy the discrepancy  $\Delta a_{\mu}=\,268(63)\,\times10^{-11}$. Each panel has three contours corresponding to $m_{A^0}$ = 200, 400 and 600 GeV respectively.}} 
 \label{fig:contmh2vsmh3}
\end{figure}

\subsection{Constraints from Higgs decay-width}
\label{section Constraints from SM Higgs decay}
Recently   CMS   analysed   the  partial decay widths of the off-shell Higgs Boson produced   through gluon fusion decaying to $W^+W^-$ Bosons \cite{Khachatryan:2016ctc} and then combined  the analysis with that for $Z\,Z$  \cite{Khachatryan:2015mma} vector Bosons  to obtain  95 \% C.L. upper   limit on the total observed  Higgs decay width of $ 2.4\times \Gamma_{\rm SM}^{h^0}$ \cite{pdg2018Higgs, Khachatryan:2016ctc}, where $\Gamma_{\rm SM}^{h^0}\simeq$ 4.1 MeV.  The authors have also investigated these decay channels for an off-shell Higgs Boson produced from the vector Boson fusion channels and obtained the upper bound on the total observed  Higgs decay width of $ 19.3\times \Gamma_{\rm SM}^{h^0}$ \cite{pdg2018Higgs,Khachatryan:2016ctc}. ATLAS also analysed the Higgs decay width  assuming that there are no anomalous couplings of the Higgs boson to vector Bosons, and obtained 95\% CL observed  upper limit on the total width of $ 6.7\times \Gamma_{\rm SM}^{h^0}$ \cite{Aad:2015xua}. However, we have used the conservative upper limit on the total observed decay width of Higgs Boson  of $ 2.4\times \Gamma_{\rm SM}^{h^0}$ for rest of the analysis in our study.

\par However, in the present model, the scalar identified with SM Higgs Boson $h^0$ is in addition likely to decay  into two light  singlet   {\it scalar portals} $h^0\to S^0\,S^0$ for  $m_{S^0}\leq \frac{m_{h^0}}{2}$. The partial decay width 
$\Gamma_{h^0\to S^0S^0}$ is given as \begin{eqnarray}
\Gamma_{h^0\to S^0S^0} = \frac{C_{h^0S^0S^0}^2}{32\,\pi\,m_{h^0}} \sqrt{1-\frac{4\, m_{S^0}^2}{m_{h^0}^2}}
\end{eqnarray}
The tri-scalar coupling $C_{h^0S^0S^0}$ is given in  equation \ref{Higgsdecay}. 
\par As total Higgs decay width is known with a fair accuracy, any contribution coming from other than SM particles should fit into the combined theoretical and experimental uncertainty. Thus, using the LHC data on the total observed Higgs decay-width,  we can put an upper limit on the tri-scalar coupling $C_{h^0S^0S^0}$. This upper limit is then used to constrain the parameter space of the model. 
\par Even restricting the parameter sets to satisfy the anomalous magnetic moment and LEP observations, the model parameter $m_{12}^2$ remains unconstrained. However, for a given choice of $\delta_{13}$, $m_{H^0}$, $m_{A^0}$ and $m_{S^0}$ an upper limit on $\left\vert C_{h^0S^0S^0}\right\vert$ constrains   $ m_{12}^2$ and thus fixes the model for further validation at colliders.
\par We study the  partial decay-width $\Gamma_{h^0\to S^0S^0}$ {\it w.r.t.}  $m_{S^0}$ for   five chosen values of the free parameter $m_{12}^2$ = 10, 20, 30, 50 and 100 GeV$^2$. We depict the variation of  the partial decay width  $\Gamma_{h^0\to S^0S^0}$  corresponding to   four different combinations of  $\left(m_{H^0},\, m_{A^0}\right)$ in GeV: $\left(400, 200\right)$, $\left(400, 400\right)$, $\left(400, 600\right)$ and $\left(600, 400\right)$ in figures  \ref{fig:Higgs_Decay_Widtha_d13_p2},   \ref{fig:Higgs_Decay_Widthb_d13_p2}, \ref{fig:Higgs_Decay_Widthc_d13_p2}, \ref{fig:Higgs_Decay_Widthd_d13_p2} respectively for $\delta_{13}$ = 0.2 and in figures \ref{fig:Higgs_Decay_Widtha},   \ref{fig:Higgs_Decay_Widthb}, \ref{fig:Higgs_Decay_Widthc}, \ref{fig:Higgs_Decay_Widthd} respectively for $\delta_{13}$ = 0.4. The top horizontal band in all the four panels in figures \ref{fig:Higgs_Decay_Widthp2} and \ref{fig:Higgs_Decay_Width} corresponds to the forbidden region arising from the observed total  Higgs decay width  at LHC. In figure \ref{fig:Higgs_Decay_Widthp2} the parameter region for $m_{S^0}$ $\geq 24$ GeV is forbidden by non-perturbativity of $H^0 \tau^+ \tau^-$ couplings. We observe that the   constraints from the total Higgs decay width  further shrinks the parameter space allowed  by $\Delta a_\mu$ between  10 GeV $\leq m_{S^0}\le 62 $ GeV  for $\delta_{13}$ = 0.4 corresponding  to 100 GeV$^2\geq m_{12}^2 \geq$ 10 GeV$^2$.  
\par To have better insight of the bearings on the model from the observed total Higgs decay width   we 
plot the contours on the $m_{S^0}-m_{H^0}$ plane for mixing angle $\delta_{13}$ =  0.4  in figures \ref{fig:contmh2vsmh3a}, \ref{fig:contmh2vsmh3b}, \ref{fig:contmh2vsmh3c} and \ref{fig:contmh2vsmh3d}  satisfying the upper bound of the total observed Higgs decay width obtained by CMS \cite{Khachatryan:2016ctc}.   We have considered four choices of $m^2_{12}$ respectively. In each panel , three curves depict the  upper limits on the partial widths which are derived from the constraints on the total observed decay width  from LHC corresponding to three chosen values $m_{A^0}$ = 200, 400 and 600 GeV respectively. We note  that with increasing $m^2_{12}$ the allowed dark shaded region shrinks and remains confined towards a lighter $m_{S^0}$.

\subsection{Lepton non-Universality and Precision Constraints}
Recently  HFAG collaboration \cite{Amhis:2014hma} provided stringent constraints  on the departure   of SM predicted universal lepton-gauge couplings.   Non universality of the lepton-gauge couplings can be parameterized as deviation from the ratio of the lepton-gauge couplings of any two different generations from unity  and  is defined as $\delta_{ll^\prime}\equiv \left(g_l/g_{l^\prime}\right) -1$. For example, the said deviation for $\tau^\pm$ and $\mu^\pm$ can be  extracted from the  measured respective  pure leptonic decay modes and is defined as
\begin{eqnarray}
\delta_{\tau\mu}\equiv\left(g_{\tau^-}/g_{\mu^-}\right) -1 = \frac{\sqrt{\Gamma\left(\tau^-\to e^-\,\bar\nu_e\,\nu_\tau\right)}}{\sqrt{\Gamma\left(\mu^-\to e^-\,\bar\nu_e\,\nu_\mu\right)}} -1. 
\end{eqnarray}
The  measured deviations of the three  different ratios are found to be  \cite{Amhis:2014hma}
\begin{eqnarray}
\delta^l_{\tau\mu} =  0.0011 \pm 0.0015; \,\,\,\, \delta^l_{\tau e} = 0.0029 \pm 0.0015, \quad
{\rm and}\,\,\, \, \delta^l_{\mu e} = 0.0018 \pm 0.0014,
\end{eqnarray}
out of which only two ratios are independent \cite{Chun:2015hsa}.
\par The implication of these data on lepto-philic type {\bf X} 2-HDM models have been studied in great detail in reference \cite{Chun:2015hsa} and are shown as  contours  in $m_{H^\pm}-\tan \beta$ and $m_{A^0}-\tan \beta$ planes,  based on $\chi^2$ analysis of  non-SM additional  tree  $\delta_{\rm tree}$ and loop   $\delta_{\rm loop}$ contributions to the lepton decay process in the leptonic mode  \cite{Krawczyk:2004na}. We find that the additional scalar in lepto-philic  2-HDM + singlet scalar model contribute to $\delta_{\tau\mu}$, $\delta_{\tau e}$ and $\delta_{\mu e}$ at the one loop level which is $\delta_{13}^2$ suppressed. However, they make a negligibly  small correction and render  the $\delta_{\rm loop}$ more negative.

\par Further we  constraint the model from the experimental bound on the  $S,\, T$ and $U$ \cite{Peskin:1990zt} oblique parameters. Constrains from these parameters for all variants of  2-HDM models  have been extensively studied in the literature \cite{Funk:2011ad}. We compute the additional contribution due to the  singlet scalar at one loop for $\Delta S$  and $\Delta T$   in  2-HDM + singlet scalar model and find that they are suppressed by the square of the mixing angle $\delta_{13}^2$ and are therefore consistent with the experimental observations as long as $m_{H^\pm}$ is   degenerate either with $m_{A^0}$ or $m_{H^0}$ for large $\tan\beta$ region to a range within $\sim$ 50 GeV \cite{Batell:2016ove}.

\section{Dark matter Phenomenology} 
 \label{section_ Dark_ matter}
 We introduce a spin 1/2 fermionic dark matter particle $\chi$ which is taken to be a SM singlet with zero-hyper-charge and is odd under a discrete $Z_2$ symmetry. The DM $\chi$ interacts with the SM particle through the {\it scalar portal} $S^0$. The interaction Lagrangian ${\cal L}_{DM}$ is given as
\begin{equation}
{\cal L}_{DM} = i \bar{\chi}\gamma^\mu \partial_\mu \chi -m_\chi\bar{\chi}\chi +g_{\chi S^0} \bar{\chi}\chi S^0
\end{equation}
We are now equipped to compute the relic density of the DM, the scattering cross-section of such DM with the nucleon and its indirect detection annihilation cross-section.
\subsection{Computation of the Relic Density}
\label{subsection_Relic_Density}
In early universe, when the temperature of the thermal bath was much greater than the corresponding mass of the particle species, the particles were in thermal equilibrium with the background. This equilibrium was maintained through interactions such as annihilation and scattering with other SM particles, such that the interaction rate remained greater than the expansion rate of the universe. As the Universe cooled, massive particles such as our DM candidate $\chi$, became non-relativistic and the interaction rate with other particles became lower than the expansion rate of the universe, hence decoupling the DM and giving us the relic abundance 0.119 \cite{Ade:2015xua,Komatsu:2014ioa} we observe today.
Evolution of the number  density of the DM $n_\chi$  is governed by the Boltzmann equation:
\begin{figure}[h!]
\centering 
\begin{subfigure}{.45\textwidth}\centering
  \includegraphics[width=\columnwidth]{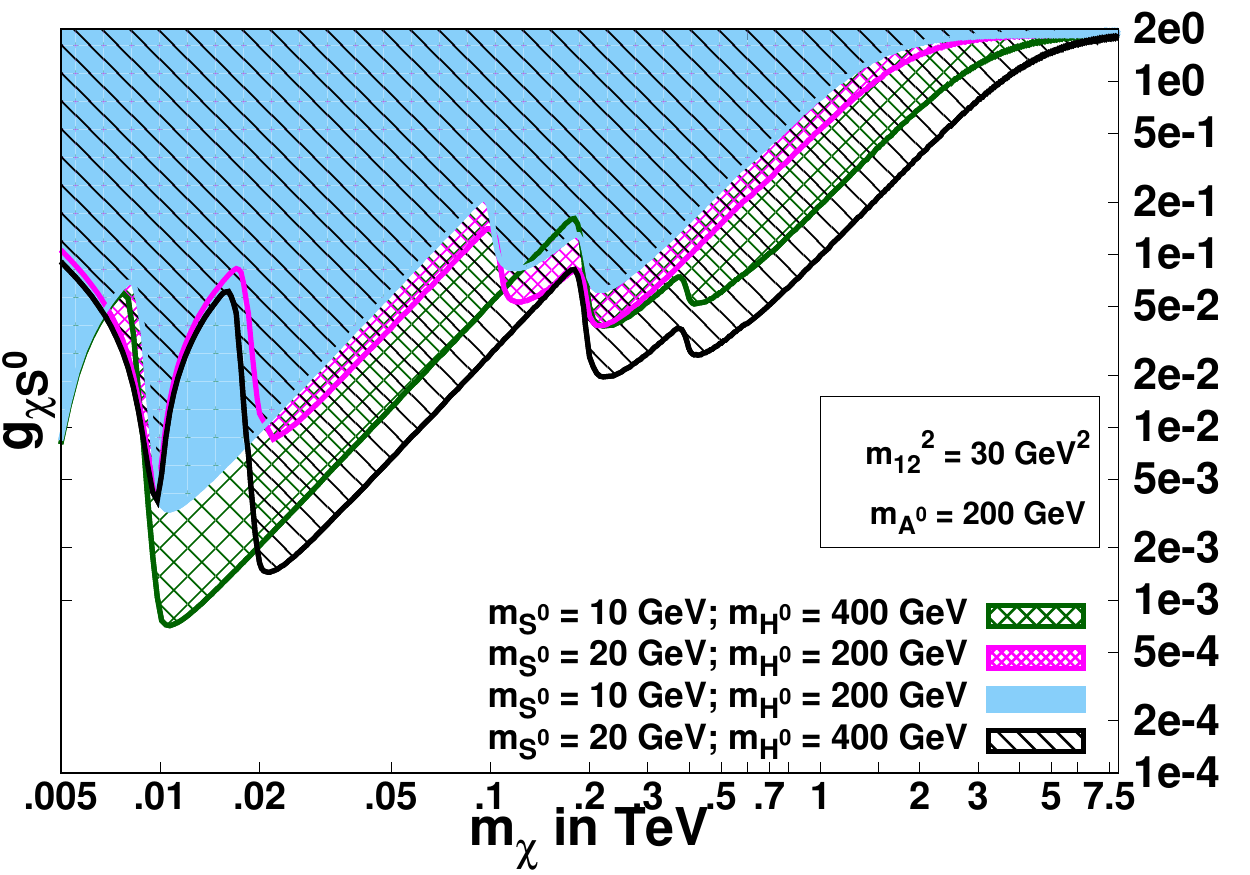}
  \caption{}
  \label{fig:Relic_Densityp2a}
\end{subfigure}
\begin{subfigure}{.45\textwidth}\centering
  \includegraphics[width=\columnwidth]{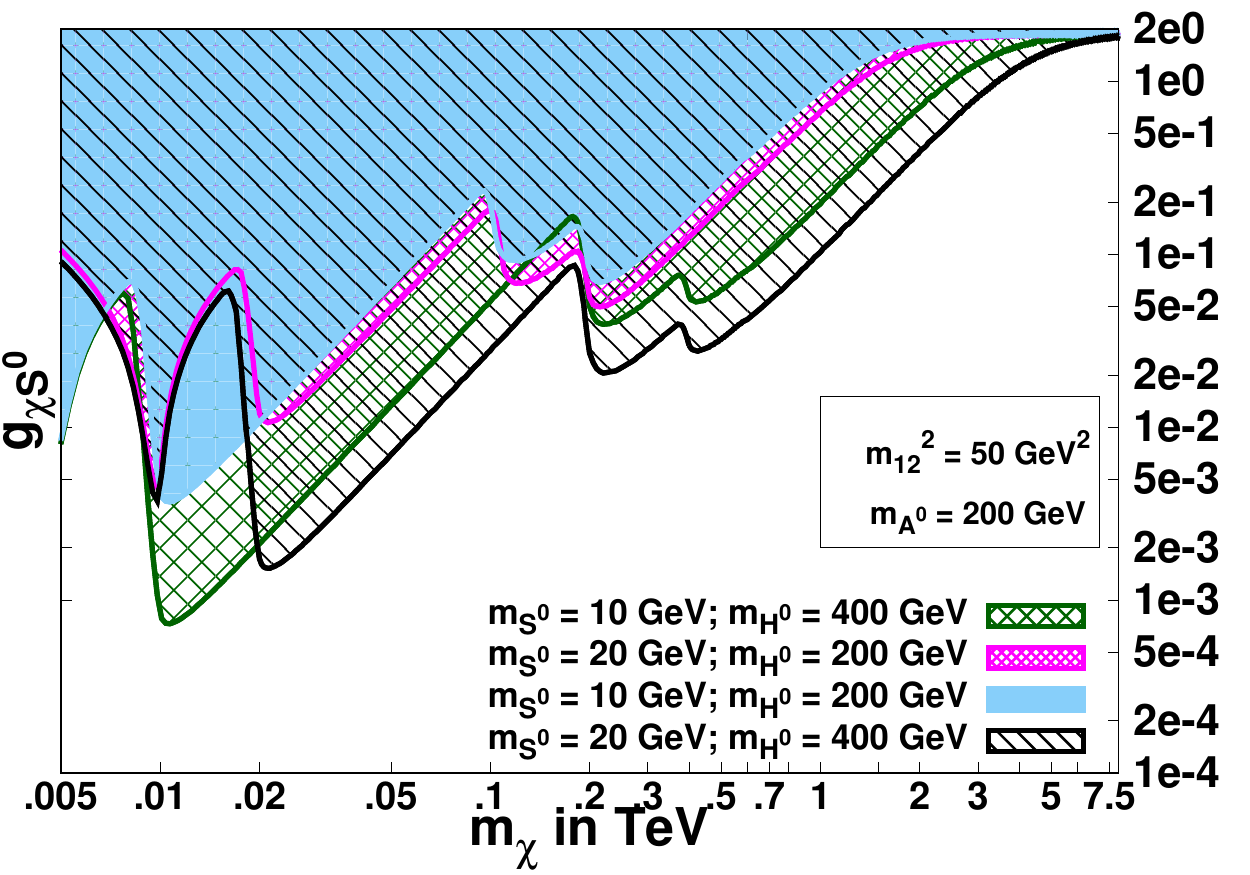}
  \caption{}
  \label{fig:Relic_Densityp2b}
\end{subfigure}%
\\
\begin{subfigure}{.45\textwidth}\centering
  \includegraphics[width=\columnwidth]{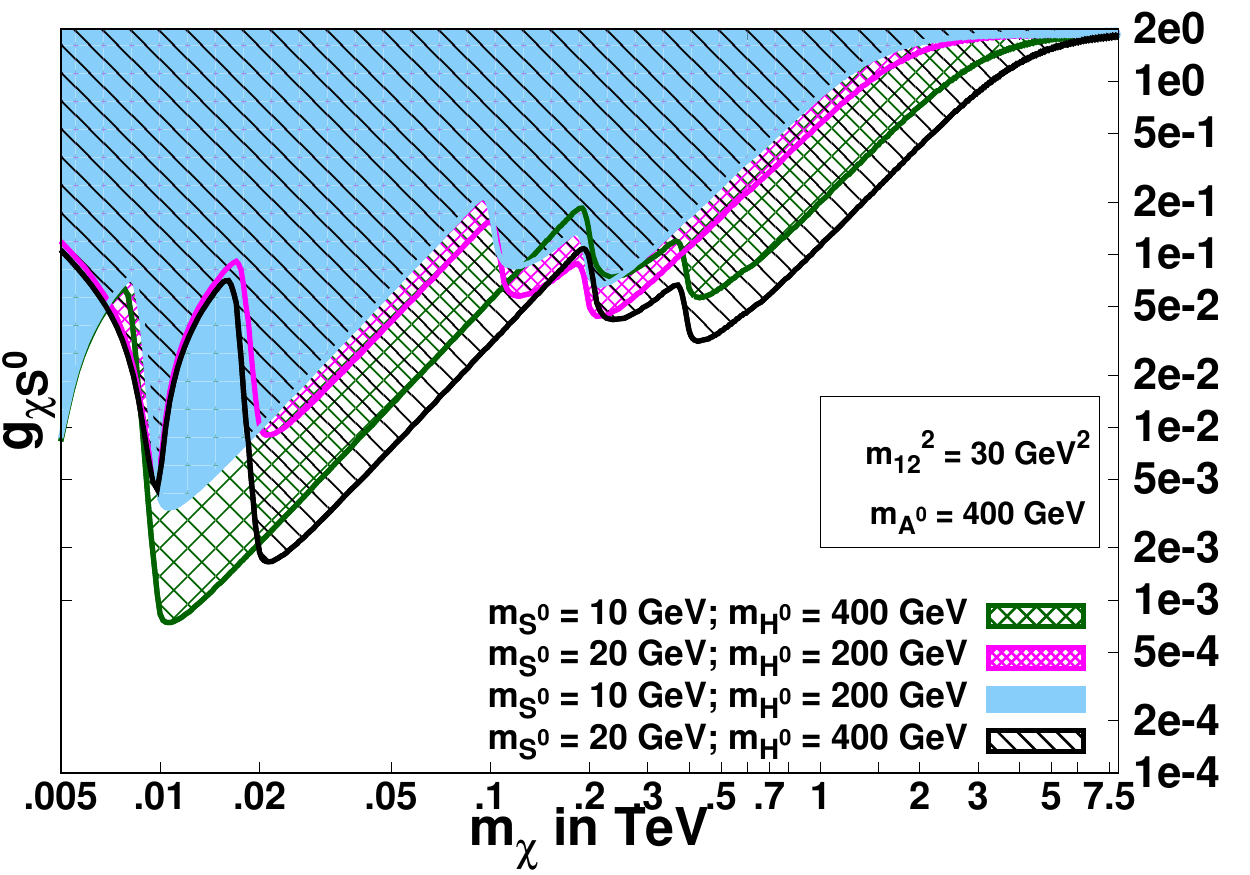}
  \caption{}
  \label{fig:Relic_Densityp2c}
\end{subfigure}
\begin{subfigure}{.45\textwidth}\centering
  \includegraphics[width=\columnwidth]{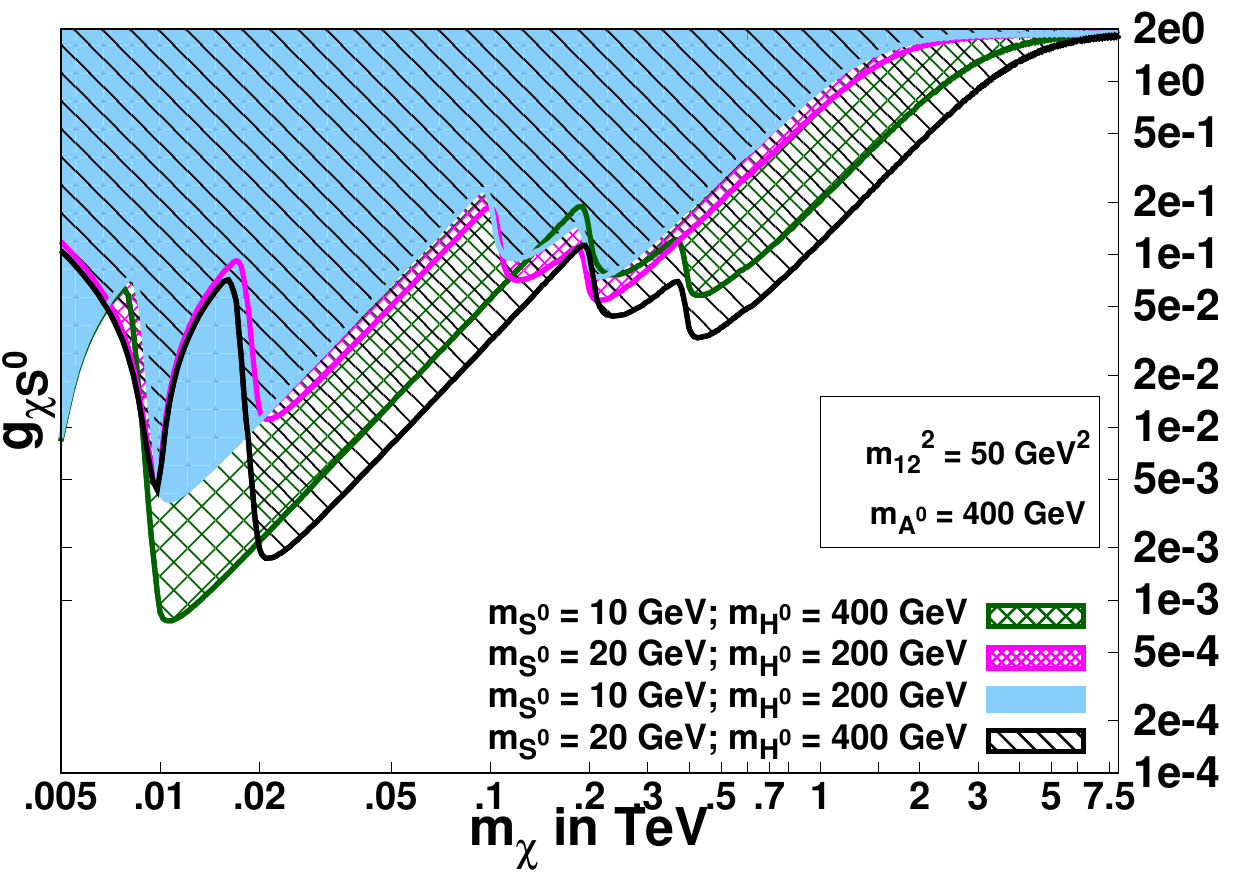}
  \caption{}
  \label{fig:Relic_Densityp2d}
\end{subfigure}%

\begin{subfigure}{.45\textwidth}\centering
  \includegraphics[width=\columnwidth]{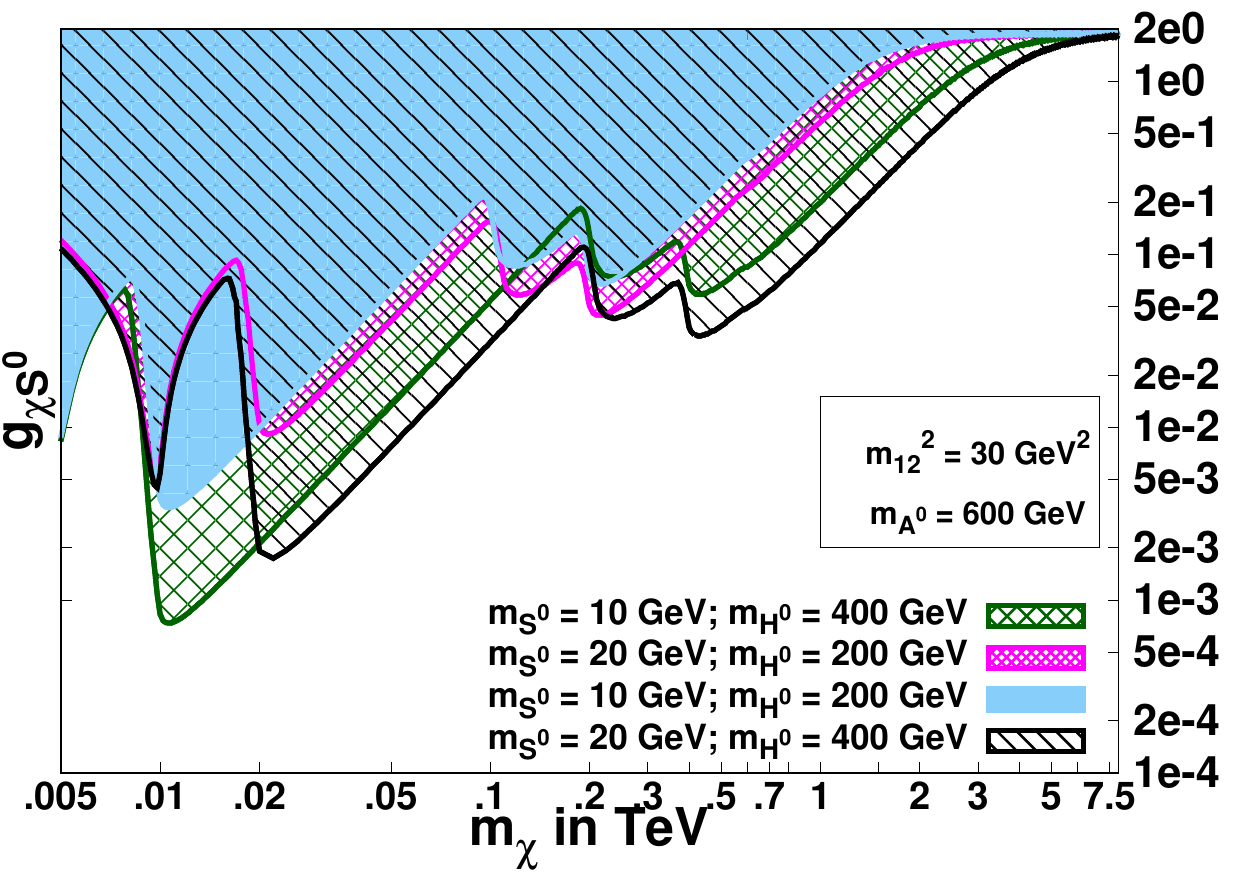}
  \caption{}
  \label{fig:Relic_Densityp2e}
\end{subfigure}
\begin{subfigure}{.45\textwidth}\centering
  \includegraphics[width=\columnwidth]{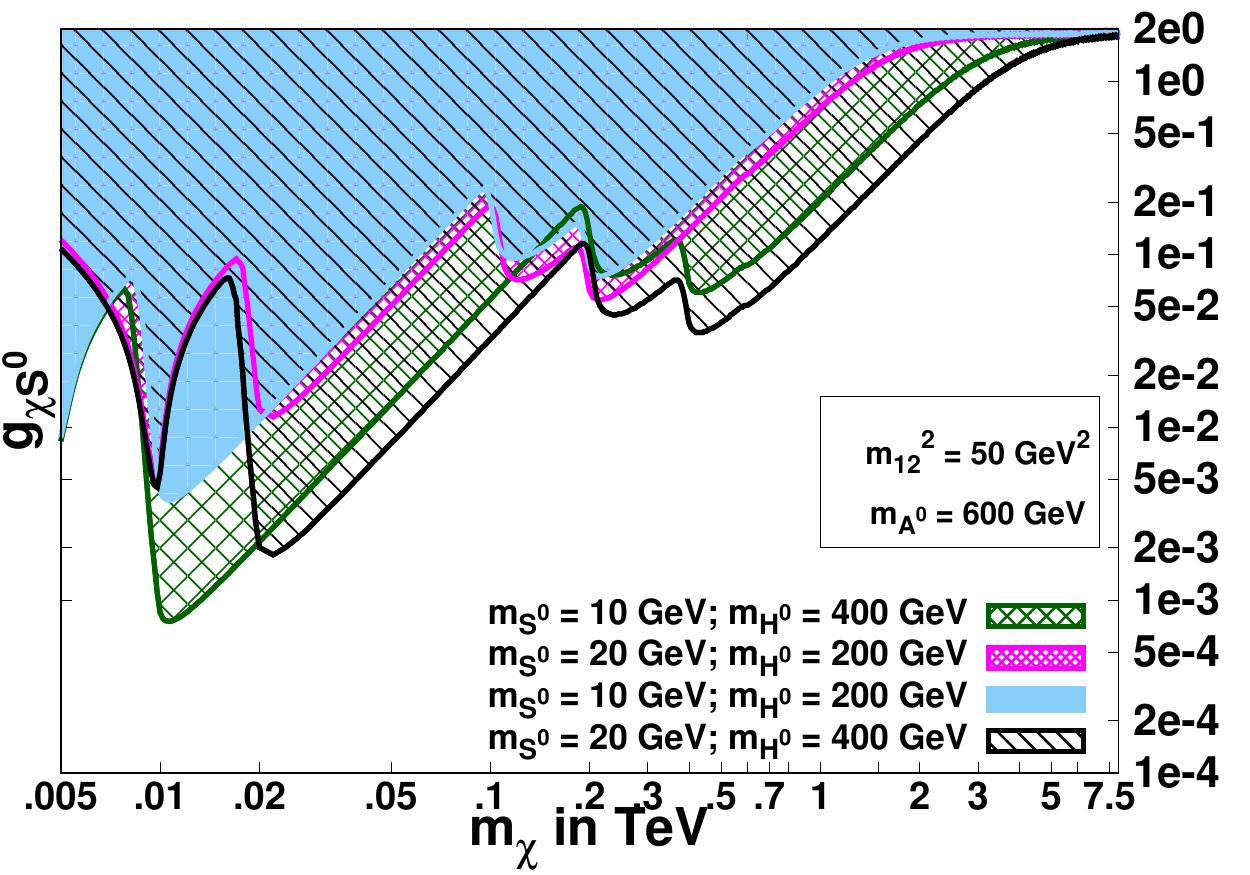}
  \caption{}
  \label{fig:Relic_Densityp2f}
\end{subfigure}%
\caption{\small \em {Figures \ref{fig:Relic_Densityp2a} to \ref{fig:Relic_Densityp2f} show contours on the $m_{\chi}$  - $g_{\chi S^0}$ plane satisfying  the relic density 0.119 \cite{Ade:2015xua,Komatsu:2014ioa} for fixed $m_{H^\pm}$ = 600 GeV, $\delta_{13}$ = 0.2  and  different choices of $m_{12}^2$ and $m_{A^0}$. All points on the contours satisfy the discrepancy  $\Delta a_{\mu}=\,268(63)\,\times10^{-11}$. In the left and right panels, we show allowed (shaded) regions for four and five combinations of $m_{S^0}$, $m_{H^0}$ respectively.}}
\label{fig:Relic_Densityp2}
\end{figure}
\begin{figure}[h!]
\centering 
\begin{subfigure}{.45\textwidth}\centering
  \includegraphics[width=\columnwidth]{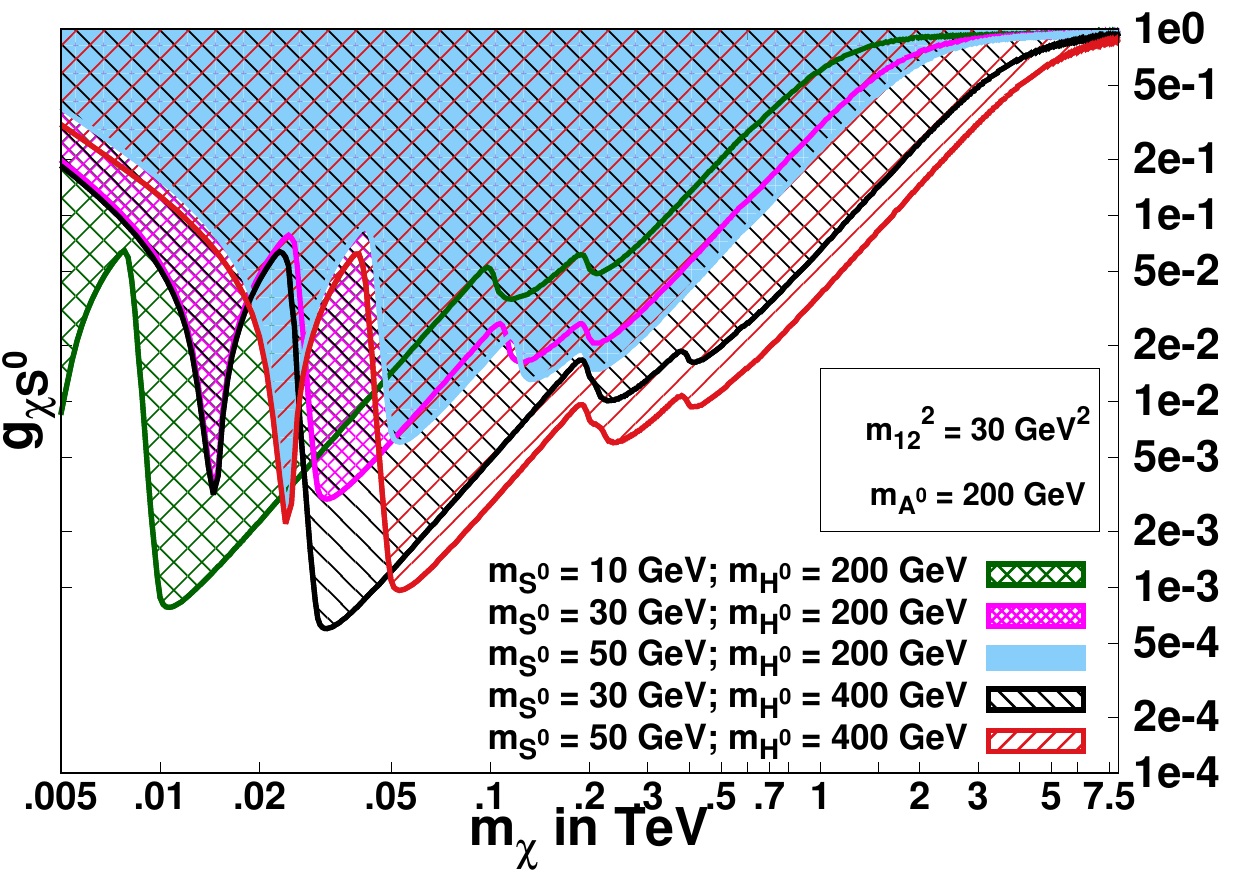}
  \caption{}
  \label{fig:Relic_Densityp4a}
\end{subfigure}
\begin{subfigure}{.45\textwidth}\centering
  \includegraphics[width=\columnwidth]{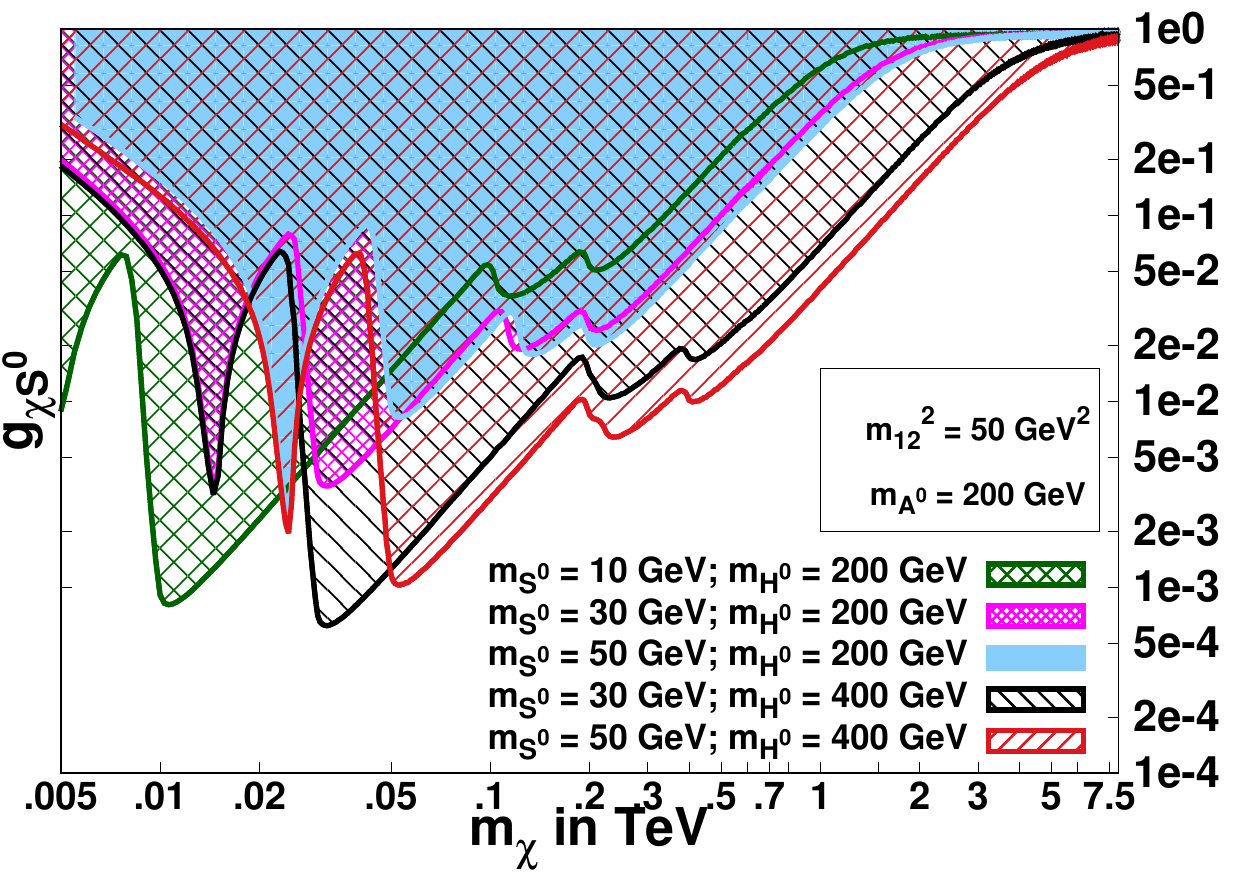}
  \caption{}
  \label{fig:Relic_Densityp4b}
\end{subfigure}%
\\
\begin{subfigure}{.45\textwidth}\centering
  \includegraphics[width=\columnwidth]{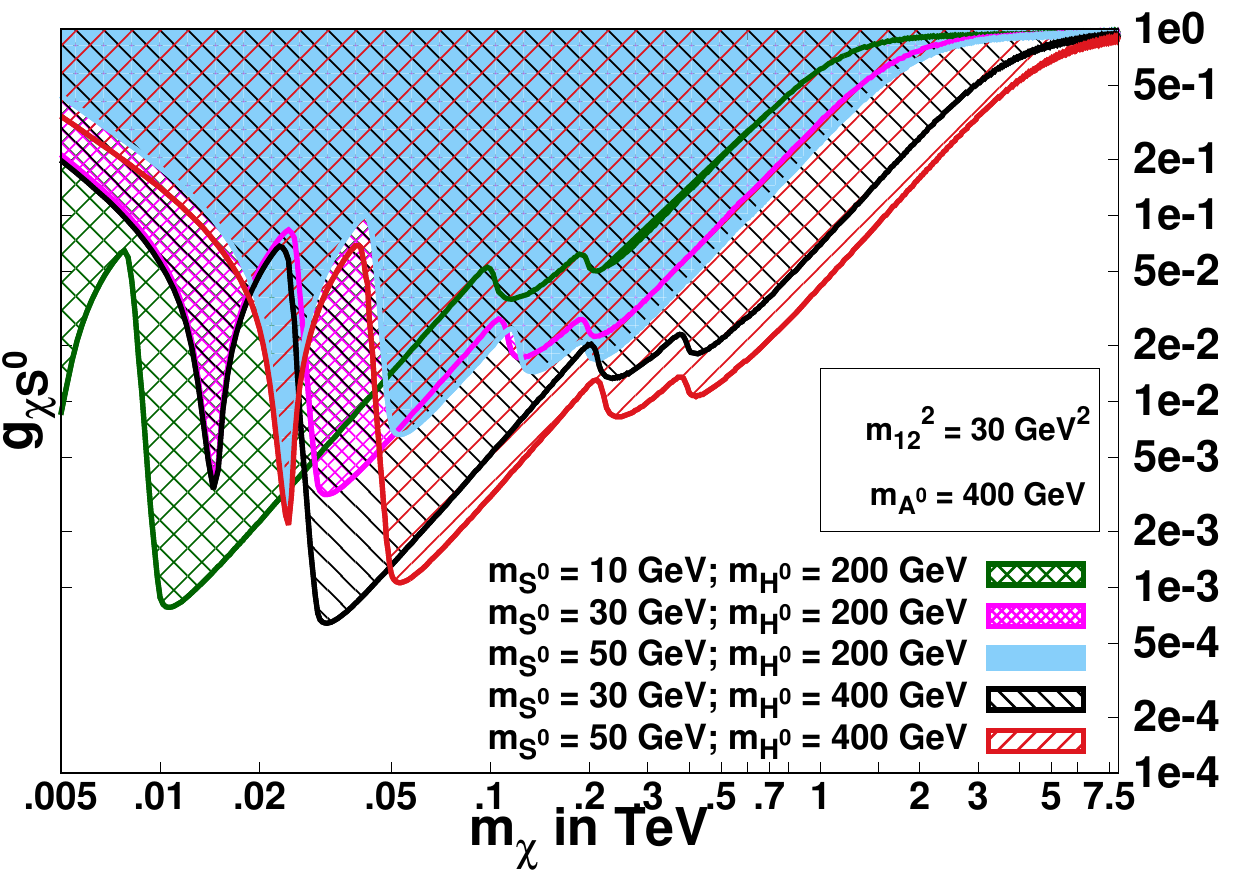}
  \caption{}
  \label{fig:Relic_Densityp4c}
\end{subfigure}
\begin{subfigure}{.45\textwidth}\centering
  \includegraphics[width=\columnwidth]{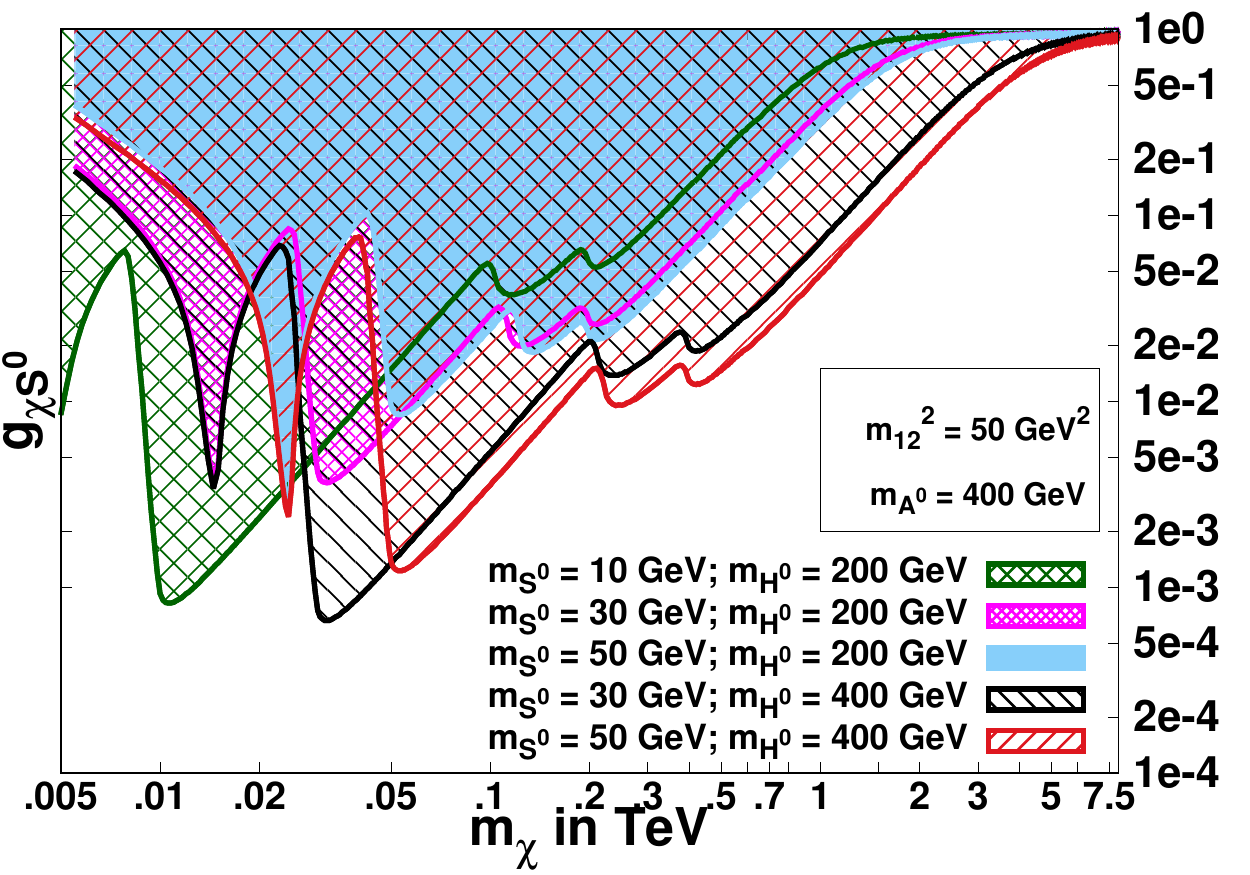}
  \caption{}
  \label{fig:Relic_Densityp4d}
\end{subfigure}%

\begin{subfigure}{.45\textwidth}\centering
  \includegraphics[width=\columnwidth]{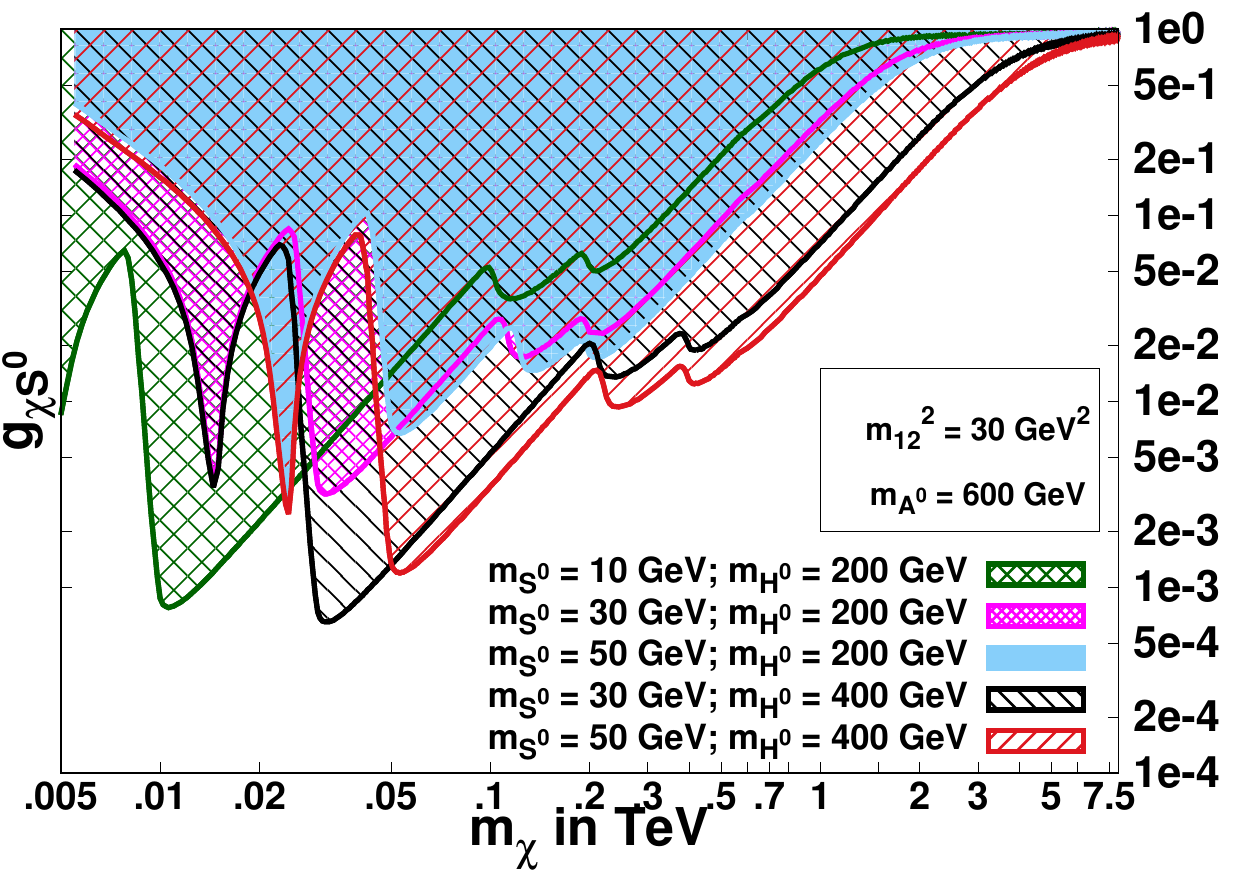}
  \caption{}
  \label{fig:Relic_Densityp4e}
\end{subfigure}
\begin{subfigure}{.45\textwidth}\centering
  \includegraphics[width=\columnwidth]{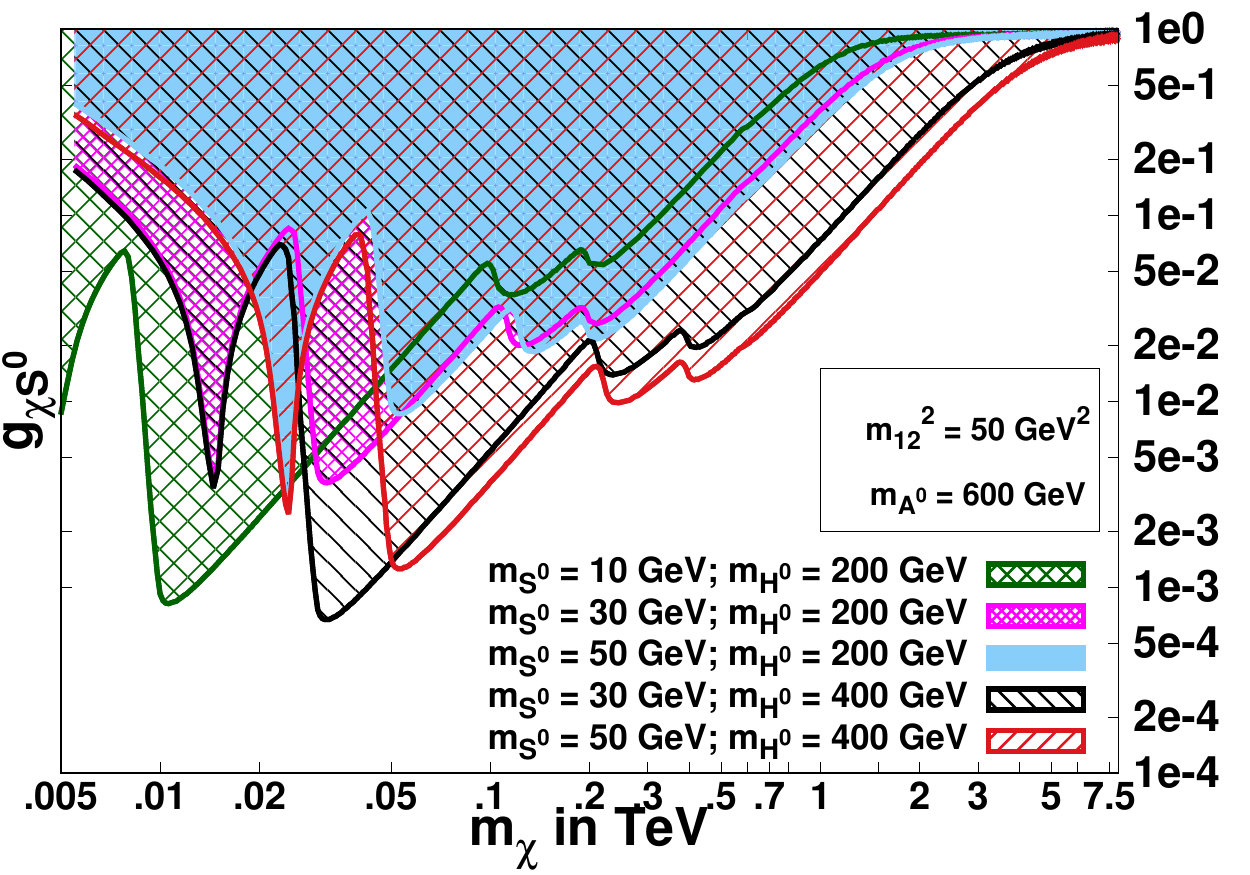}
  \caption{}
  \label{fig:Relic_Densityp4f}
\end{subfigure}%
\caption{\small \em {Figures \ref{fig:Relic_Densityp4a} to \ref{fig:Relic_Densityp4f} show contours on the $m_{\chi}$  - $g_{\chi S^0}$ plane satisfying  the relic density 0.119 \cite{Ade:2015xua,Komatsu:2014ioa} for fixed $m_{H^\pm}$ = 600 GeV, $\delta_{13}$ = 0.4  and  different choices of $m_{12}^2$ and $m_{A^0}$. All points on the contours satisfy the discrepancy  $\Delta a_{\mu}=\,268(63)\,\times10^{-11}$. In the left and right panels, we show allowed (shaded) regions for four and five combinations of $m_{S^0}$, $m_{H^0}$ respectively.}}
\label{fig:Relic_Densityp4}
\end{figure}
\begin{equation}
\frac{dn_\chi}{dt} + 3 \frac{\dot{a}}{a}\,n_\chi = -\langle\sigma\left\vert \vec  v\right\vert\rangle\left(n^2_\chi - n^2_{\chi eq}   \right)
\end{equation}
 where $\frac{\dot{a}}{a} = \sqrt{\frac{8\pi\rho}{3\,M_{Pl}}}$ , $\langle\sigma\left\vert \vec  v
\right\vert\rangle$ is thermally averaged cross-section and 

$n^2_{\chi eq}={\mathbf g}\left(\frac{m_\chi T}{2\pi}\right)^{\frac{3}{2}}\exp\left[\frac{-m_\chi}{T}\right] $ where ${\mathbf g}$ is the degrees of freedom, and it is 2 for fermions. As for a massive thermal relics, freeze-out occurs when the species is non-relativistic $\left\vert\vec v\right\vert <<c$.  Therefore, we expand  $\langle\sigma\left\vert \vec  v
\right\vert\rangle$ as $\langle\sigma\left\vert \vec  v\right\vert\rangle = a + b\left \vert\vec v\right\vert^2 + \mathcal{O}(\left\vert\vec v\right\vert^4)$.
The Boltzmann equation can be solved to give the thermal relic density \cite{F.Tanedo}
 
 \begin{equation}
 \Omega_\chi {\mathbf h}^2 \simeq \frac{1.07 \times 10^9 x_F}{M_{Pl} \,\,\sqrt{{\mathbf g^*} (x_F)} (a + \frac{6b}{x_F})}
 \end{equation}
 where ${\mathbf h}$ is dimensionless Hubble parameter, ${\mathbf g^*} (x_F)$ is total number of dynamic degrees of freedom near freeze-out temperature $T_F$ and $x_F=\frac{m_\chi}{T_F}$ is given by 
 \begin{equation}
 x_F =\ln\left[c\,(c+2) \sqrt{\frac{45}{8}}\, \frac{{\mathbf g}\,M_{Pl}\,\, m_\chi \left(a+ \frac{6\,b}{x_F}\right)}{ 2 \pi^3\sqrt{{\mathbf g^*} \left(x_F\right)}\sqrt{x_F}} \right]
 \end{equation}
 where $c$ is of the order 1. The thermal-averaged scattering cross-sections  as a function of DM mass $m_\chi $ are given in the Appendix \ref{thermalAveragedCrosssection}.

 \par To compute relic density numerically, we have used MadDM \cite{Ambrogi:2018jqj} and MadGraph \cite{Alwall:2014hca}. We have generated the input model file required by  MadGraph using FeynRules \cite{Alloul:2013bka}, which calculates all the required couplings and Feynman rules by using the full Lagrangian. 

\par For a given  charged Higgs mass of 600 GeV we depict the contours of constant relic density $\simeq$ 0.119  \cite{Ade:2015xua,Komatsu:2014ioa} in  $g_{\chi S^0}$ (DM coupling) and $m_\chi$ (DM mass) plane in figure \ref{fig:Relic_Densityp2} corresponding to  two choices of singlet scalar masses of 10 and 20 GeV for  $\delta_{13}$ = 0.2 and in figure  \ref{fig:Relic_Densityp4} corresponding to three choices of singlet scalar masses of 10, 30 and 50 GeV  for  $\delta_{13}$ = 0.4. The six different panels in figures \ref{fig:Relic_Densityp2} and \ref{fig:Relic_Densityp4} correspond to the following six different  combinations of $\left(m_{A^0}, m_{12}^2\right):$
\begin{eqnarray}
&&\left(200\, {\rm GeV}, \, 30\,{\rm GeV}^2\right),   \,\left(200\, {\rm GeV}, \, 50\,{\rm GeV}^2\right),\, \left(400\, {\rm GeV}, \, 30\,{\rm GeV}^2\right),\nonumber\\
&&\left(400\, {\rm GeV}, \, 50\,{\rm GeV}^2\right),\, \left(600\, {\rm GeV}, \, 30\,{\rm GeV}^2\right)\,\, {\rm and} \,\,\left(600\, {\rm GeV}, \, 50\,{\rm GeV}^2\right).\nonumber 
\end{eqnarray}
The un-shaded  regions in $g_{\chi S^0}-m_\chi$  plane in  figures corresponding to over closing of the Universe by DM relic density contribution.  The successive dips in the relic density contours  arise due to opening up of additional DM annihilation channel   with the increasing DM mass. Initial dip is caused by {\it s-channel} propagator. Dip observed around 0.2 TeV and 0.4 TeV are caused by opening of $\bar{\chi} \chi \rightarrow S^0 H^0$ and $\bar{\chi} \chi \rightarrow H^0 H^0\, (A^0 A^0)$ channels. The parameter sets  chosen for the calculation of the relic density are consistent with the observed value of $\Delta a_\mu$ and measured total Higgs decay width.

 \begin{figure}[h!]
\centering 
\begin{subfigure}{.45\textwidth}\centering
  \includegraphics[width=\columnwidth]{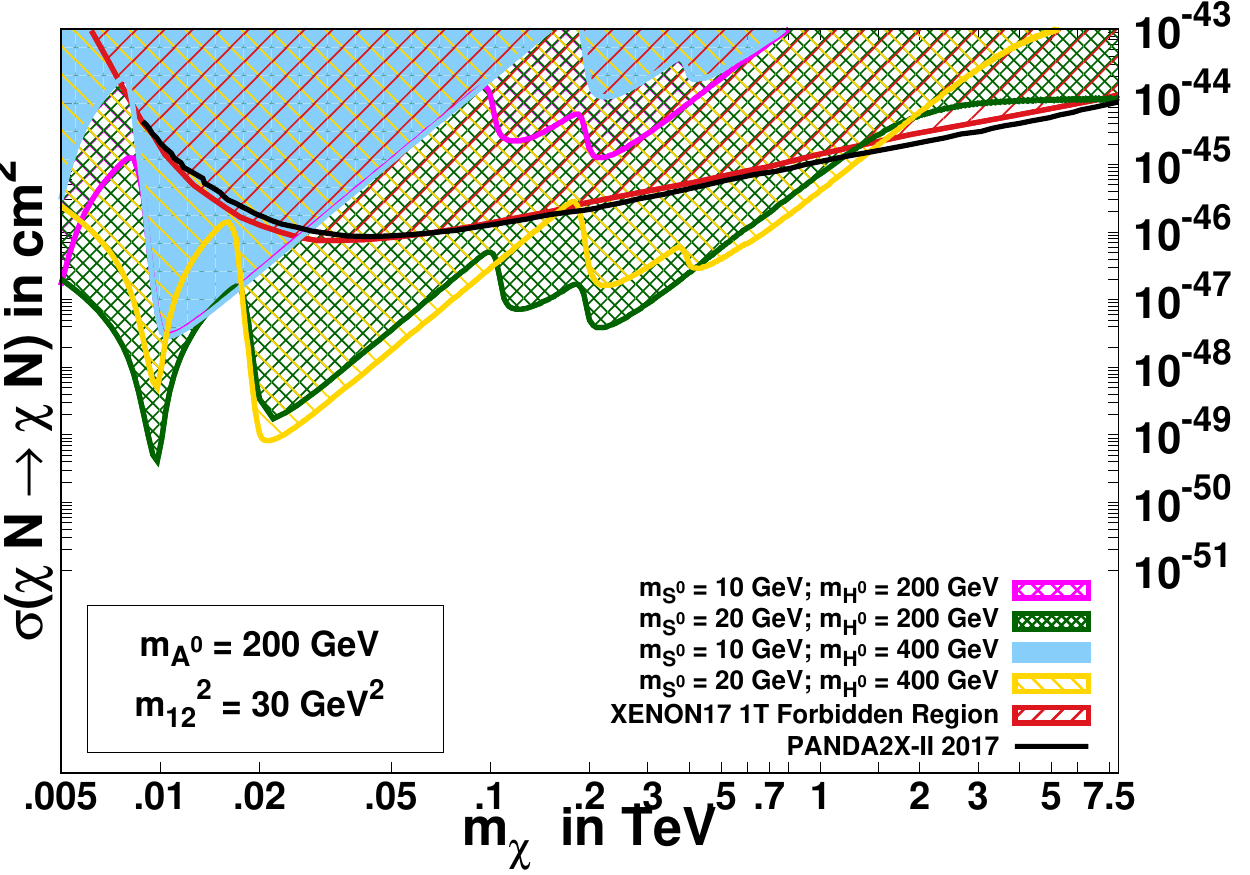}
  \caption{}
  \label{fig:Direct_Detectionp2a}
\end{subfigure}%
\begin{subfigure}{.45\textwidth}\centering
  \includegraphics[width=\columnwidth]{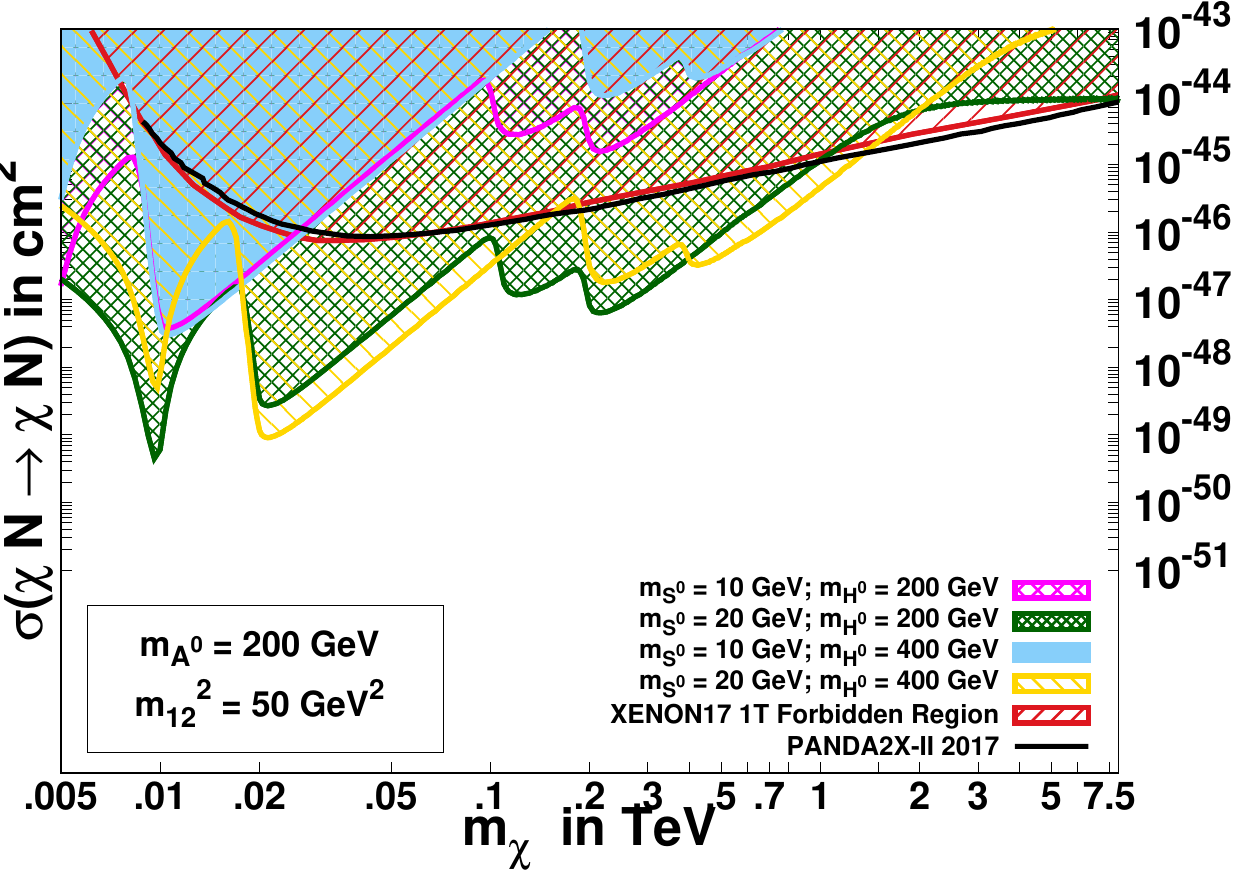}
  \caption{}
  \label{fig:Direct_Detectionp2b}
\end{subfigure}%

\begin{subfigure}{.45\textwidth}\centering
  \includegraphics[width=\columnwidth]{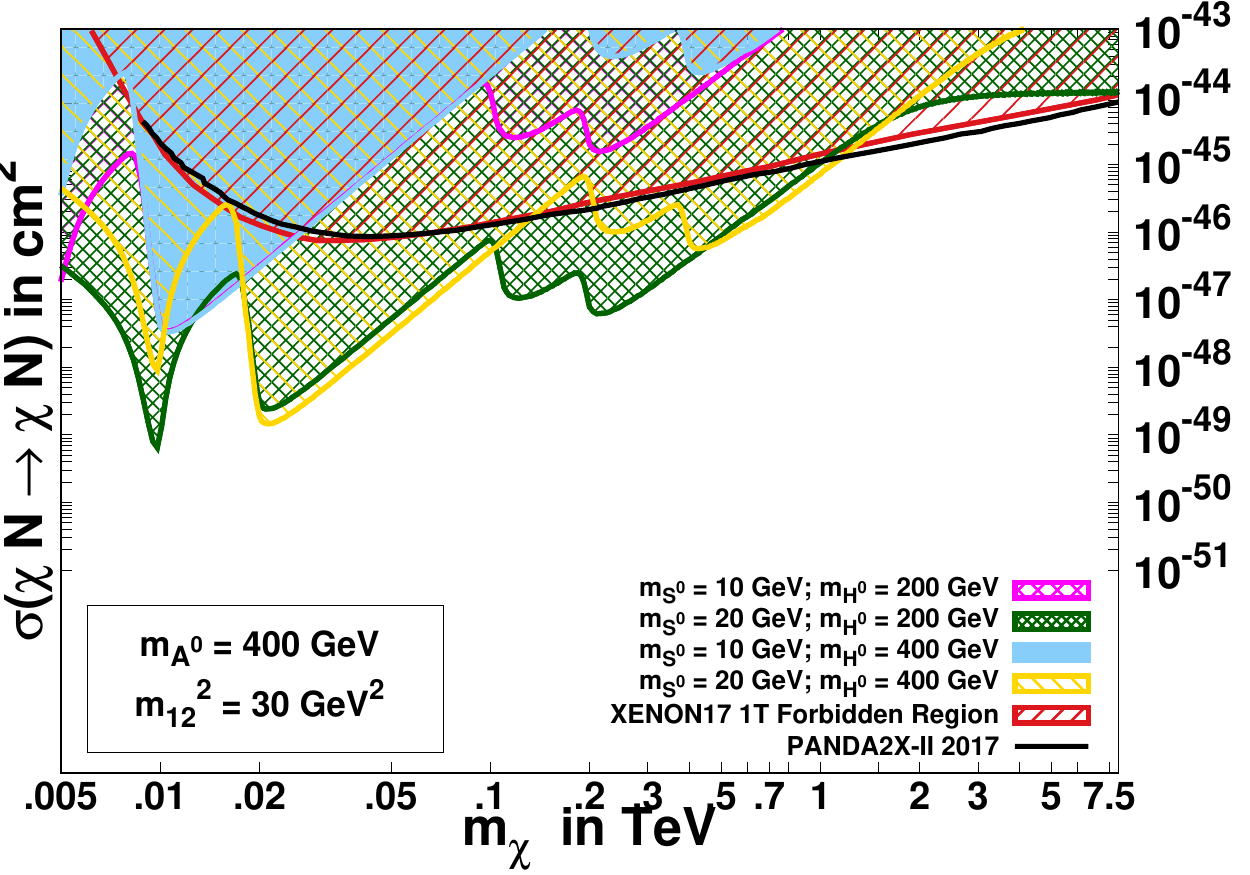}
  \caption{}
  \label{fig:Direct_Detectionp2c}
\end{subfigure}%
\begin{subfigure}{.45\textwidth}\centering
  \includegraphics[width=\columnwidth]{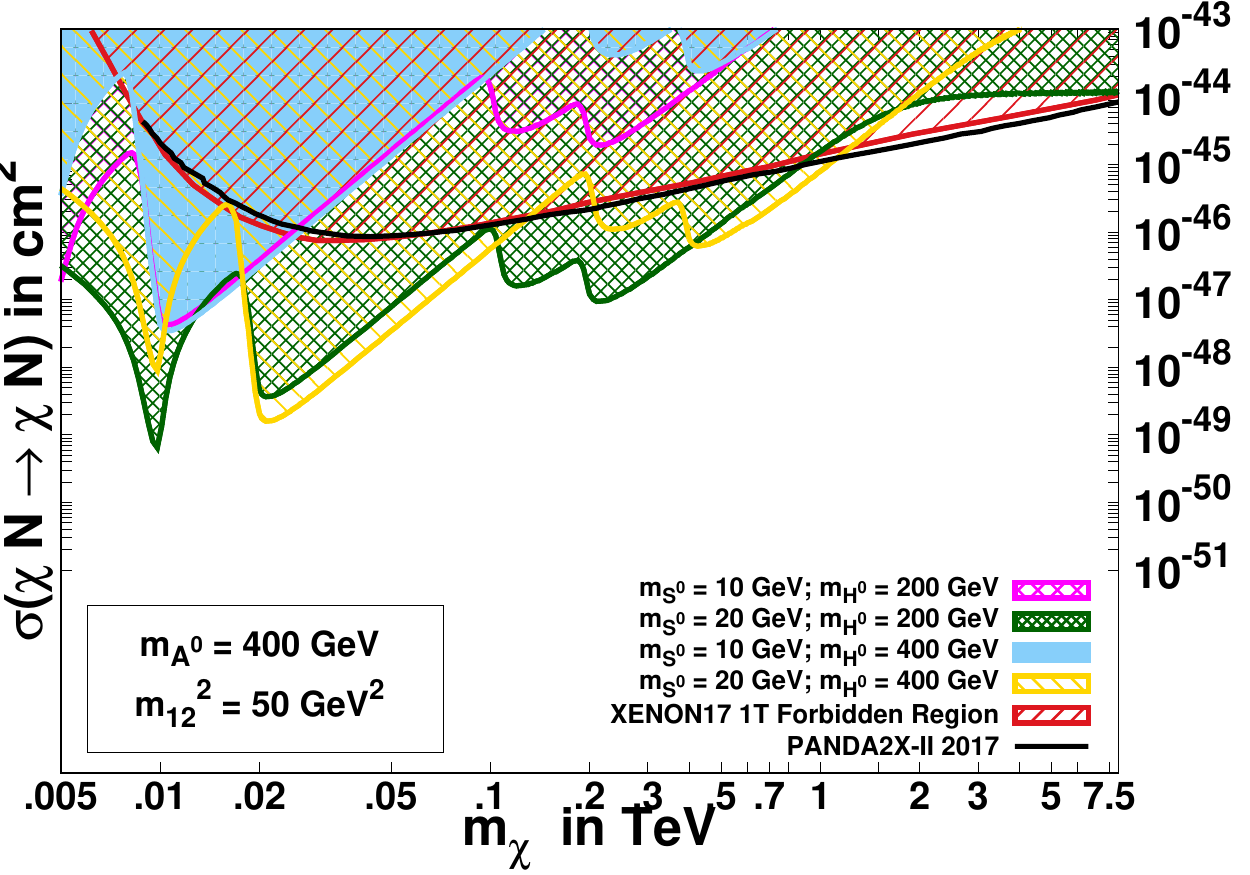}
  \caption{}
  \label{fig:Direct_Detectionp2d}
\end{subfigure}%

\begin{subfigure}{.45\textwidth}\centering
  \includegraphics[width=\columnwidth]{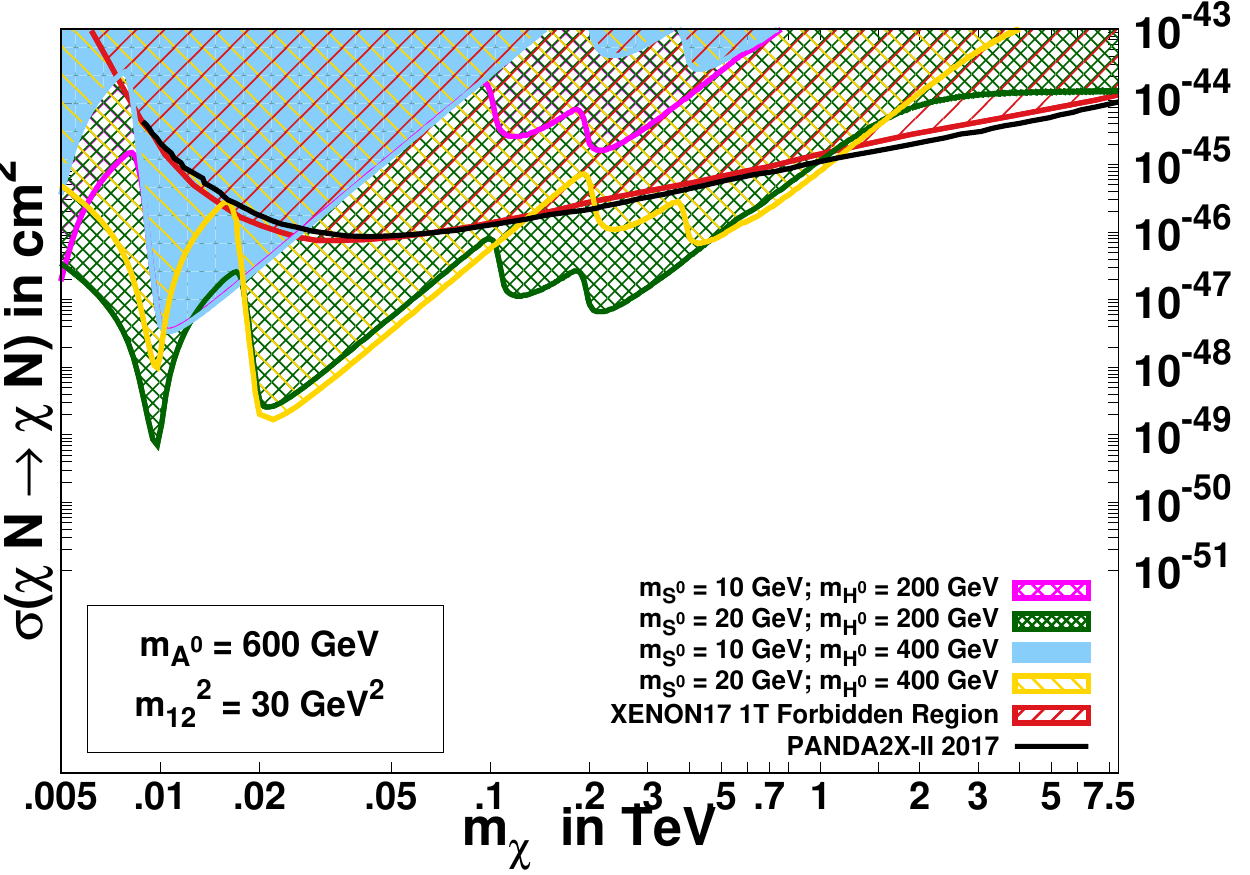}
  \caption{}
  \label{fig:Direct_Detectionp2e}
\end{subfigure}%
\begin{subfigure}{.45\textwidth}\centering
  \includegraphics[width=\columnwidth]{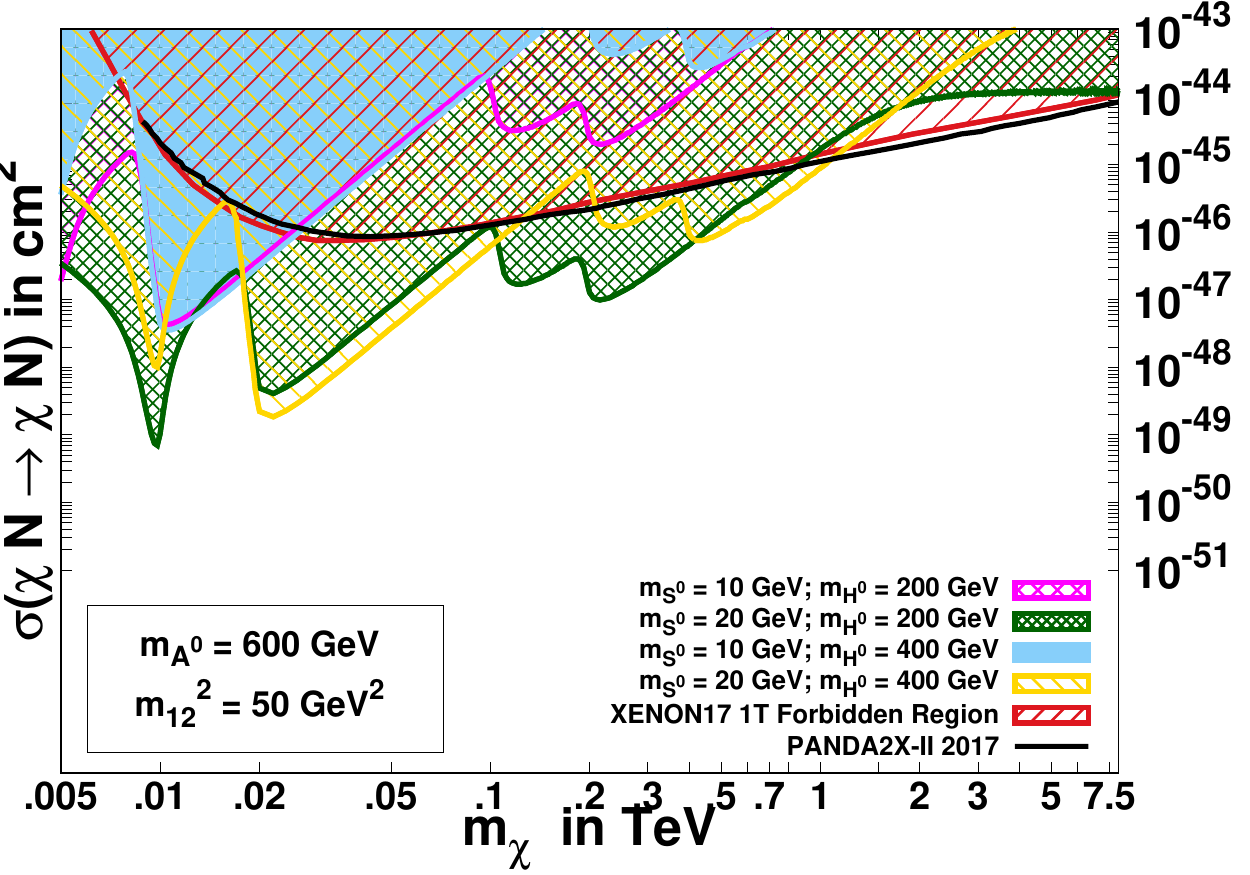}
  \caption{}
  \label{fig:Direct_Detectionp2f}
\end{subfigure}%
\caption{\small \em{Figures \ref{fig:Direct_Detectionp2a} to  \ref{fig:Direct_Detectionp2f} show the spin-independent DM-Nucleon cross-section variation with the $m_{\chi}$ for fixed $m_{H^\pm}$ = 600 GeV
, $\delta_{13}$ = 0.2 and  different choices of $m_{12}^2$ and $m_{A^0}$. All points on the contours satisfy the relic density 0.119 and also explain the discrepancy  $\Delta a_{\mu}=\,268(63)\,\times10^{-11}$. In the left and right panels, we plot the variation curves (bold lines) and  allowed (shaded) regions for five combinations of $m_{S^0}$ and $m_{H^0}$. The  upper limit from PANDA 2X-II 2017 \cite{Cui:2017nnn} and  XENON-1T \cite{Aprile:2015uzo,Aprile:2017aty} are also shown along with the forbidden region shaded in red.}}
\label{fig:Direct_Detectionp2}
\end{figure}
\begin{figure}[h!]
\centering 
\begin{subfigure}{.45\textwidth}\centering
  \includegraphics[width=\columnwidth]{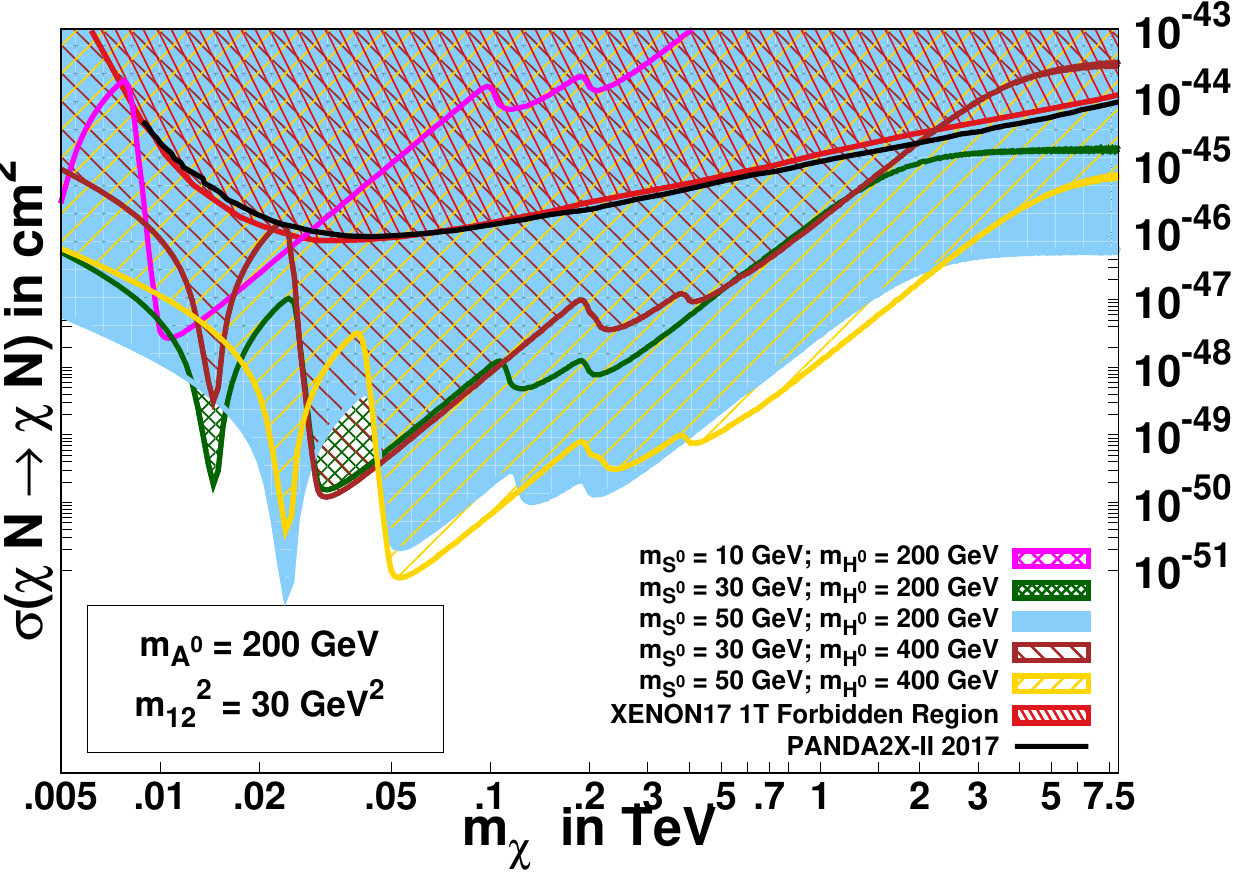}
  \caption{}
  \label{fig:Direct_Detectionp4a}
\end{subfigure}%
\begin{subfigure}{.45\textwidth}\centering
  \includegraphics[width=\columnwidth]{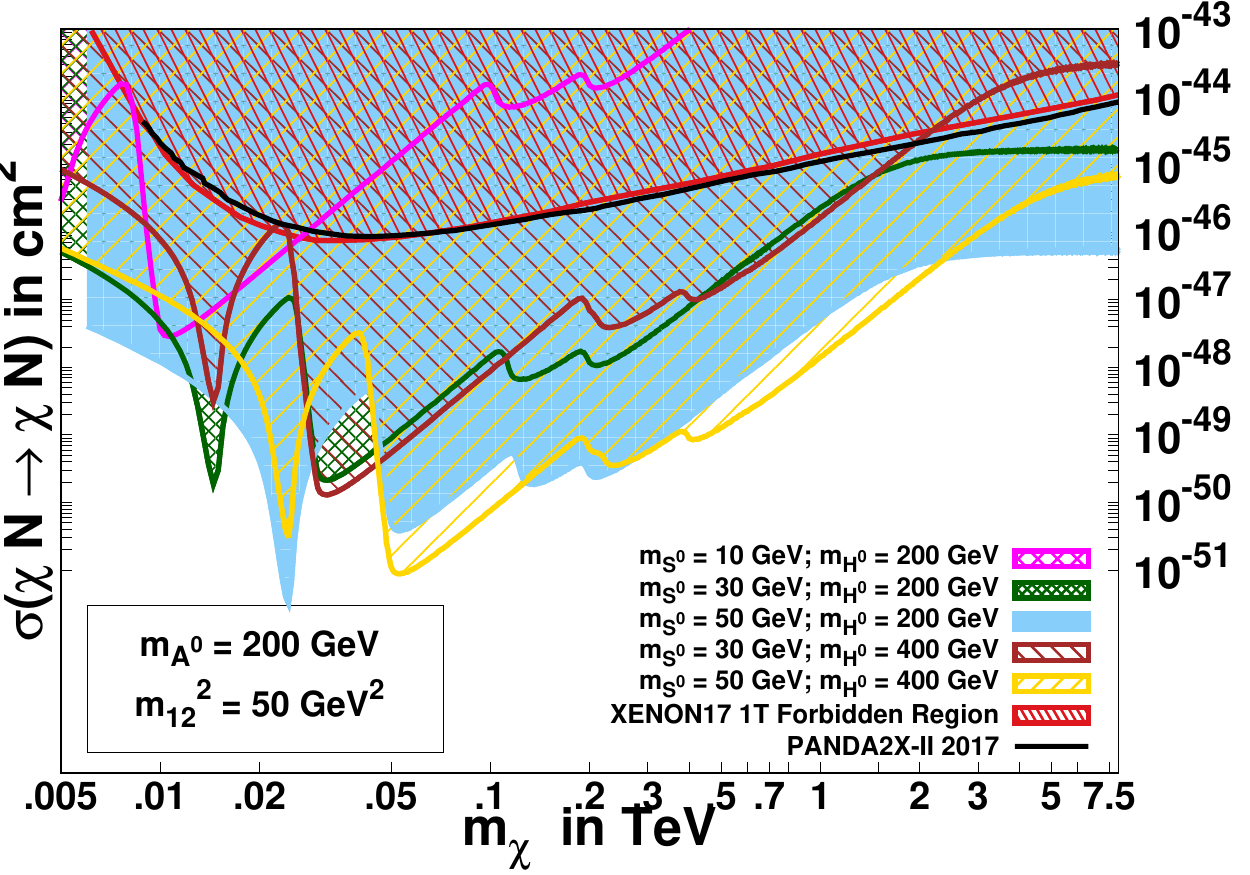}
  \caption{}
  \label{fig:Direct_Detectionp4b}
\end{subfigure}%

\begin{subfigure}{.45\textwidth}\centering
  \includegraphics[width=\columnwidth]{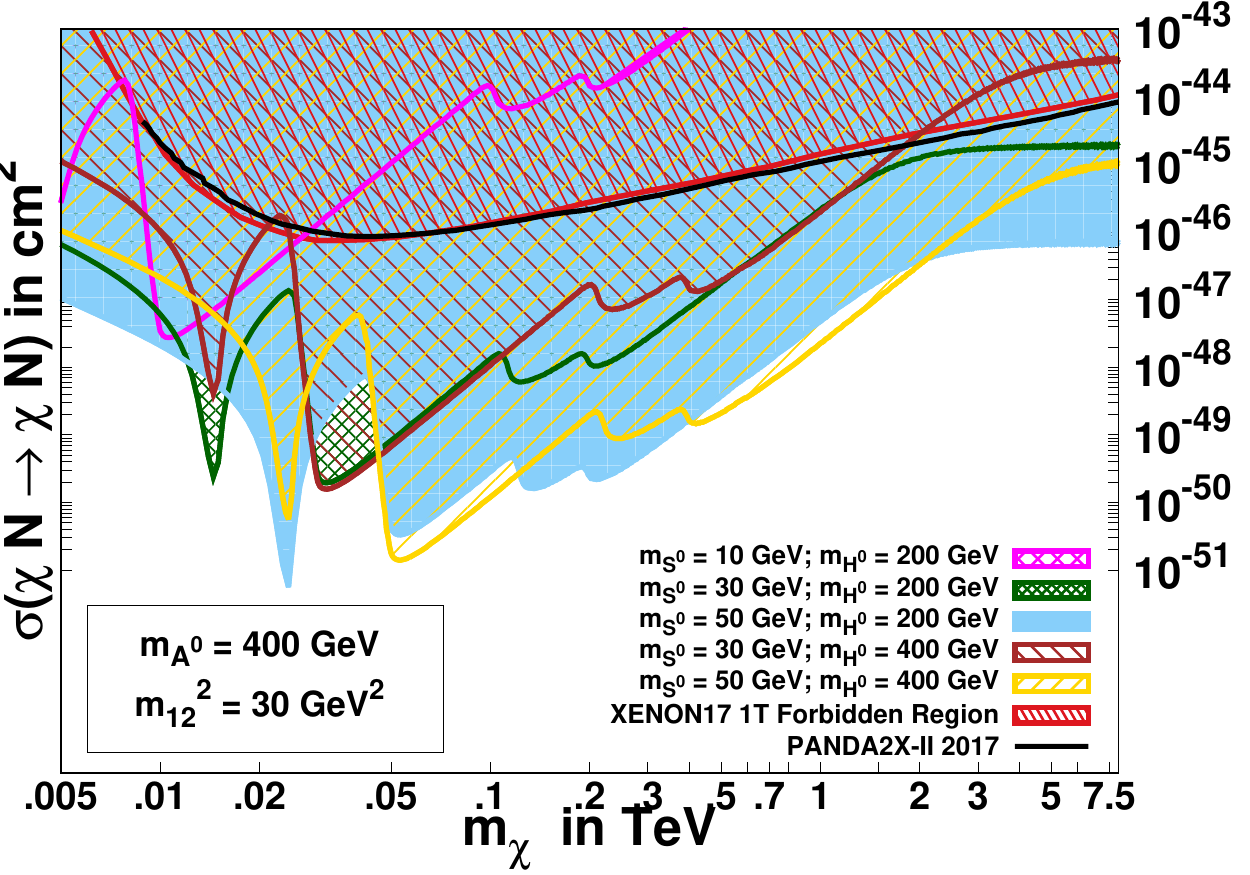}
  \caption{}
  \label{fig:Direct_Detectionp4c}
\end{subfigure}%
\begin{subfigure}{.45\textwidth}\centering
  \includegraphics[width=\columnwidth]{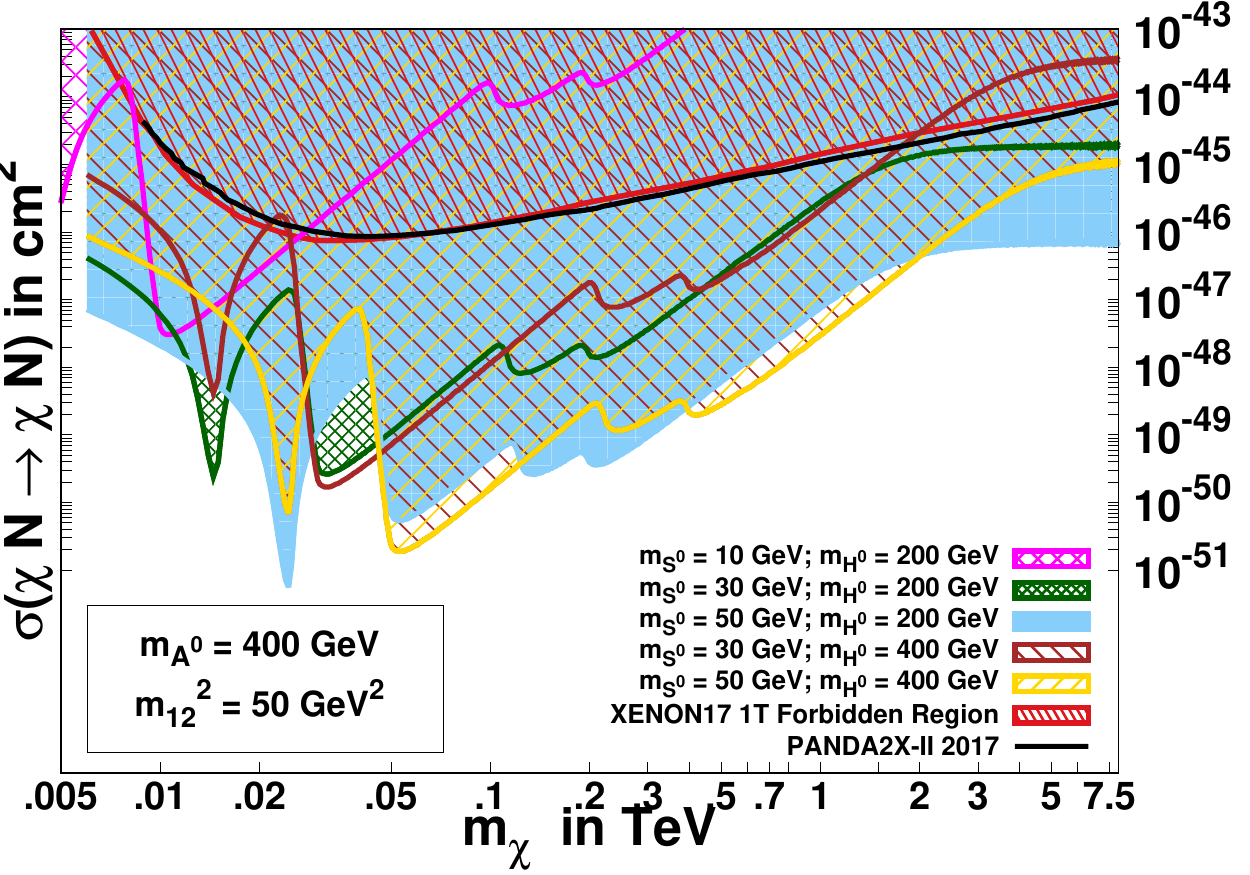}
  \caption{}
  \label{fig:Direct_Detectionp4d}
\end{subfigure}%

\begin{subfigure}{.45\textwidth}\centering
  \includegraphics[width=\columnwidth]{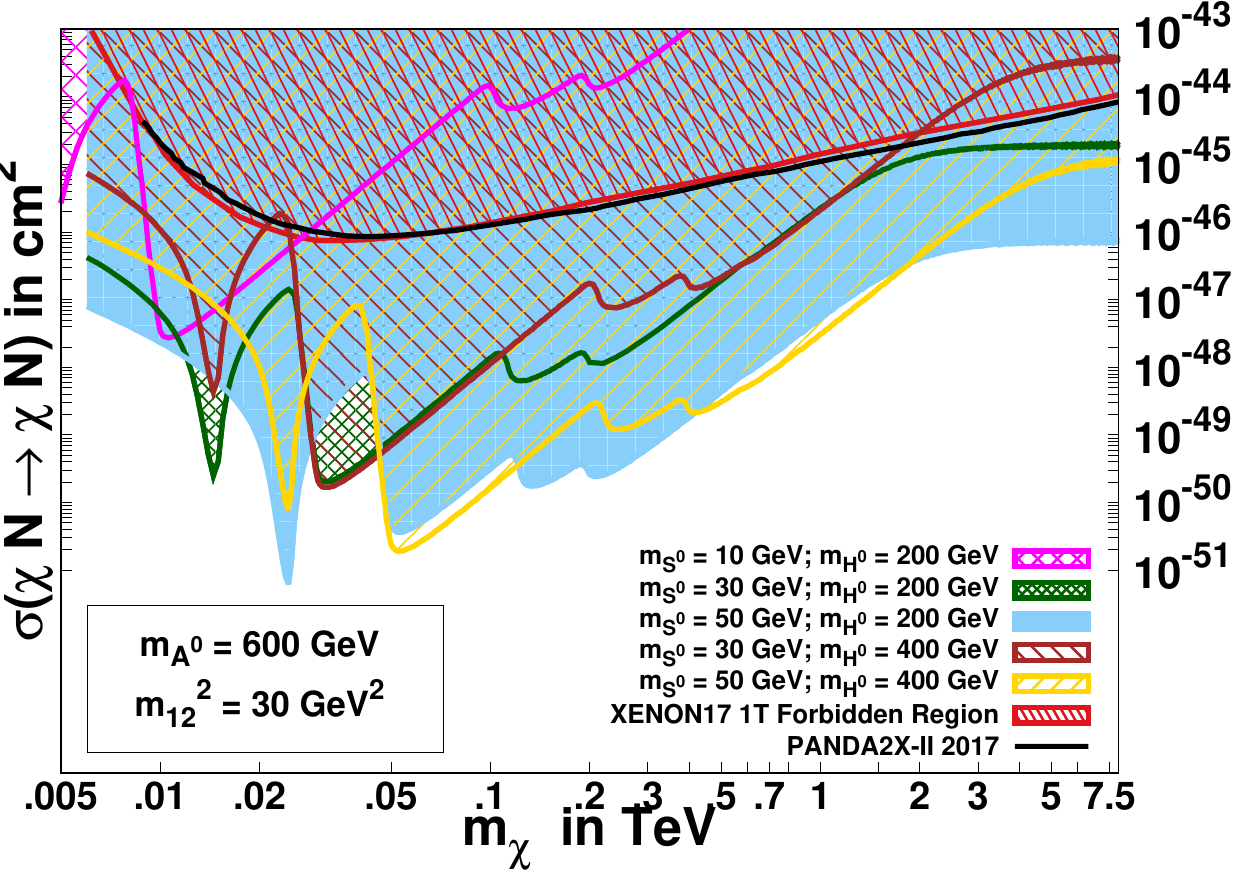}
  \caption{}
  \label{fig:Direct_Detectionp4e}
\end{subfigure}%
\begin{subfigure}{.45\textwidth}\centering
  \includegraphics[width=\columnwidth]{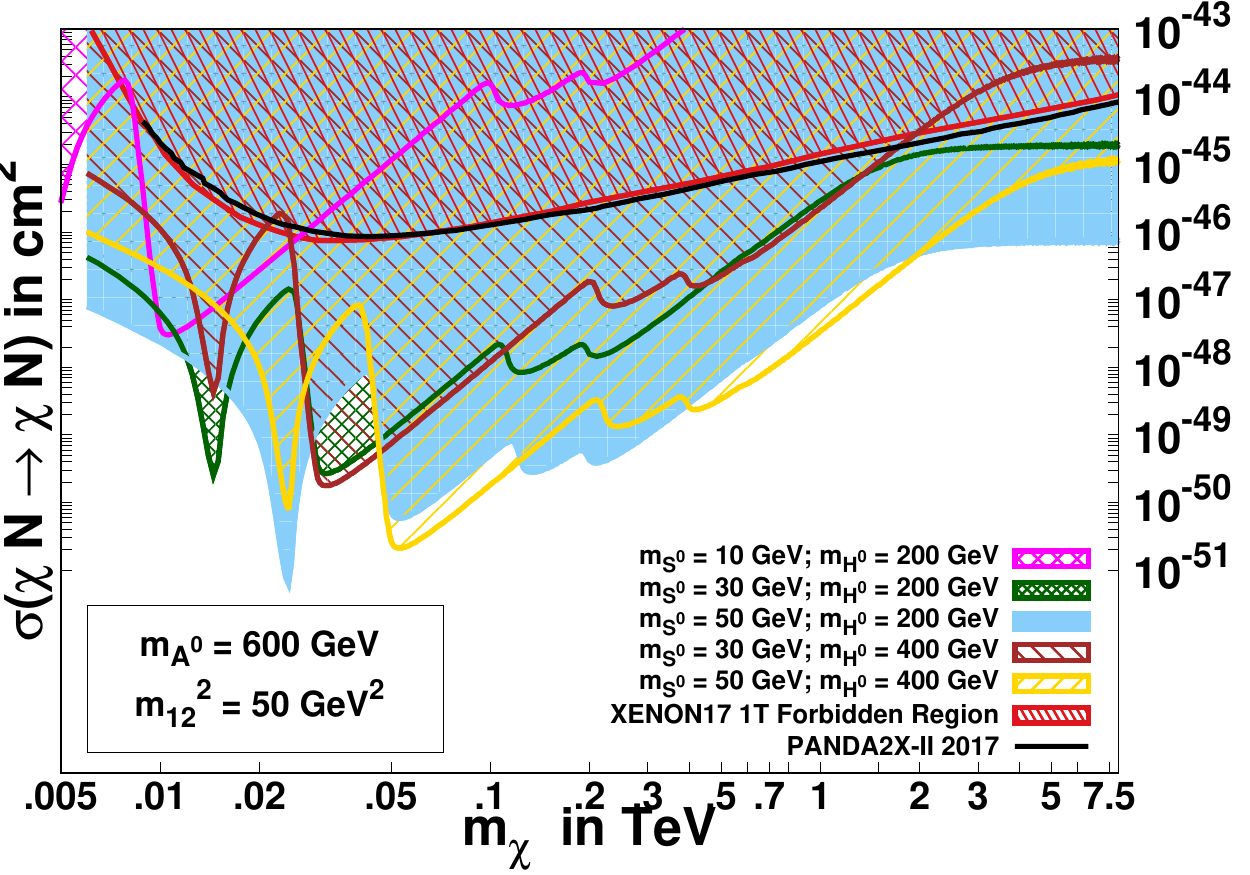}
  \caption{}
  \label{fig:Direct_Detectionp4f}
\end{subfigure}%
\caption{\small \em{Figures \ref{fig:Direct_Detectionp4a} to  \ref{fig:Direct_Detectionp4f} show the spin-independent DM-Nucleon cross-section variation with the $m_{\chi}$ for fixed $m_{H^\pm}$ = 600 GeV
, $\delta_{13}$ = 0.4 and  different choices of $m_{12}^2$ and $m_{A^0}$. All points on the contours satisfy the relic density 0.119 and also explain the discrepancy  $\Delta a_{\mu}=\,268(63)\,\times10^{-11}$. In the left and right panels, we plot the variation curves (bold lines) and  allowed (shaded) regions for five combinations of $m_{S^0}$ and $m_{H^0}$. The  upper limit from PANDA 2X-II 2017 \cite{Cui:2017nnn} and  XENON-1T \cite{Aprile:2015uzo,Aprile:2017aty} are also shown along with the forbidden region shaded in red.}}
\label{fig:Direct_Detectionp4}
\end{figure}

\subsection{Direct Detection}
\label{subsection_Direct_Detection}
Direct detection of DM measures the recoil generated by DM interaction with matter. For the case of lepto-philic DM, we have tree level DM-Electron interaction, where DM can scatter with electron in-elastically, leading to ionization of the atom to which it is bound or elastically, where excitation of atom is succeeded by de-excitation, releasing a photon. The DM-Nucleon scattering in this model occurs at the loop level and though suppressed by one or two powers of respective coupling strengths and  the loop factor, it vastly dominates over the DM-Electron and DM-Atom scattering \cite{Kopp:2009et,Kopp:2014tsa,DEramo:2017zqw}.

\par The scalar spin-independent DM-Nucleon scattering are induced through    the effective DM-photon, DM-quark and DM-gluon interactions which are mediated by the singlet scalar portal of the model. Following  reference \cite{Kopp:2009et},  we approximate the DM-Nucleon scattering cross-section through two photons by integrating out the contributions of heavier  fermions running in the loop. The total cross-section Spin-Independent DM-Nucleon in this case is given as
\begin{eqnarray}
\sigma^{\gamma \gamma}_N = \left(\frac{\alpha_{em} Z}{\pi}\right)^2 \left[\frac{\mu^2_N}{\pi} \left(\frac{\alpha_{em} Z}{\pi m^2_{S^0}}\right)^2\right]\left( \frac{\pi^2}{12}\right)^2 \left( \frac{\mu_N v}{m_{\tau}}\right)^2 2\left(  g_{\chi S^0} \xi^{S^0}_l \frac{m_\tau}{v_0}\right)^2\label{photoncrosssec}
\end{eqnarray} 
where $Z$ is the atomic number of the detector material,  $\mu_N$ is the reduced mass of the DM-Nucleon system  and $v$ is the   DM velocity  of the order of 10$^{-3}$. 

\par The effective DM-gluon interactions are induced through a quark triangle loop, where, the negligible contribution of light quarks  $u$, $d$ and $s$  to the loop integral can be dropped.  In this approximation,  the effective Lagrangian for singlet scalar-gluon interactions  can be derived by integrating out  contributions from heavy quarks $c$, $b$ and $t$ in the triangle loop and  can be written as
\begin{equation}
{\cal L}_{\rm eff.}^{S^0gg}= - \frac{\xi^{S^0}_q}{12 \pi}\frac{\alpha_s}{v_o}\left\{ \sum_{q=c,b,t} I_q \right\} G_{\mu\nu}^a G^{\mu\nu a} S^0
\end{equation}
 where the loop integral $I_q$ is given in Appendix \ref{loopIntegral}. The DM-gluon effective Lagrangian is the given as
 \begin{equation}
 {\cal L}_{\rm eff.}^{\chi \chi gg} = \frac{\alpha_s(m_{S^0})}{12\pi}\frac{\xi^{S^0}_q g_{\chi S^0}}{v_o m^2_{S^0}}\left\{ \sum_{q=c,b,t} I_q \right\} \bar{\chi}\chi G_{\mu\nu}^a G^{\mu\nu a}.\label{lagchchigg}
 \end{equation}
Using \eqref{lagchchigg}, the  DM-gluon scattering cross-section  can be computed and given as:
\begin{eqnarray}\label{sigma_N_gluons}
 \sigma^{gg}_N  = \left(\frac{ 2\xi^{S^0}_q g_{\chi S^0} m_N}{m^2_{S^0} 27 v_o}\right)^2 \left\vert\sum_{q=c,b,t} I_q\right\vert^2 \frac{2}{\pi ( m_\chi + m_{N} )^2 } m_N ^2 m^2_\chi \label{gluoncrosssec}
 \end{eqnarray}
To compare the  cross-sections given in \eqref{photoncrosssec} and \eqref{gluoncrosssec}, we evaluate the ratio  
\begin{equation}
\frac{\sigma^{\gamma \gamma}_N}{ \sigma^{gg}_N}\simeq (\alpha_{em})^4 \frac{\mu^2_N}{m^2_N}\left(\frac{\xi^{S^0}_\tau}{\xi^{S^0}_q}\right)^2 \left(\frac{9}{8}\right)^2 \frac{v^2}{c^2}\simeq 10^{-6}-10^{-10}.
\end{equation}
Thus even though the effective DM-quark coupling is suppressed by $\tan^2\beta$ {\it w.r.t} that of DM-lepton coupling,  the scattering cross-sections induced {\it via} the singlet coupled to the  quark-loop  dominates over the   $\sigma^{\gamma \gamma}_N$ due to suppression resulting from the fourth power of the electromagnetic coupling. 
\par We convolute the DM-quark and DM-gluon scattering cross-sections with the quark form factor $F^{q_i/N} (q^2)$ and gluon form factor $F^{g/N} (q^2)$  respectively to compute nuclear recoil energy observed in the experiment. However, this form factor is extracted at low $q^2\ll m^2_N$ \cite{Bishara:2017nnn,DelNobile:2013sia,Dutta:2017jfj}. The form factors are defined as  
\begin{subequations}
\begin{eqnarray}
\left\langle N^\prime \left\vert \frac{\alpha_s}{12\pi}{G^a}^{\mu\nu} G^a_{\mu\nu}\right\vert N\right\rangle &= &F^{g/N} (q^2) \bar{u}^\prime_N u_N\\
\left\langle N^\prime \left\vert m_{q_i} \bar{q_i}q_i\right\vert N\right\rangle &= &F^{q_i/N} (q^2) \bar{u}^\prime_N u_N
\end{eqnarray}
\end{subequations}
Since, 
$m_N\equiv\sum_{u,d,s}\left\langle N\left\vert m_q \bar{q}q\right\vert N\right\rangle - \frac{9\alpha_S}{8\pi} \left\langle N\left\vert {G^a}^{\mu\nu}{G^a}_{\mu\nu}\right\vert N\right\rangle $, the gluon form factor can be expressed as 
\begin{eqnarray}
F^{g/N} = 1 - \sum_{u,d,s}\frac{F^{q_i/N}_S (q^2)}{m_N}=-\frac{1}{m_N}\frac{9\alpha_s}{8\pi}\left\langle  N\left\vert {G^a}^{\mu\nu}{G^a}_{\mu\nu}\right\vert N\right\rangle
\end{eqnarray}
The $F^{g/N}$ is found  to be $\approx$ 0.92 using the values for $F^{q_i/N}_S (q^2)$ as quoted in the literature  \cite{DelNobile:2013sia}. Thus, at the low momentum transfer the quartic DM-gluon   $\left(\chi \chi g g\right)$ effective interaction induced through relatively heavy quarks dominates over the quartic DM-quark  $\left(\chi \chi q q\right)$  effective interactions for light quarks in the direct-detection experiments.
 
Using the  expression \ref{sigma_N_gluons} we have plotted the spin-independent DM-Nucleon scattering cross-section as a function of the DM mass $m_\chi$. Figures \ref{fig:Direct_Detectionp2} and \ref{fig:Direct_Detectionp4} corresponding to mixing angle $\delta_{13}$=0.2 and $\delta_{13}$=0.4 respectively. The parameter sets used in the computation of direct detection cross-section are consistent with the observed relic density as given in figures \ref{fig:Relic_Densityp2} and \ref{fig:Relic_Densityp4}. Different panels in figures  \ref{fig:Direct_Detectionp2} and \ref{fig:Direct_Detectionp4} show combinations of $m_{A^0}$ and $m^2_{12}$. In each panel different combinations of $m_{S^0}$ and $m_{H^0}$ are used as shown. Current bounds on spin-independent interactions from  experiments like PANDA 2X-II 2017 \cite{Cui:2017nnn} and XENON-1T \cite{Aprile:2015uzo,Aprile:2017aty} are also shown. It can be seen that most of the parameter space for $m_{S^0}$ less than 10 GeV is ruled out by the current bounds.
 \begin{figure}[h!]
\centering 
\begin{subfigure}{.48\textwidth}\centering
  \includegraphics[width=\columnwidth]{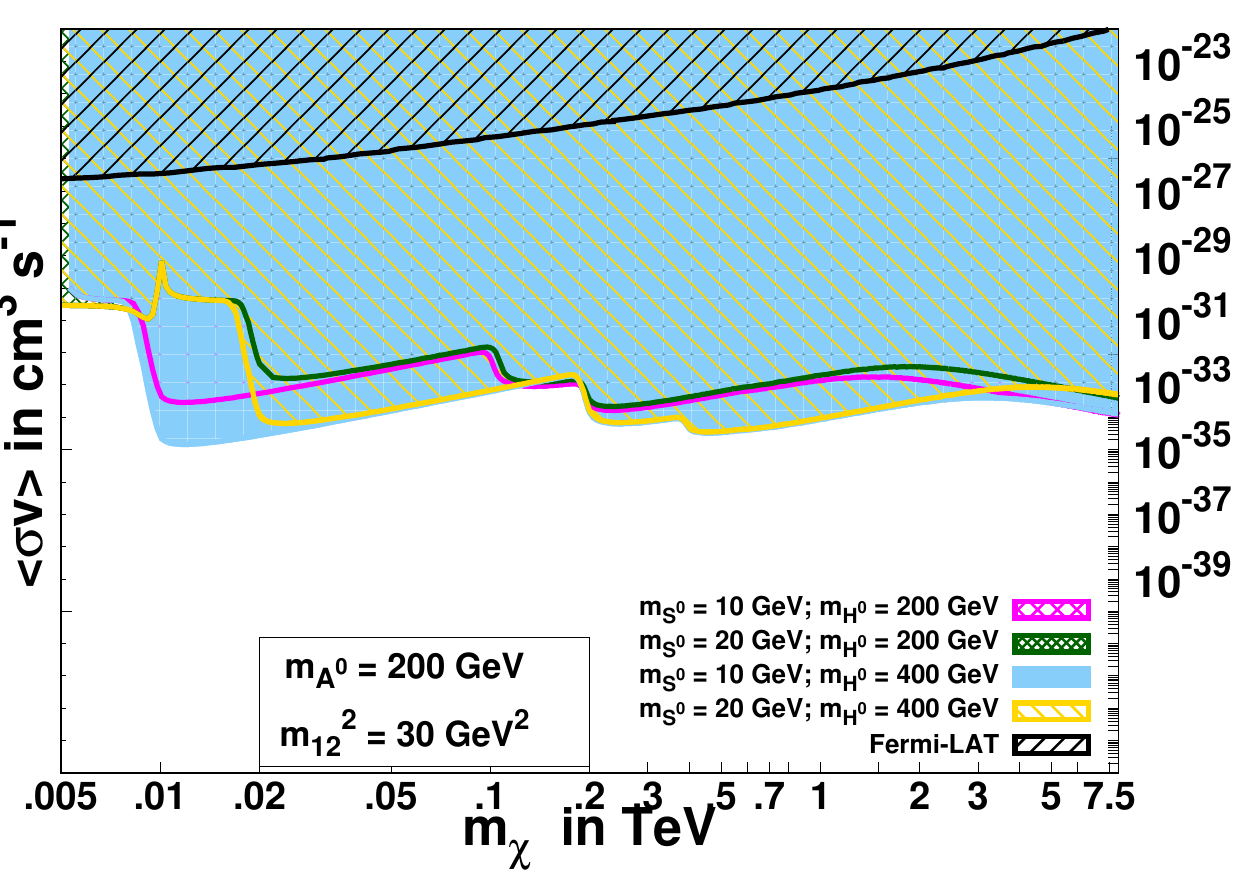}
  \caption{}
  \label{fig:Indirect_Detectionp2a}
\end{subfigure}%
\begin{subfigure}{.48\textwidth}\centering
  \includegraphics[width=\columnwidth]{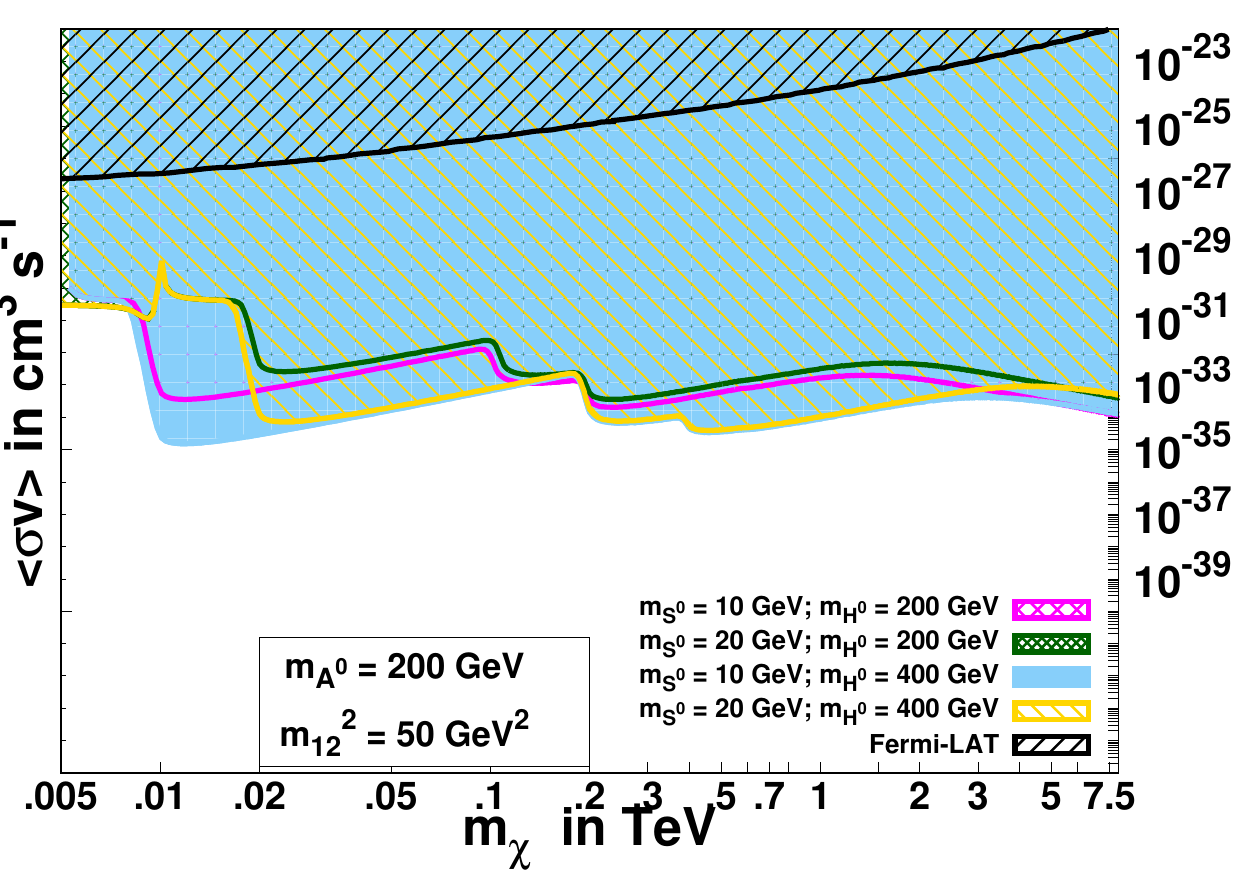}
  \caption{}
  \label{fig:Indirect_Detectionp2b}
\end{subfigure}%

\begin{subfigure}{.48\textwidth}\centering
  \includegraphics[width=\columnwidth]{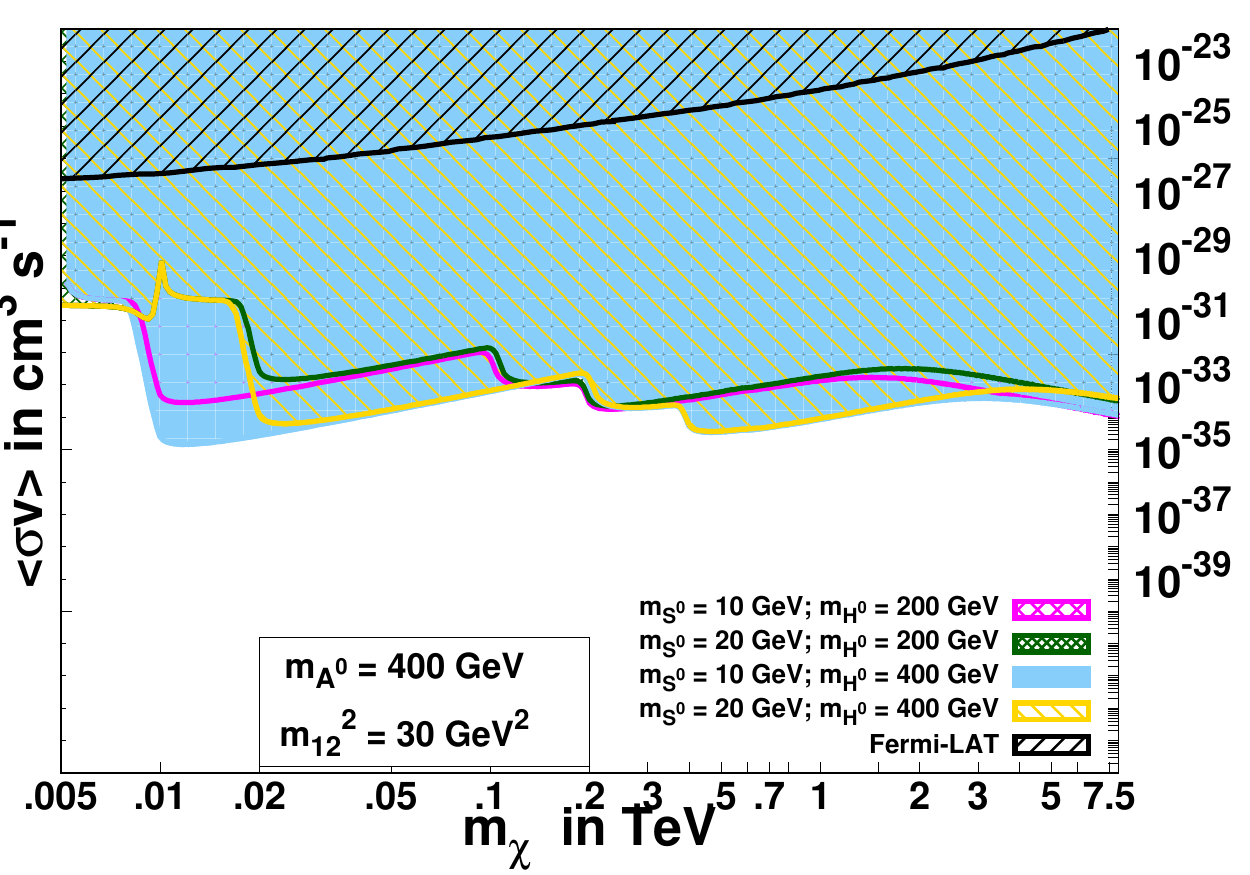}
  \caption{}
  \label{fig:Indirect_Detectionp2c}
\end{subfigure}%
\begin{subfigure}{.48\textwidth}\centering
  \includegraphics[width=\columnwidth]{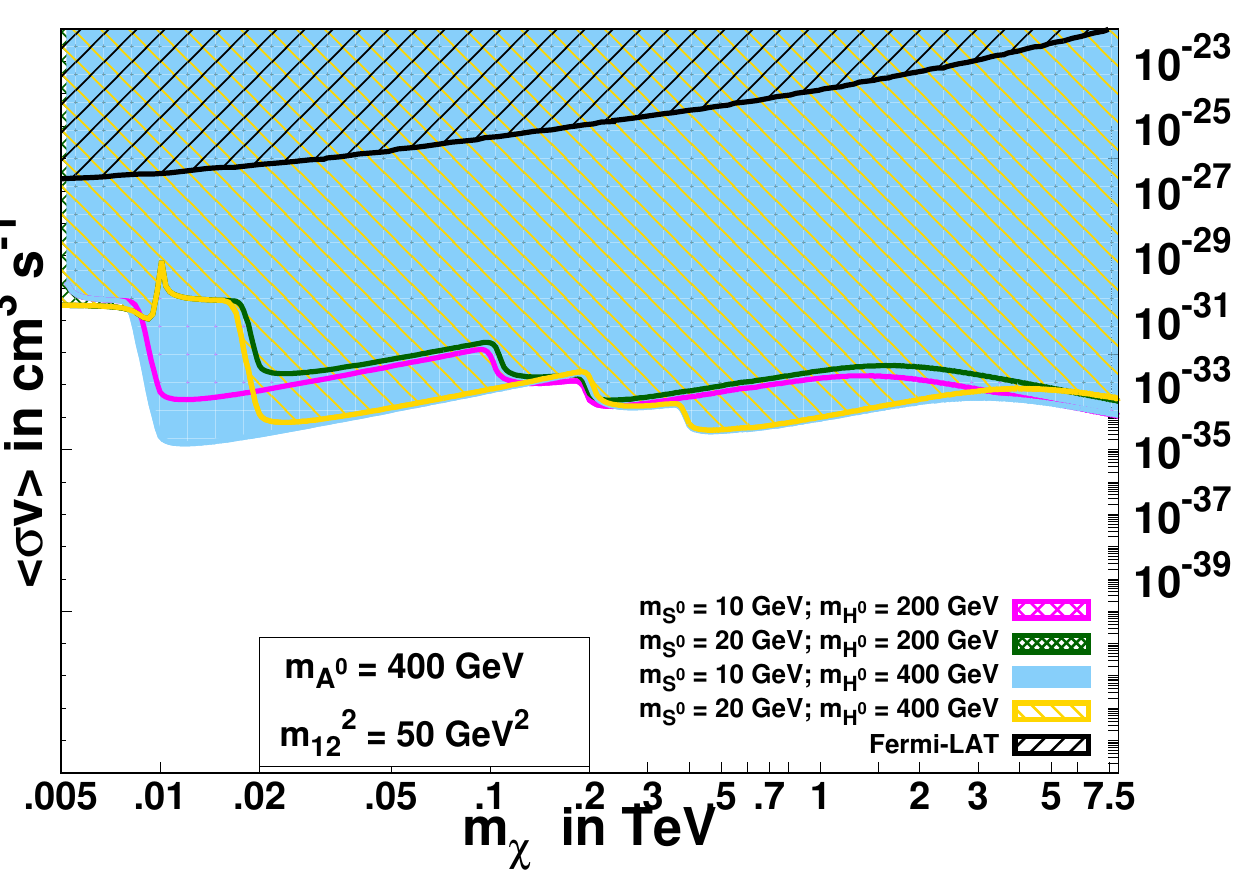}
  \caption{}
  \label{fig:Indirect_Detectionp2d}
\end{subfigure}%

\begin{subfigure}{.48\textwidth}\centering
  \includegraphics[width=\columnwidth]{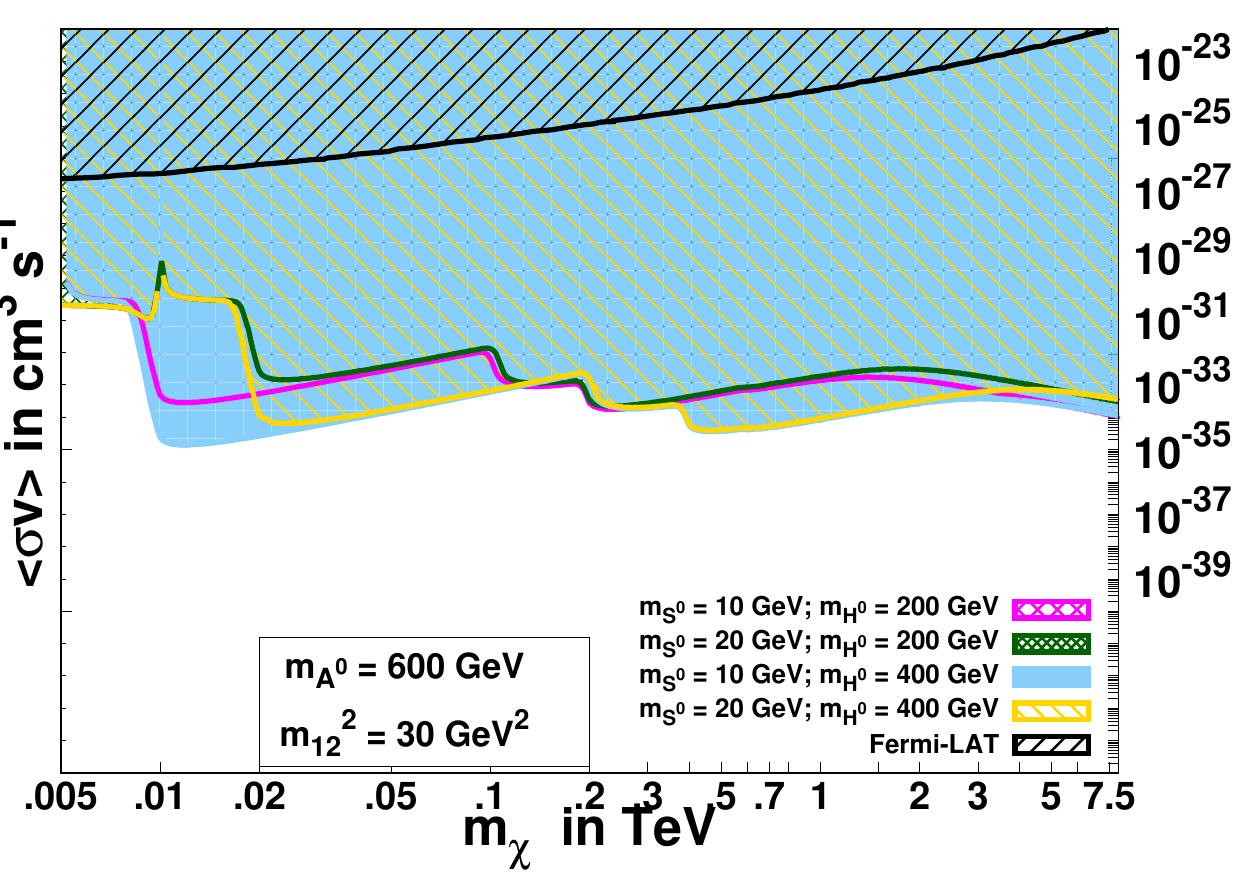}
  \caption{}
  \label{fig:Indirect_Detectionp2e}
\end{subfigure}%
\begin{subfigure}{.48\textwidth}\centering
  \includegraphics[width=\columnwidth]{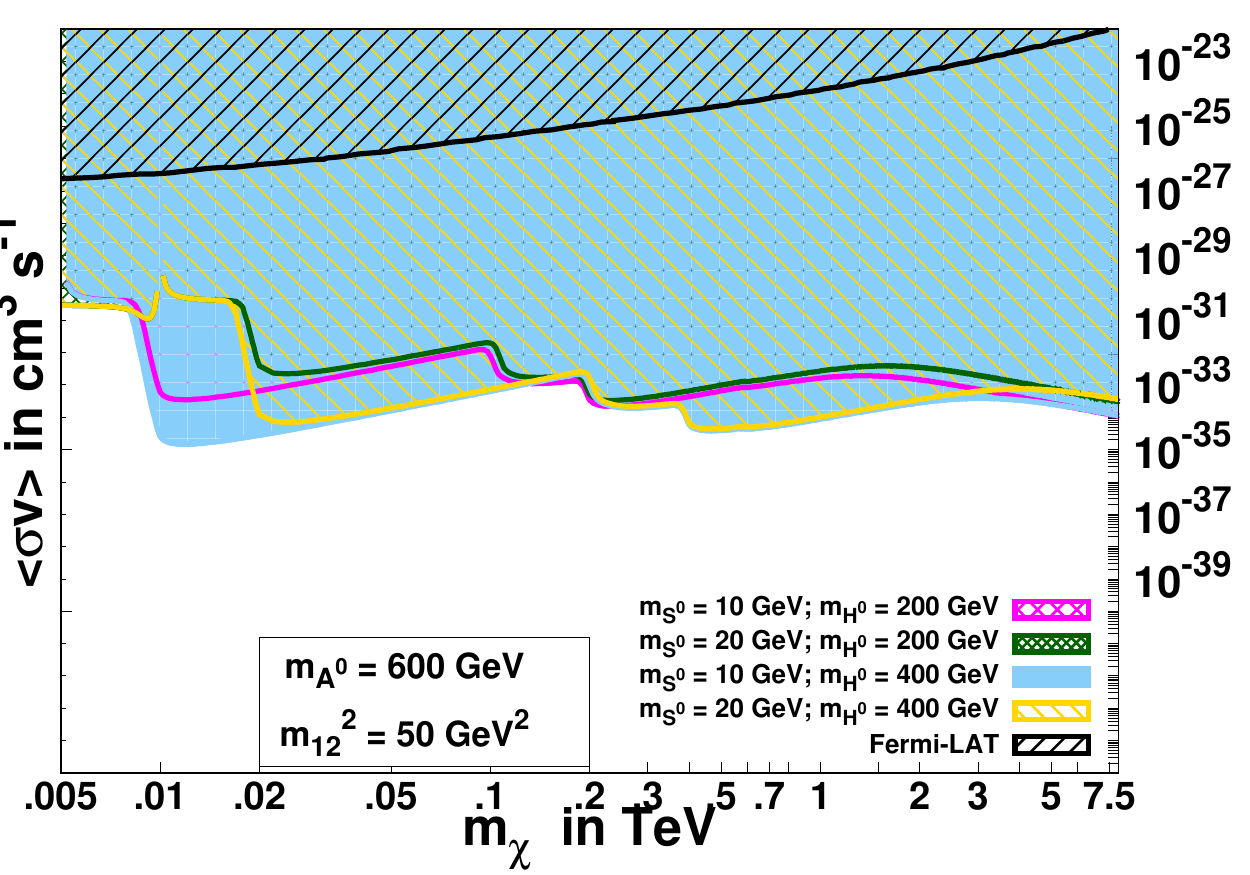}
  \caption{}
  \label{fig:Indirect_Detectionp2f}
\end{subfigure}%
\caption{\small \em{Figures \ref{fig:Indirect_Detectionp2a} to \ref{fig:Indirect_Detectionp2f} show the 
velocity-averaged scattering cross-section $<\sigma v>_{\tau^+ \tau^- }$ variation with the $m_{\chi}$ for fixed $m_{H^\pm}$ = 600 GeV, $\delta_{13}$ = 0.2  and  different choices of $m_{12}^2$ and $m_{A^0}$. All points on the contours satisfy the relic density 0.119 and also explain the discrepancy  $\Delta a_{\mu}=\,268(63)\,\times10^{-11}$. In the left and right panels, we plot the variation curves (bold lines) and  allowed (shaded) regions for four five combinations of $m_{S^0}$ and $m_{H^0}$. The  upper limit  on  velocity-averaged annihilation cross-section observed from Fermi-LAT \cite{Ackermann:2015zua} is shown. }}   
\label{fig:Indirect_Detectionp2}
\end{figure}
\begin{figure}[h!]
\centering 
\begin{subfigure}{.48\textwidth}\centering
  \includegraphics[width=\columnwidth]{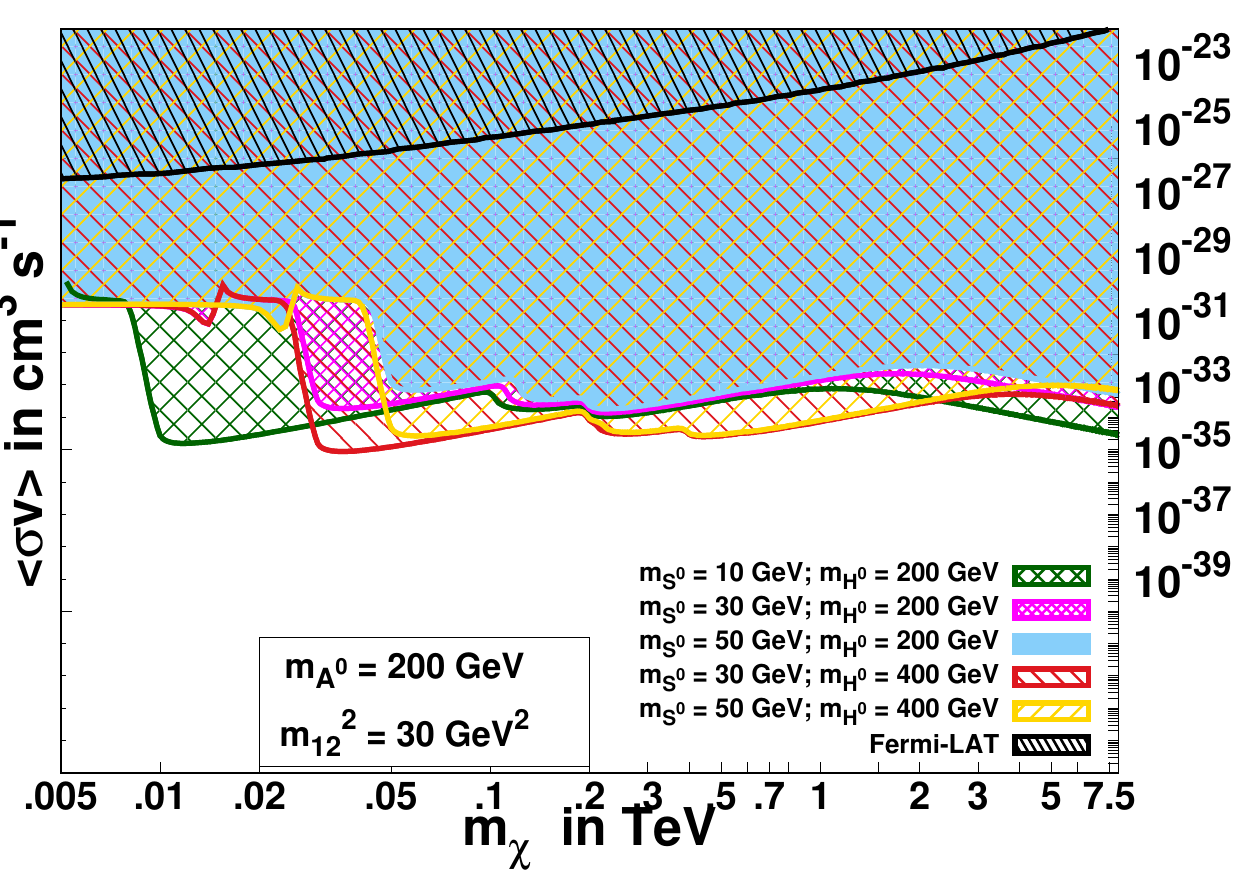}
  \caption{}
  \label{fig:Indirect_Detectionp4a}
\end{subfigure}%
\begin{subfigure}{.48\textwidth}\centering
  \includegraphics[width=\columnwidth]{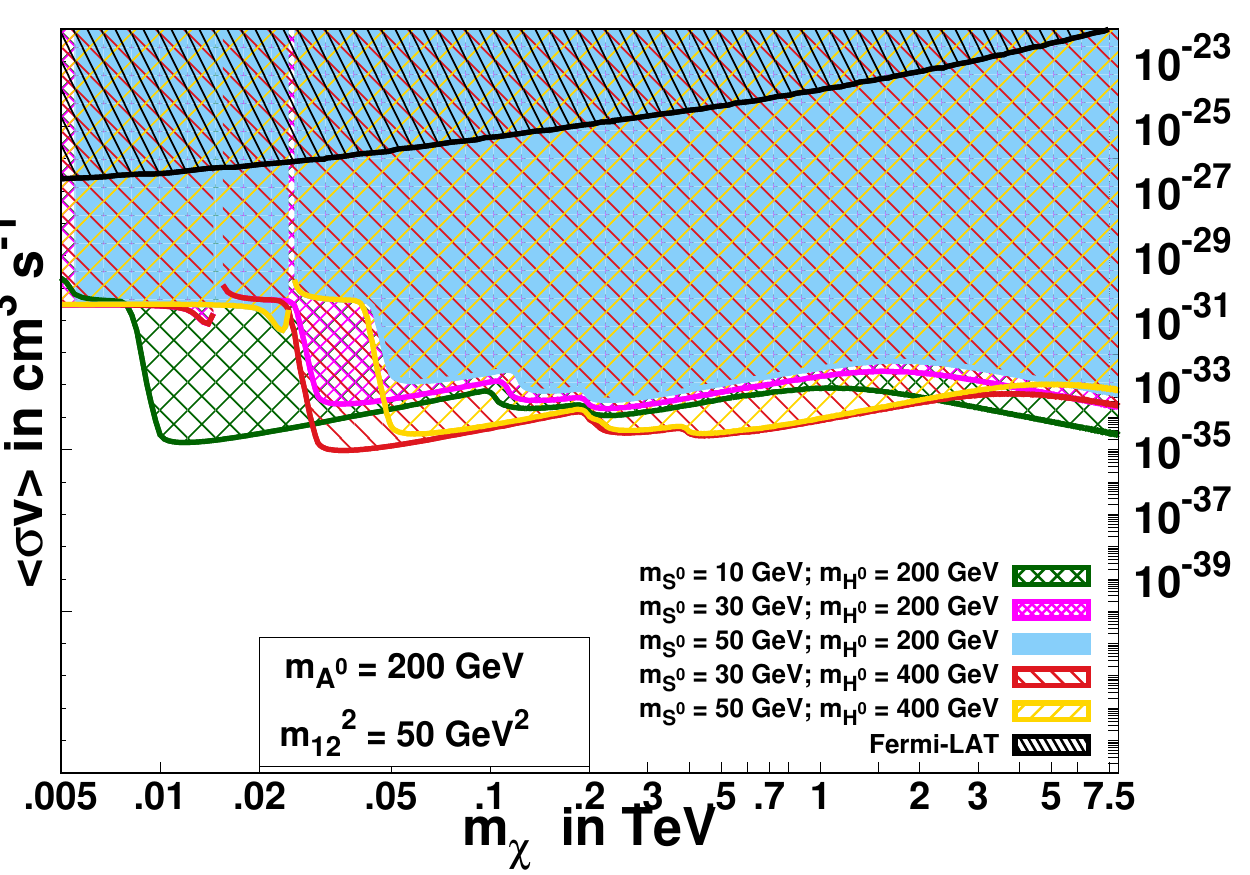}
  \caption{}
  \label{fig:Indirect_Detectionp4b}
\end{subfigure}%

\begin{subfigure}{.48\textwidth}\centering
  \includegraphics[width=\columnwidth]{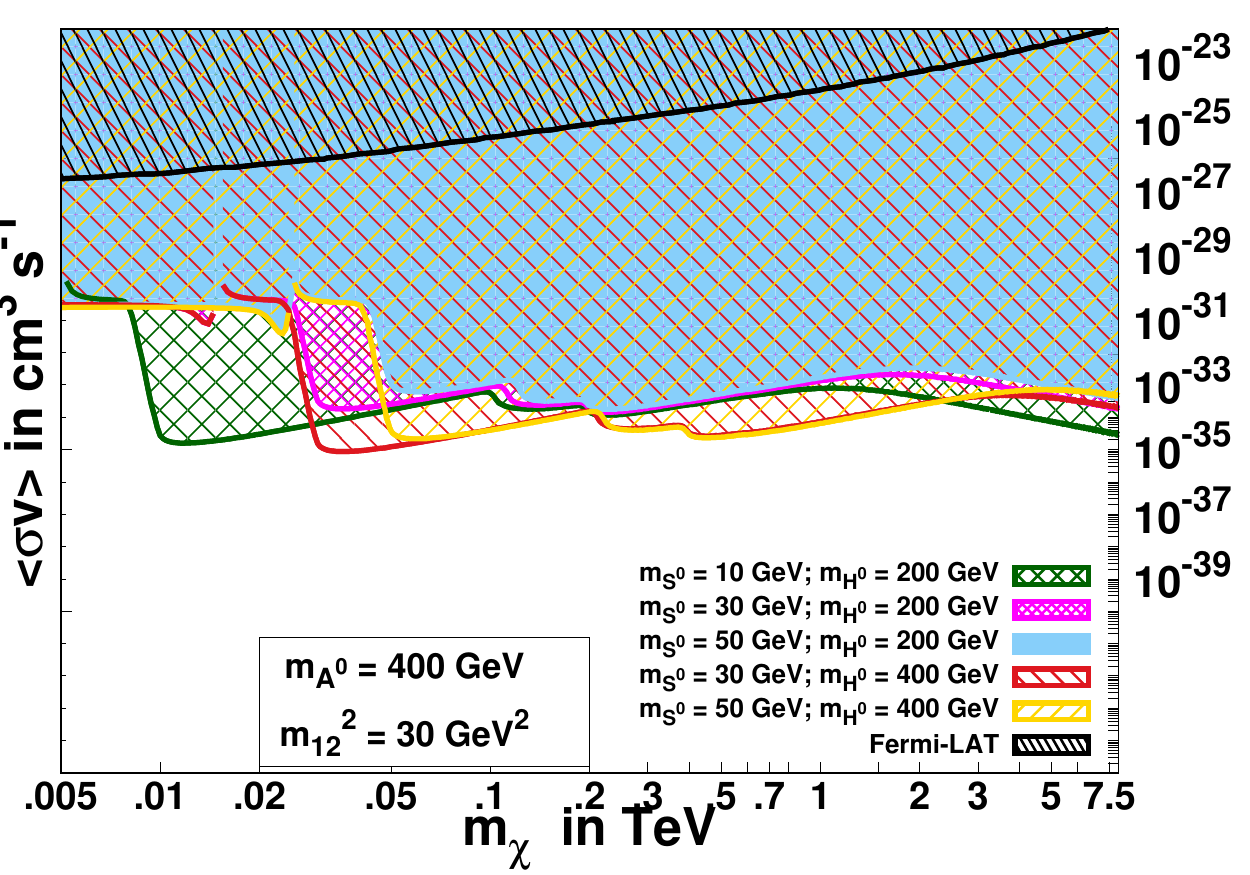}
  \caption{}
  \label{fig:Indirect_Detectionp4c}
\end{subfigure}%
\begin{subfigure}{.48\textwidth}\centering
  \includegraphics[width=\columnwidth]{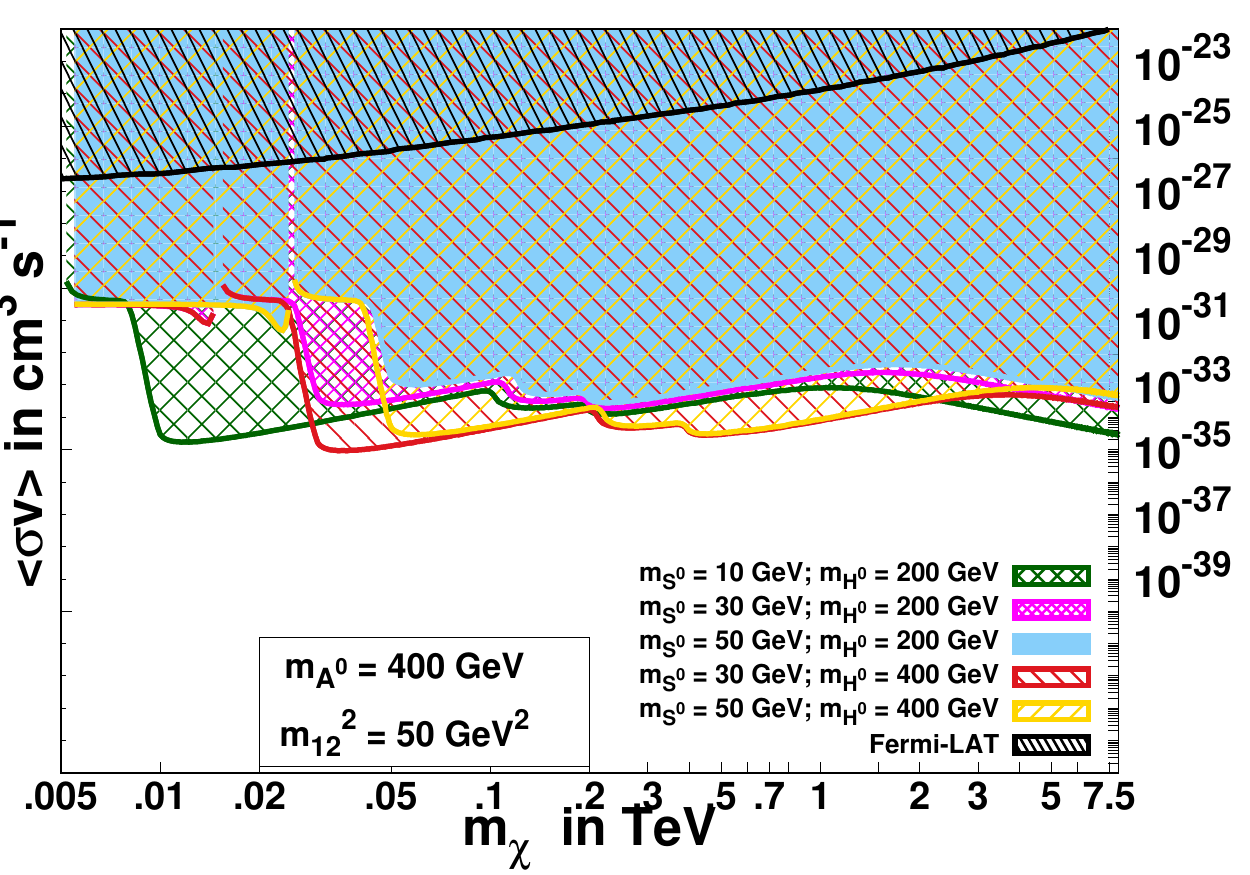}
  \caption{}
  \label{fig:Indirect_Detectionp4d}
\end{subfigure}%

\begin{subfigure}{.48\textwidth}\centering
  \includegraphics[width=\columnwidth]{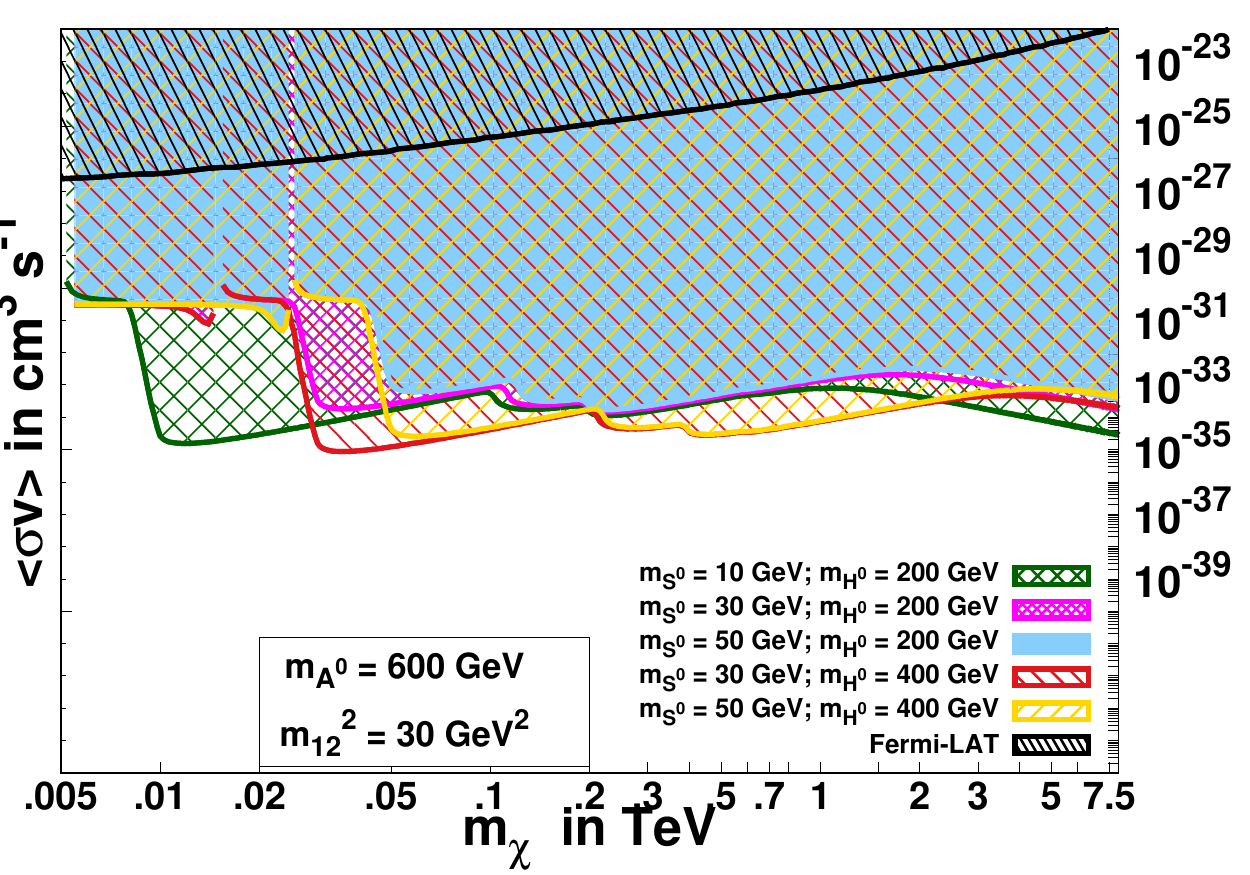}
  \caption{}
  \label{fig:Indirect_Detectionp4e}
\end{subfigure}%
\begin{subfigure}{.48\textwidth}\centering
  \includegraphics[width=\columnwidth]{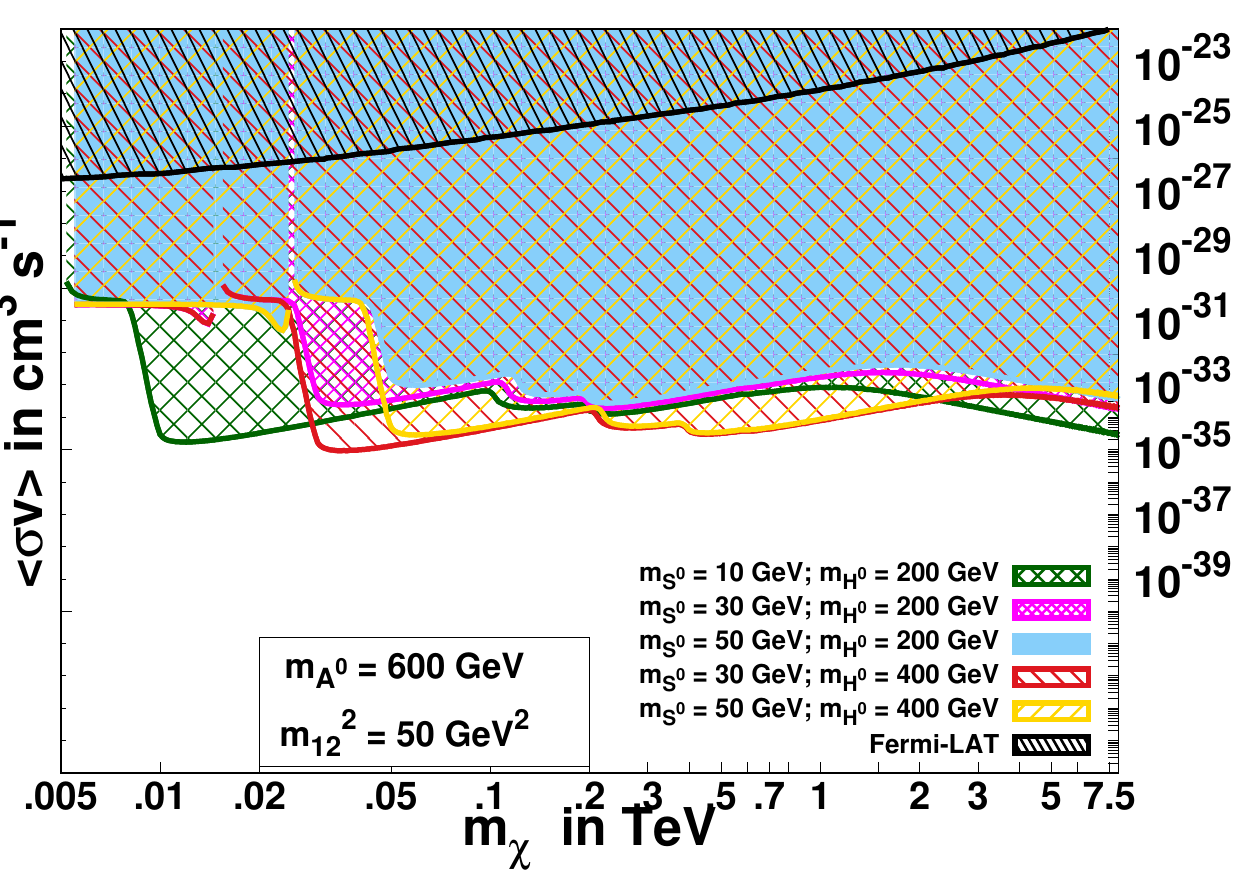}
  \caption{}
  \label{fig:Indirect_Detectionp4f}
\end{subfigure}%
\caption{\small \em{Figures \ref{fig:Indirect_Detectionp4a} to \ref{fig:Indirect_Detectionp4f} show the 
velocity-averaged scattering cross-section $<\sigma v>_{\tau^+ \tau^- }$ variation with the $m_{\chi}$ for fixed $m_{H^\pm}$ = 600 GeV, $\delta_{13}$ = 0.4  and  different choices of $m_{12}^2$ and $m_{A^0}$. All points on the contours satisfy the relic density 0.119 and also explain the discrepancy  $\Delta a_{\mu}=\,268(63)\,\times10^{-11}$. In the left and right panels, we plot the variation curves (bold lines) and  allowed (shaded) regions for four five combinations of $m_{S^0}$ and $m_{H^0}$. The  upper limit  on  velocity-averaged annihilation cross-section observed from Fermi-LAT \cite{Ackermann:2015zua} is shown. }}   
\label{fig:Indirect_Detectionp4}
\end{figure}
\begin{figure}[h!]
\centering 
\begin{subfigure}{.48\textwidth}\centering
  \includegraphics[width=\columnwidth]{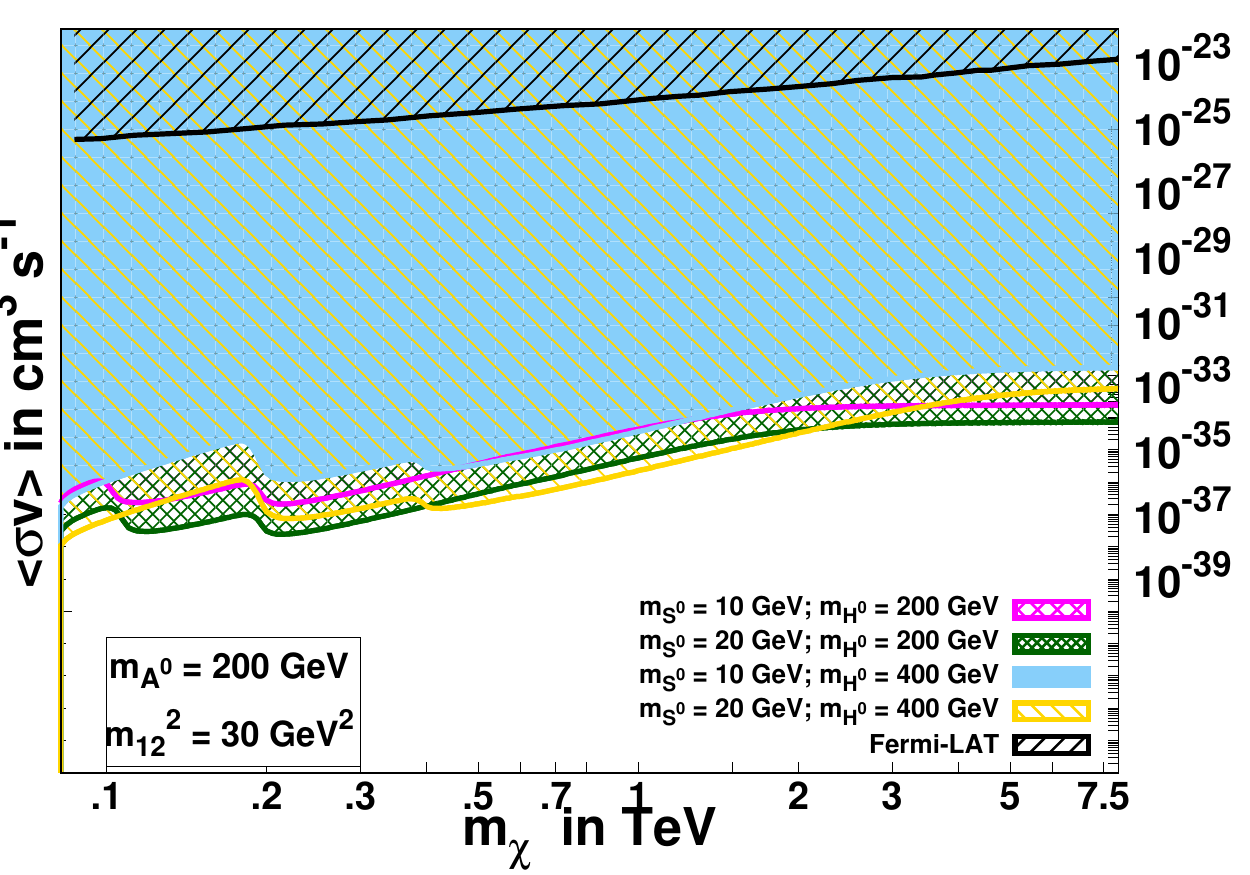}
  \caption{}
  \label{fig:Indirect_Detection_ww_ap2}
\end{subfigure}%
\begin{subfigure}{.48\textwidth}\centering
  \includegraphics[width=\columnwidth]{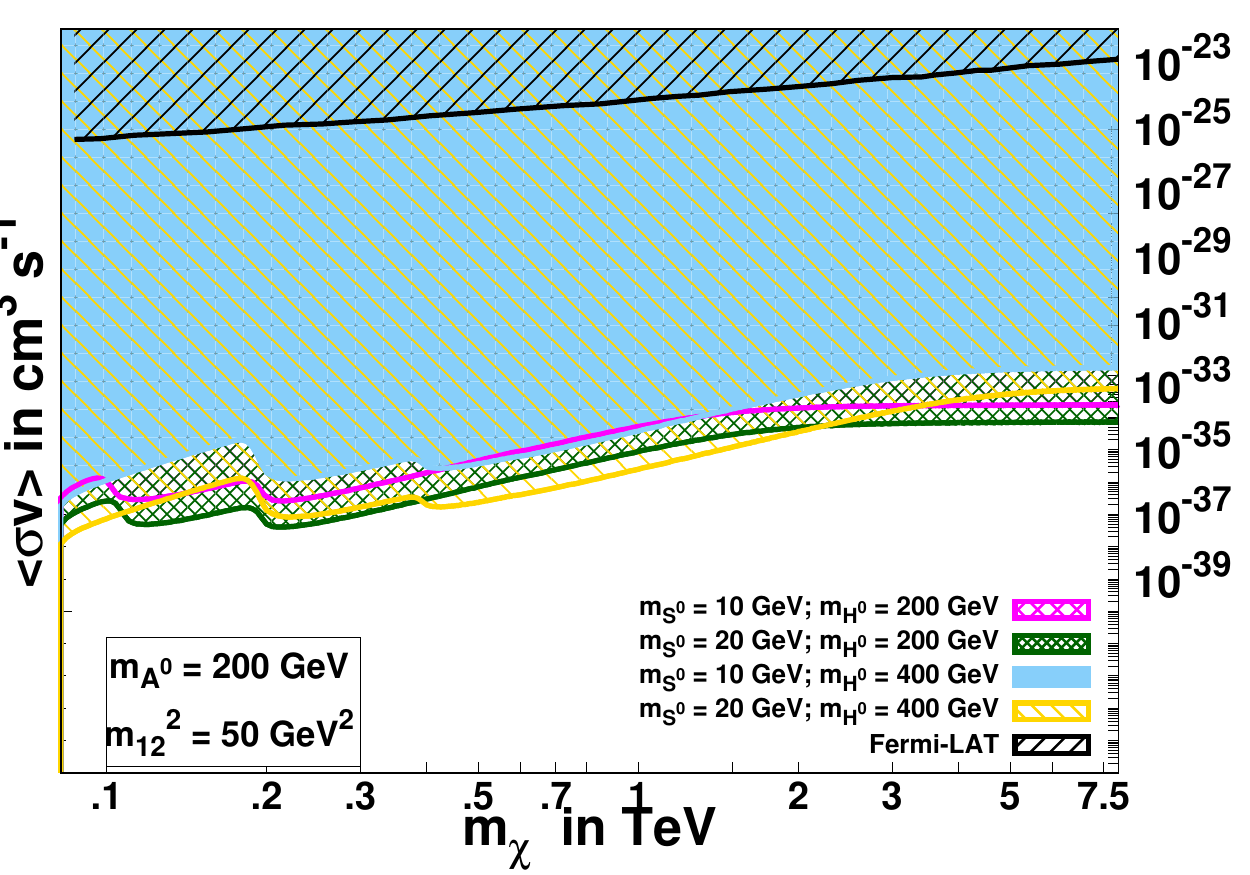}
  \caption{}
  \label{fig:Indirect_Detection_ww_bp2}
\end{subfigure}%

\begin{subfigure}{.48\textwidth}\centering
  \includegraphics[width=\columnwidth]{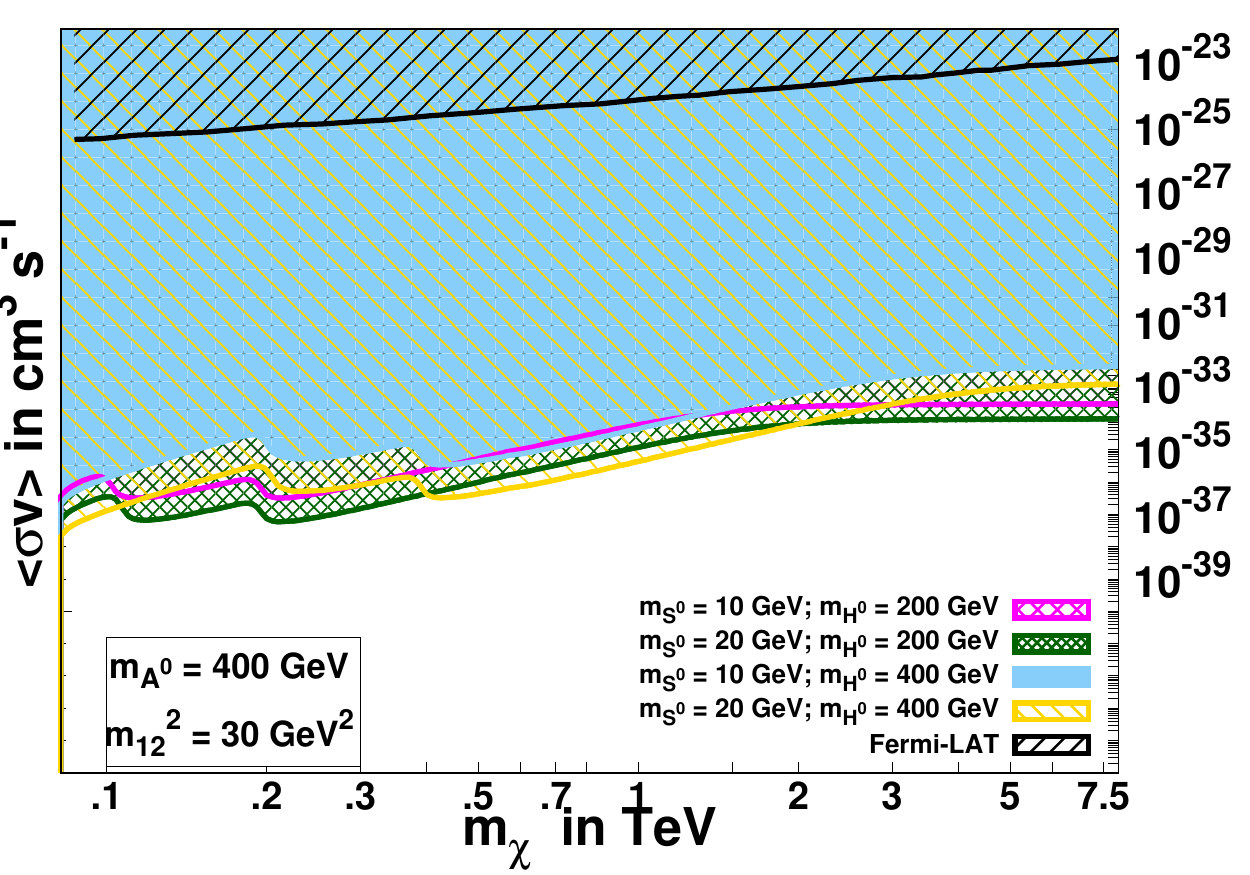}
  \caption{}
  \label{fig:Indirect_Detection_ww_cp2}
\end{subfigure}%
\begin{subfigure}{.48\textwidth}\centering
  \includegraphics[width=\columnwidth]{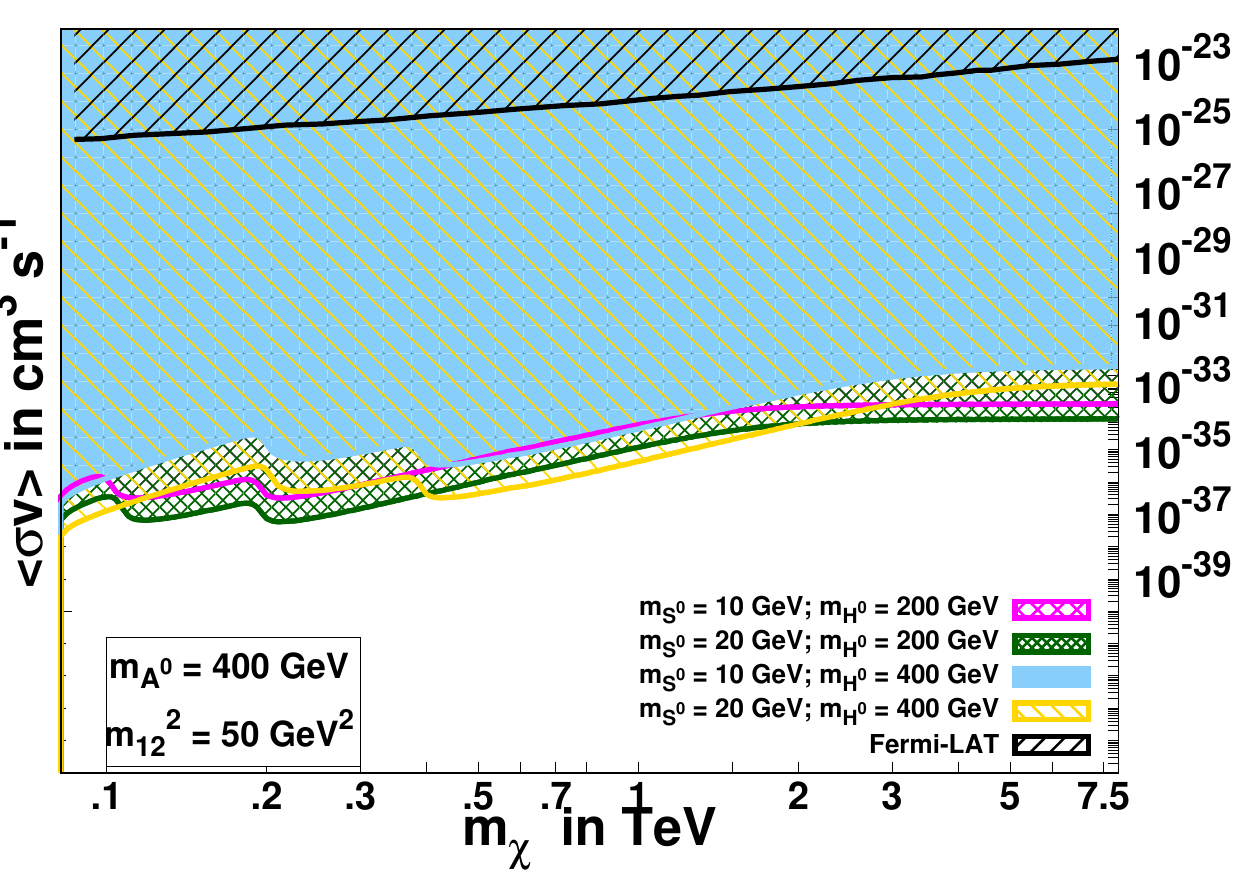}
  \caption{}
  \label{fig:Indirect_Detection_ww_dp2}
\end{subfigure}%

\begin{subfigure}{.48\textwidth}\centering
  \includegraphics[width=\columnwidth]{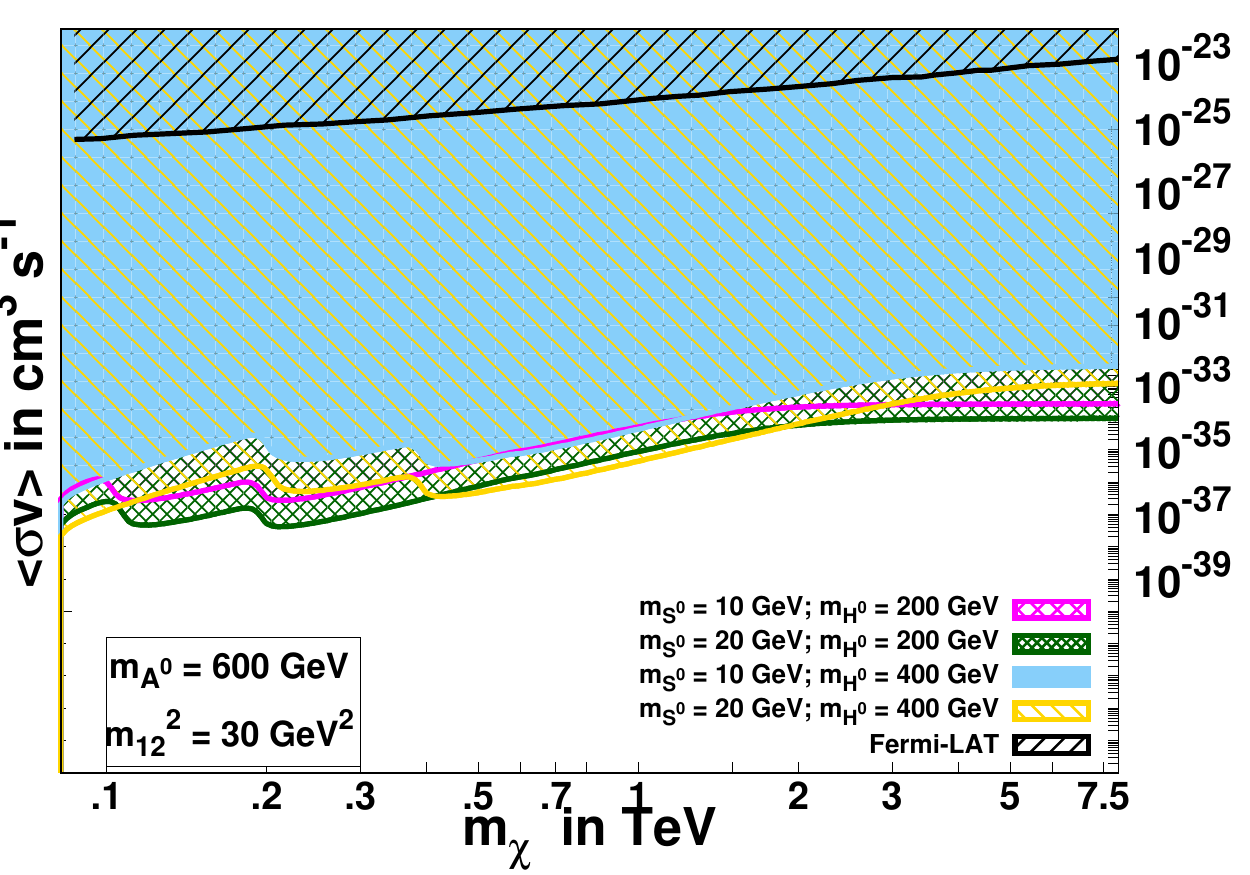}
  \caption{}
  \label{fig:Indirect_Detection_ww_ep2}
\end{subfigure}%
\begin{subfigure}{.48\textwidth}\centering
  \includegraphics[width=\columnwidth]{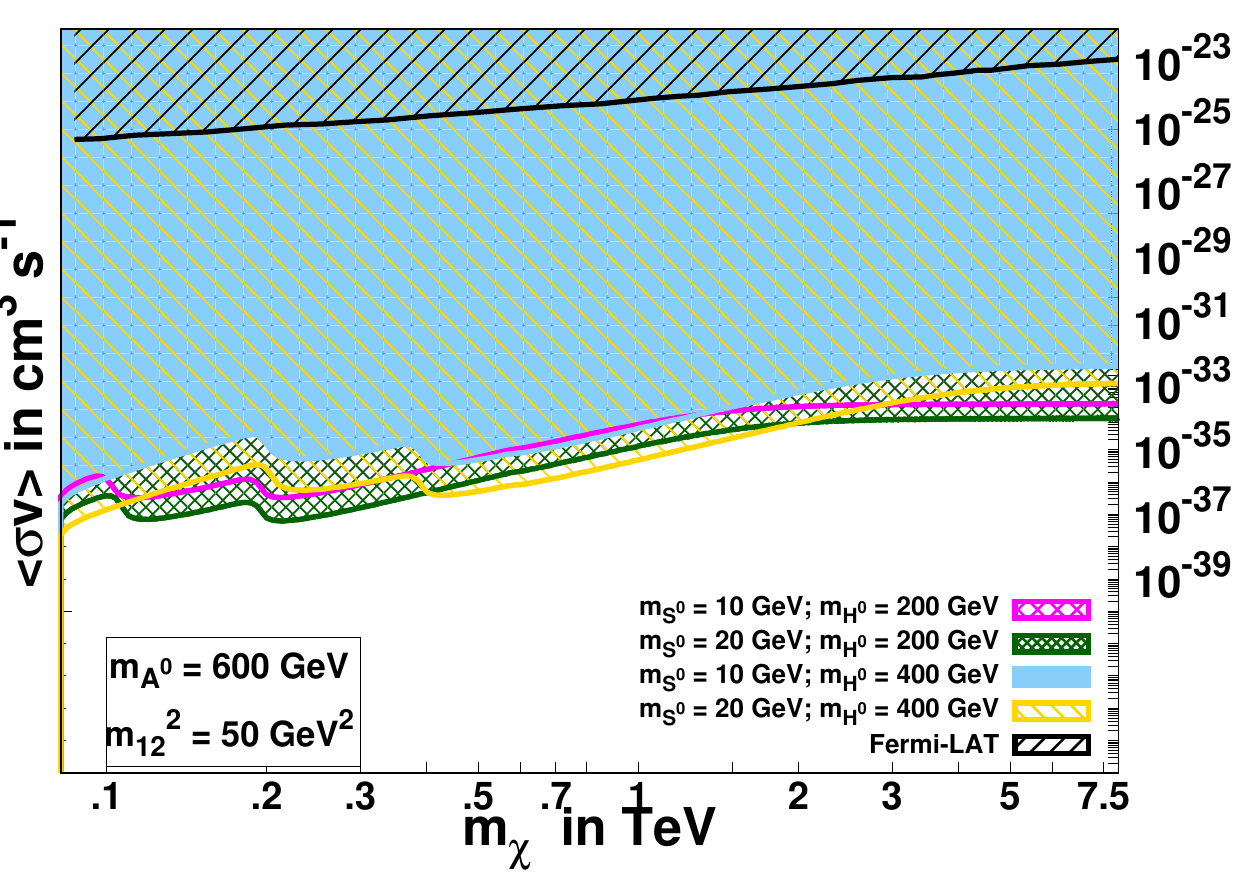}
  \caption{}
  \label{fig:Indirect_Detection_ww_fp2}
\end{subfigure}%
\caption{\small \em{Figures \ref{fig:Indirect_Detection_ww_ap2} to  \ref{fig:Indirect_Detection_ww_fp2} show the 
velocity-averaged scattering cross-section $<\sigma v>_{W^+ W^-}$ variation with the $m_{\chi}$ for fixed $m_{H^\pm}$ = 600 GeV, $\delta_{13}$ = 0.2  and   different choices of $m_{12}^2$ and $m_{A^0}$. All points on the contours satisfy the relic density 0.119 and also explain the discrepancy  $\Delta a_{\mu}=\,268(63)\,\times10^{-11}$. In the left and right panels, we plot the variation curves (bold lines) and  allowed (shaded) regions for five combinations of $m_{S^0}$ and $m_{H^0}$ . The  upper limit  on  velocity-averaged annihilation cross-section observed from Fermi-LAT \cite{Ackermann:2015zua} is shown. }}   
\label{fig:Indirect_Detection_wwp2}
\end{figure}
\begin{figure}[h!]
\centering 
\begin{subfigure}{.48\textwidth}\centering
  \includegraphics[width=\columnwidth]{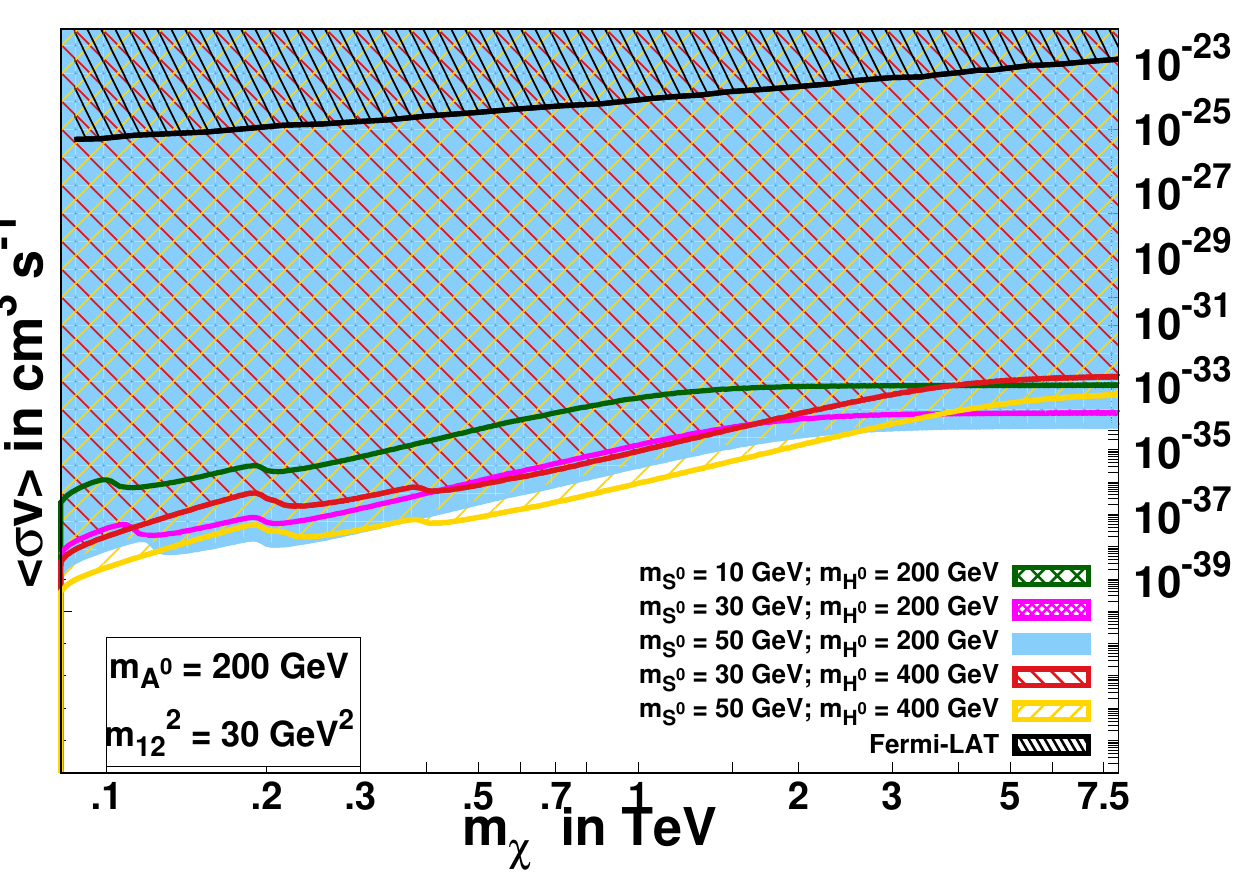}
  \caption{}
  \label{fig:Indirect_Detection_ww_a4}
\end{subfigure}%
\begin{subfigure}{.48\textwidth}\centering
  \includegraphics[width=\columnwidth]{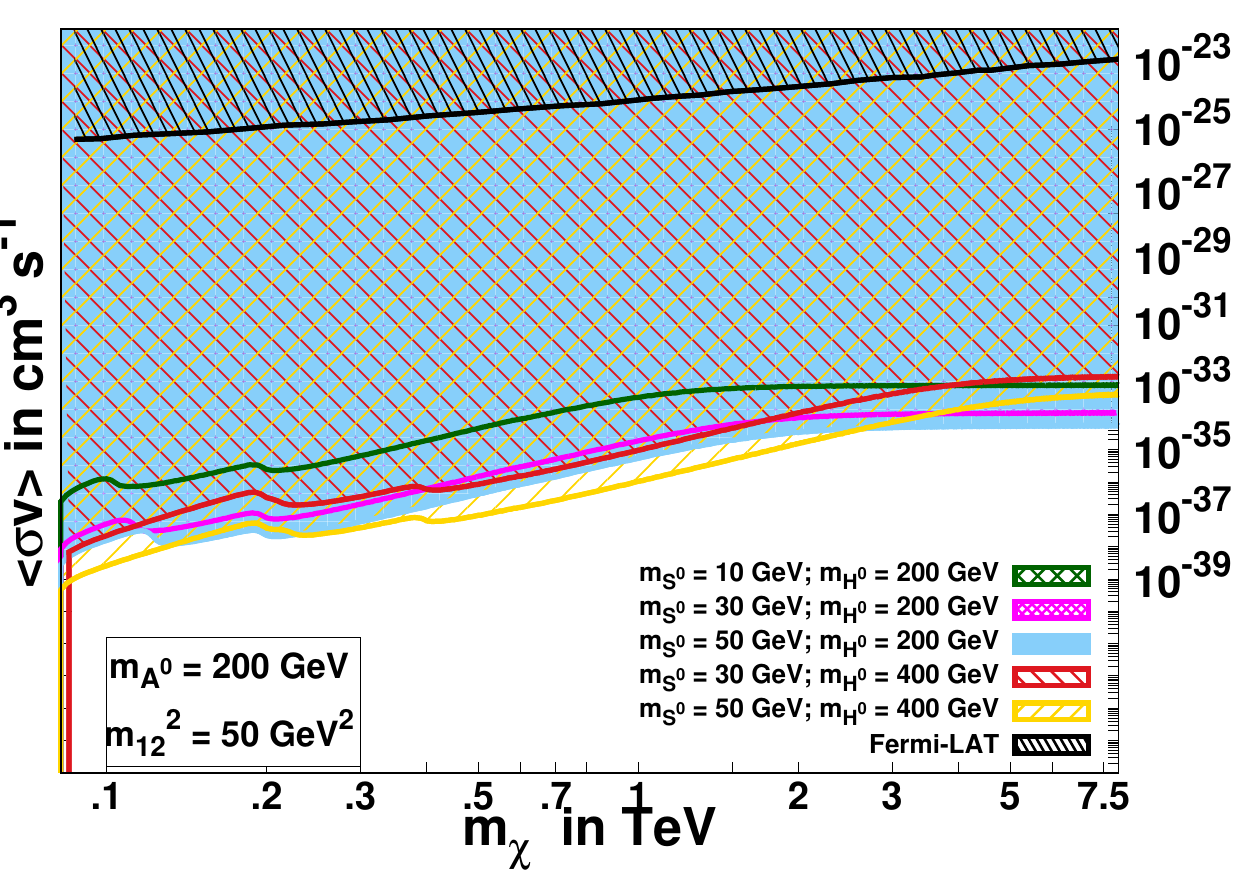}
  \caption{}
  \label{fig:Indirect_Detection_ww_b4}
\end{subfigure}%

\begin{subfigure}{.48\textwidth}\centering
  \includegraphics[width=\columnwidth]{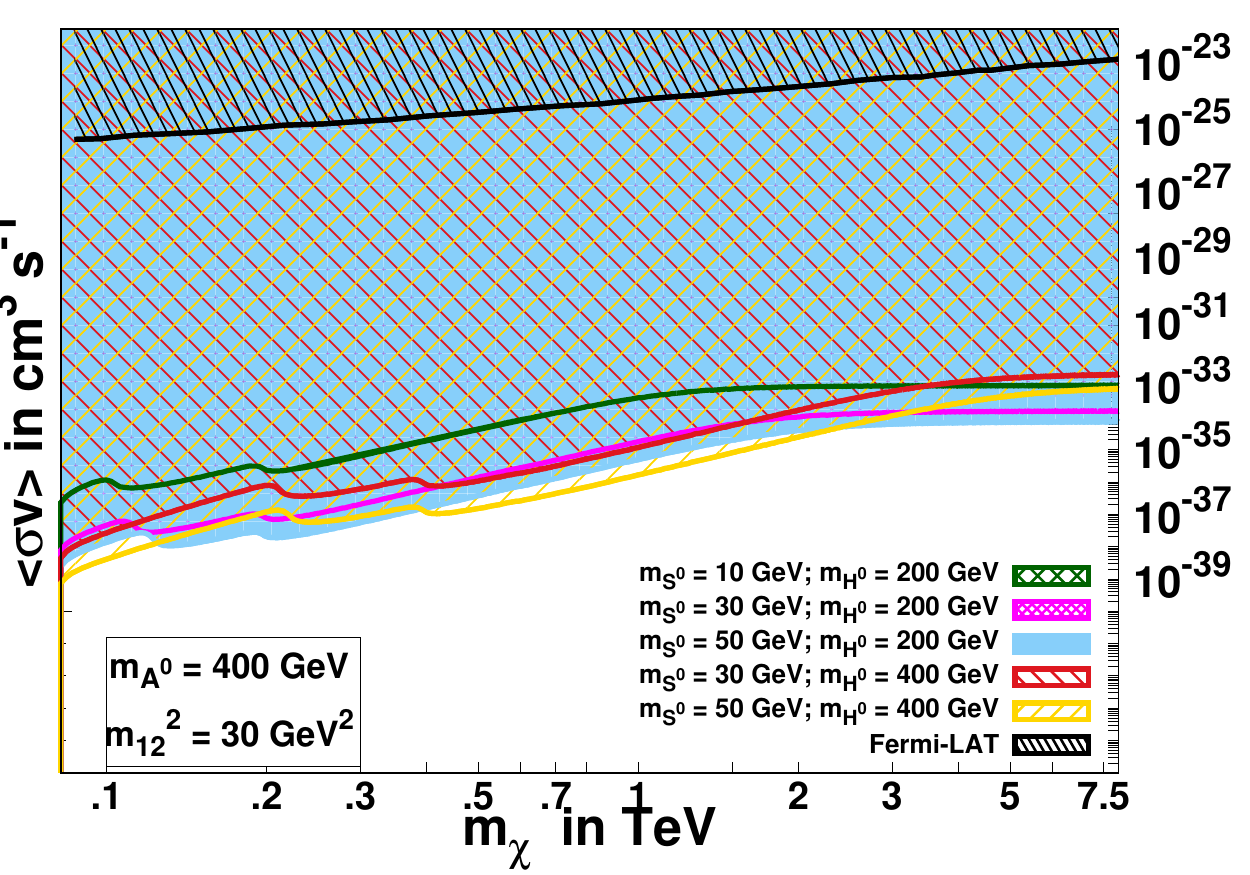}
  \caption{}
  \label{fig:Indirect_Detection_ww_cp4}
\end{subfigure}%
\begin{subfigure}{.48\textwidth}\centering
  \includegraphics[width=\columnwidth]{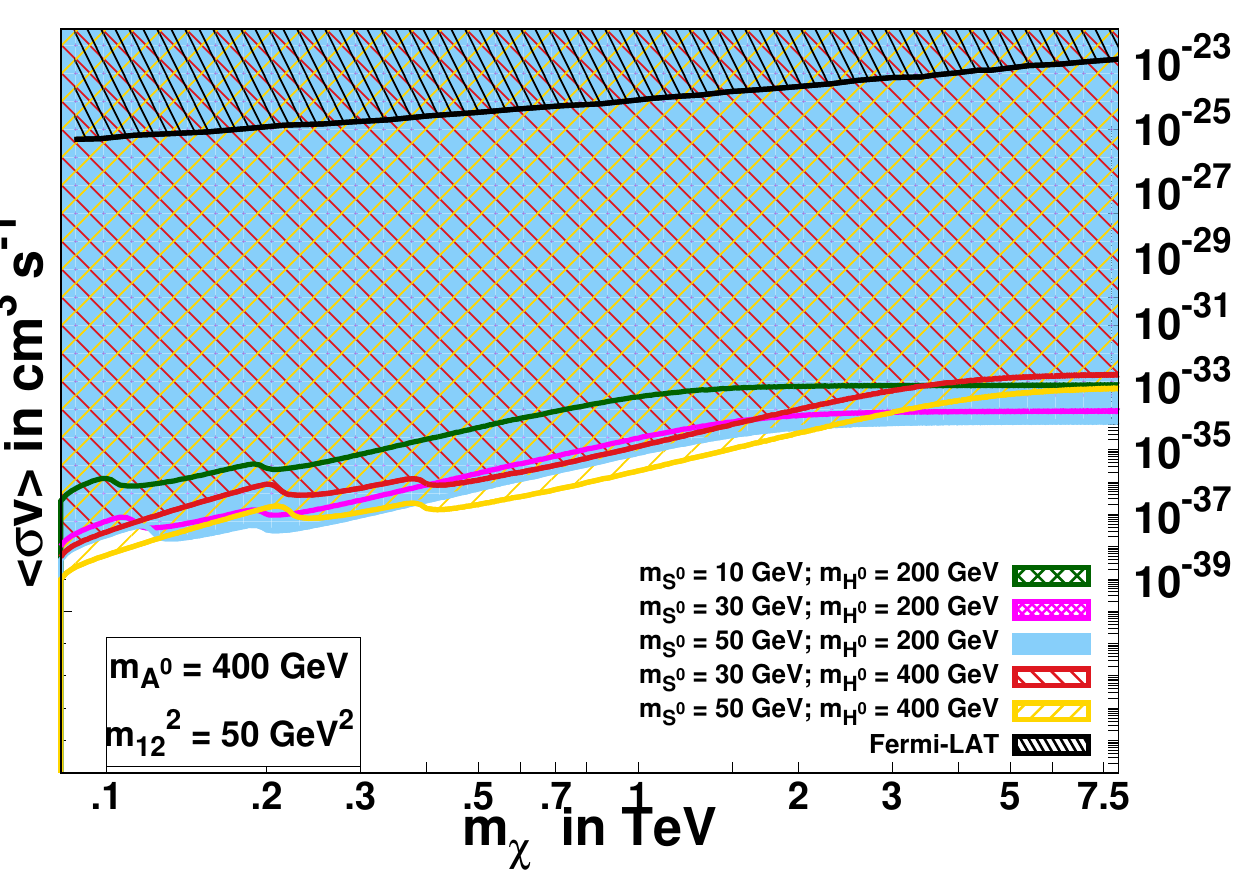}
  \caption{}
  \label{fig:Indirect_Detection_ww_dp4}
\end{subfigure}%

\begin{subfigure}{.48\textwidth}\centering
  \includegraphics[width=\columnwidth]{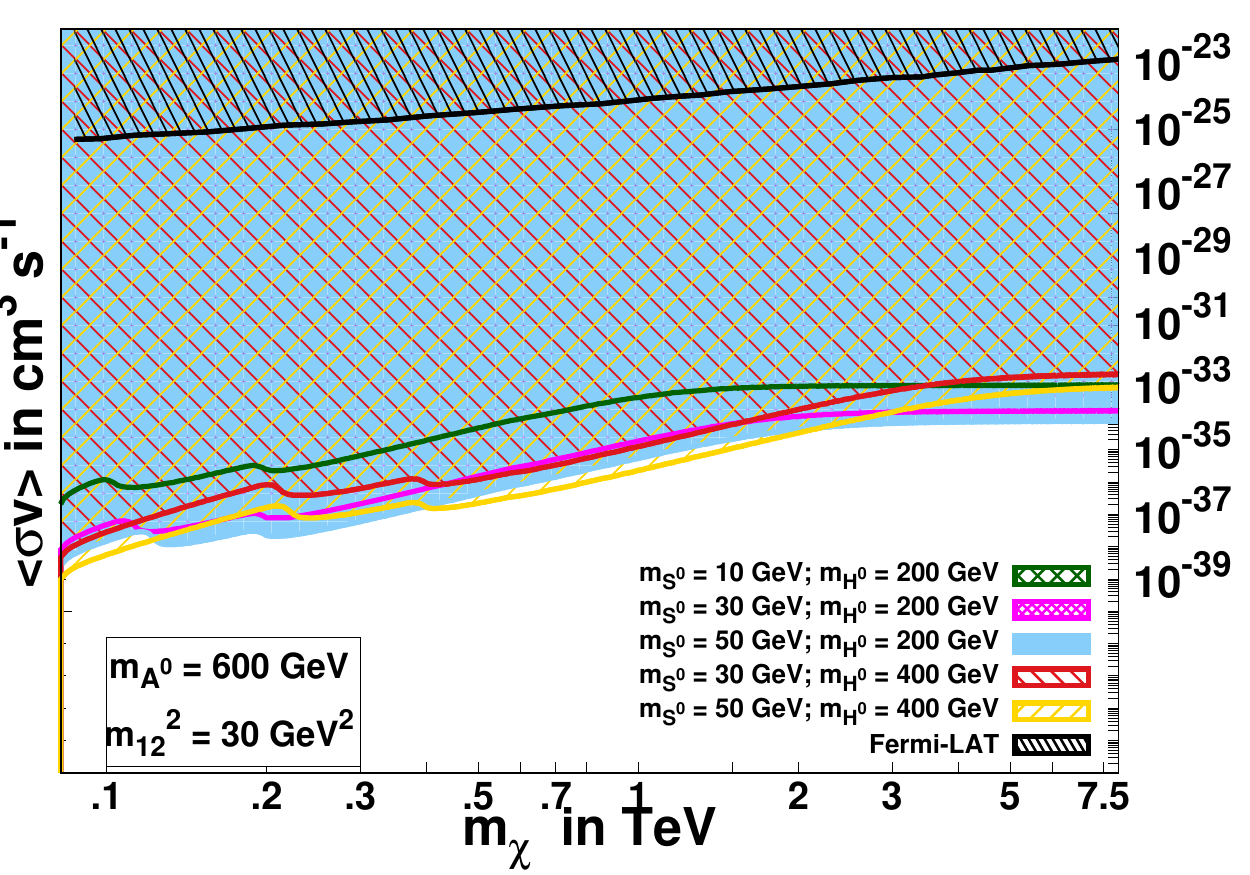}
  \caption{}
  \label{fig:Indirect_Detection_ww_ep4}
\end{subfigure}%
\begin{subfigure}{.48\textwidth}\centering
  \includegraphics[width=\columnwidth]{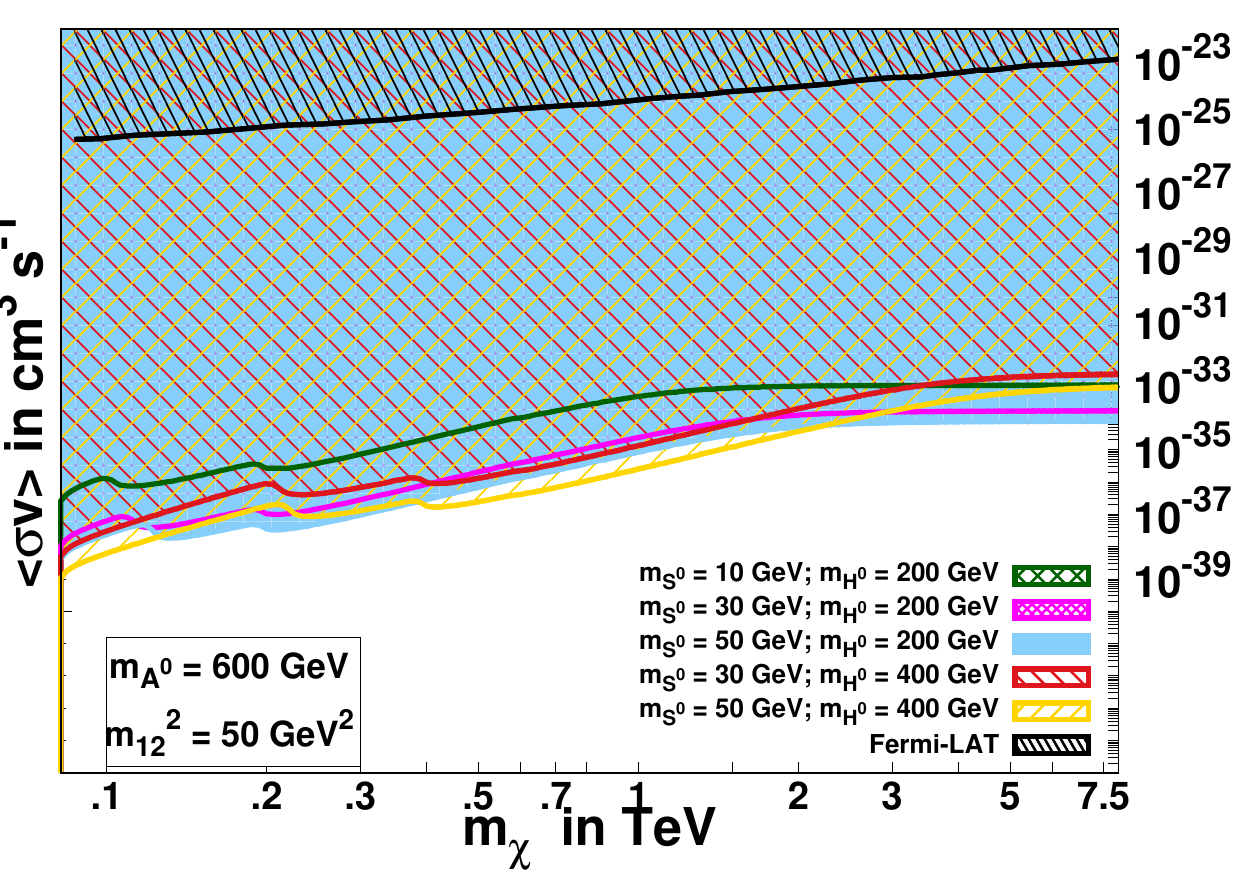}
  \caption{}
  \label{fig:Indirect_Detection_ww_fp4}
\end{subfigure}%
\caption{\small \em{Figures \ref{fig:Indirect_Detection_ww_a4} to  \ref{fig:Indirect_Detection_ww_fp4} show the 
velocity-averaged scattering cross-section $<\sigma v>_{W^+ W^-}$ variation with the $m_{\chi}$ for fixed $m_{H^\pm}$ = 600 GeV, $\delta_{13}$ = 0.4  and   different choices of $m_{12}^2$ and $m_{A^0}$. All points on the contours satisfy the relic density 0.119 and also explain the discrepancy  $\Delta a_{\mu}=\,268(63)\,\times10^{-11}$. In the left and right panels, we plot the variation curves (bold lines) and  allowed (shaded) regions for five combinations of $m_{S^0}$ and $m_{H^0}$ . The  upper limit  on  velocity-averaged annihilation cross-section observed from Fermi-LAT \cite{Ackermann:2015zua} is shown. }}   
\label{fig:Indirect_Detection_wwp4}
\end{figure}
\begin{figure}[h!]
\centering 
\begin{subfigure}{.48\textwidth}\centering
  \includegraphics[width=\columnwidth]{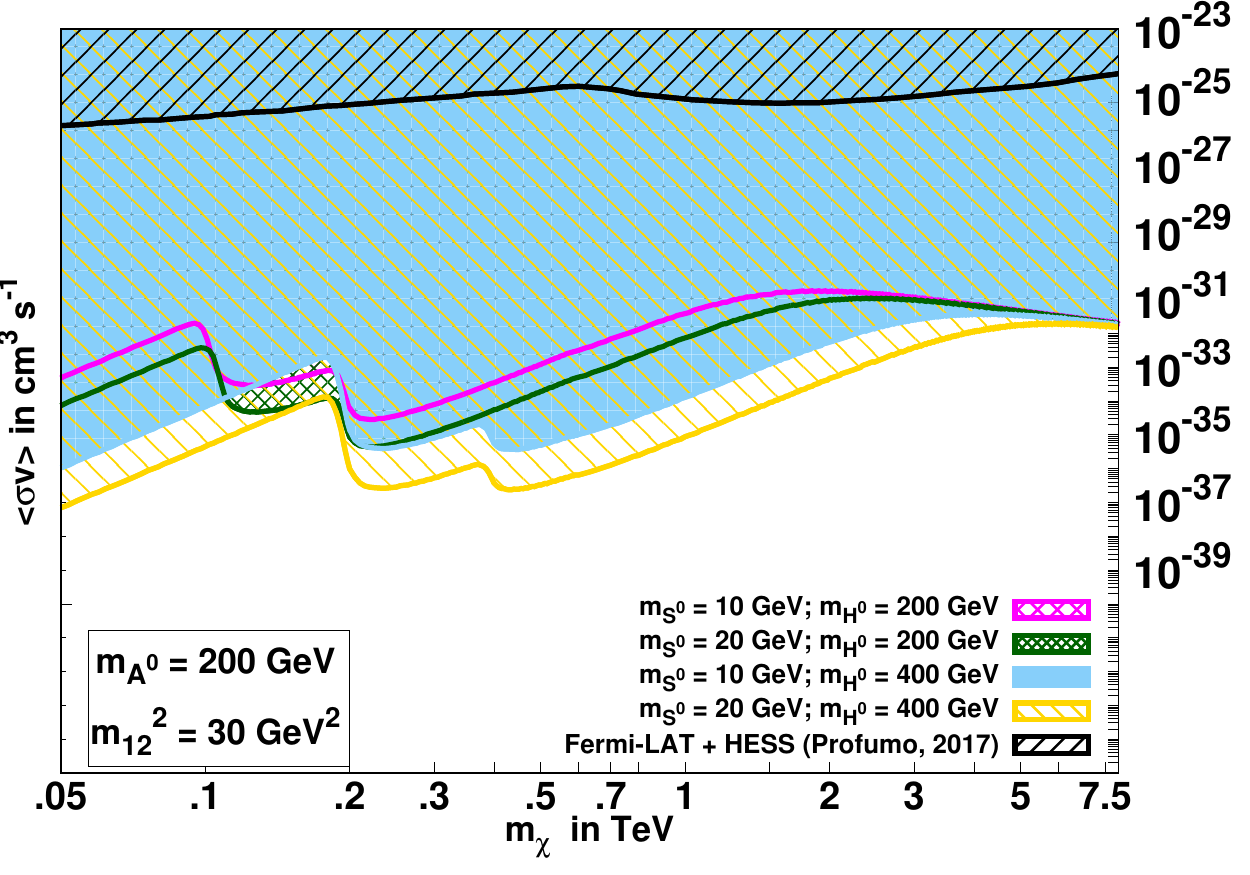}
  \caption{}
  \label{fig:Indirect_Detection_ss_p2a}
\end{subfigure}%
\begin{subfigure}{.48\textwidth}\centering
  \includegraphics[width=\columnwidth]{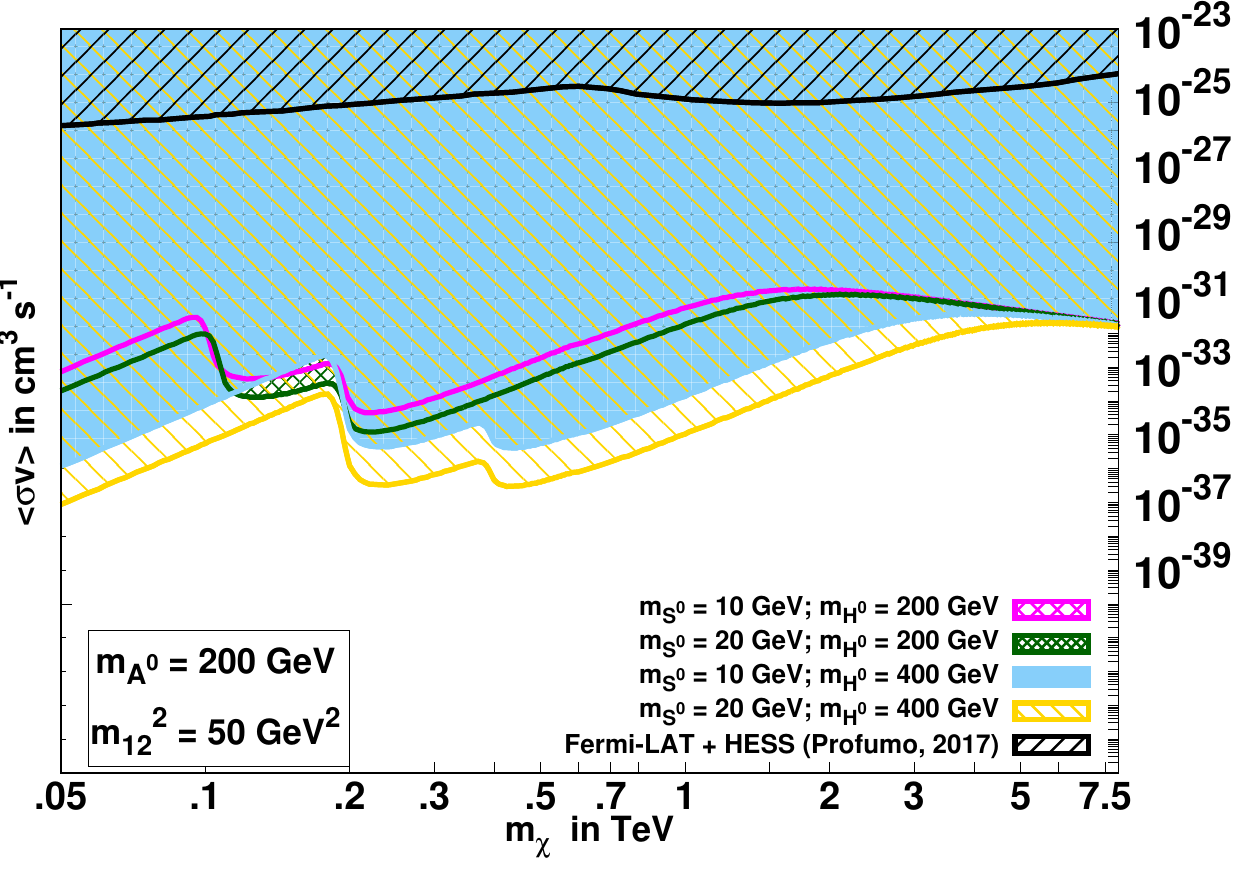}
  \caption{}
  \label{fig:Indirect_Detection_ss_p2b}
\end{subfigure}%

\begin{subfigure}{.48\textwidth}\centering
  \includegraphics[width=\columnwidth]{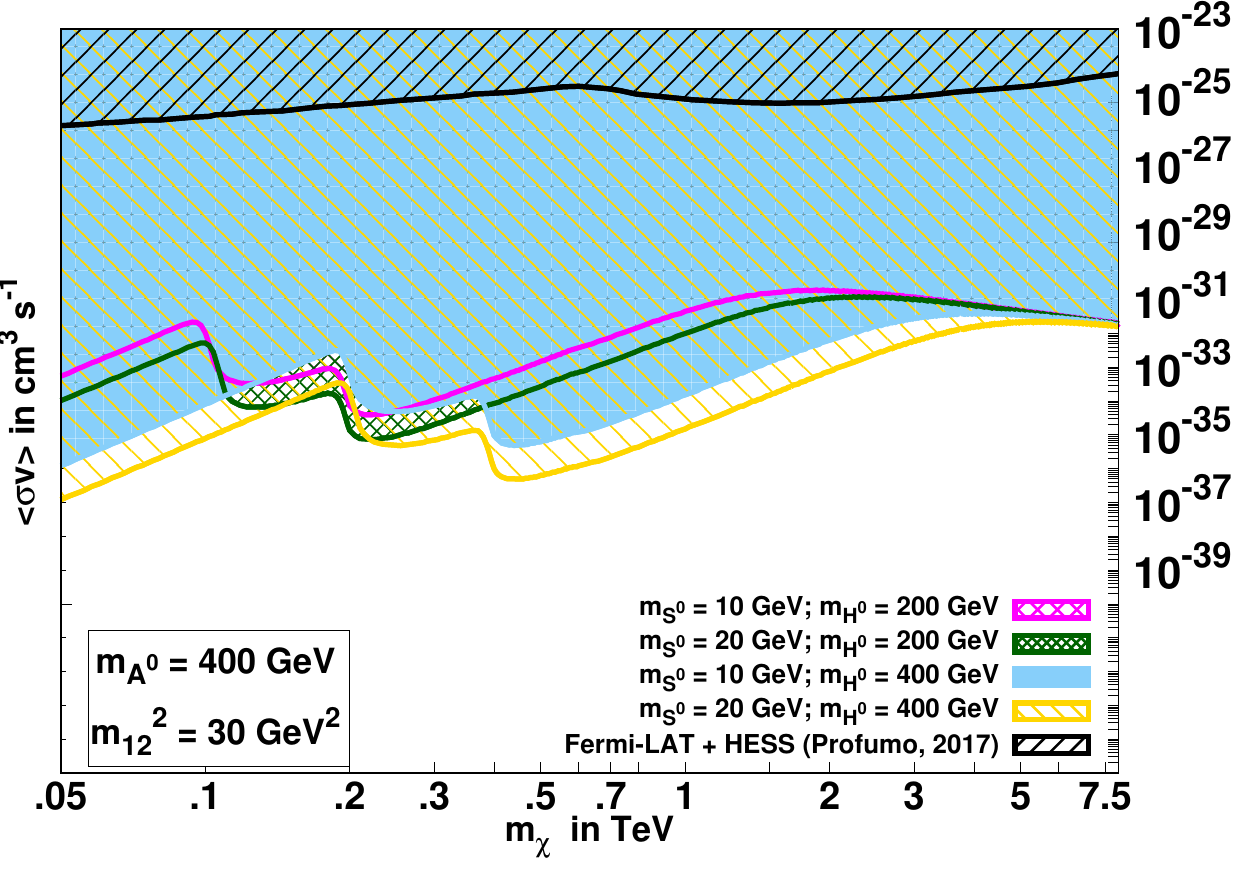}
  \caption{}
  \label{fig:Indirect_Detection_ss_p2c}
\end{subfigure}%
\begin{subfigure}{.48\textwidth}\centering
  \includegraphics[width=\columnwidth]{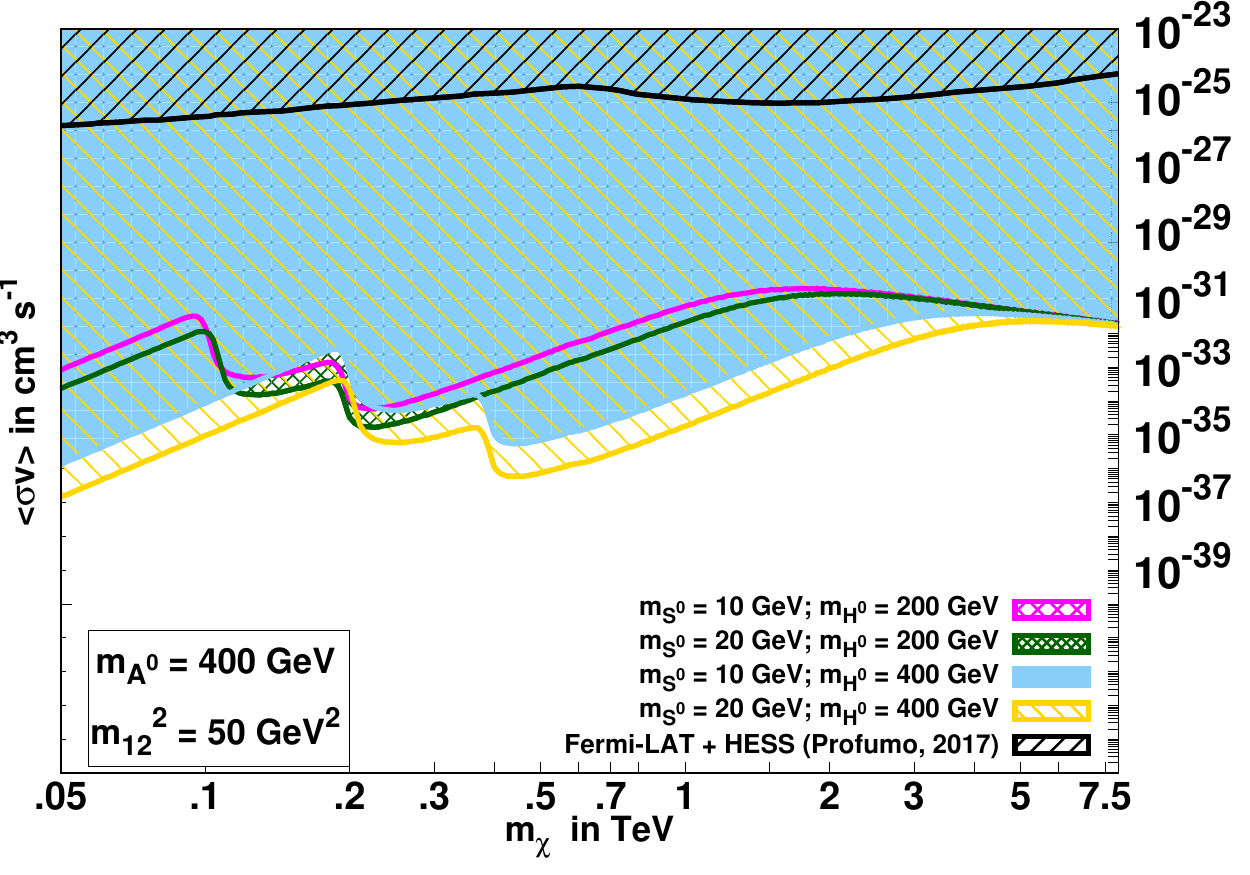}
  \caption{}
  \label{fig:Indirect_Detection_ss_p2d}
\end{subfigure}%

\begin{subfigure}{.48\textwidth}\centering
  \includegraphics[width=\columnwidth]{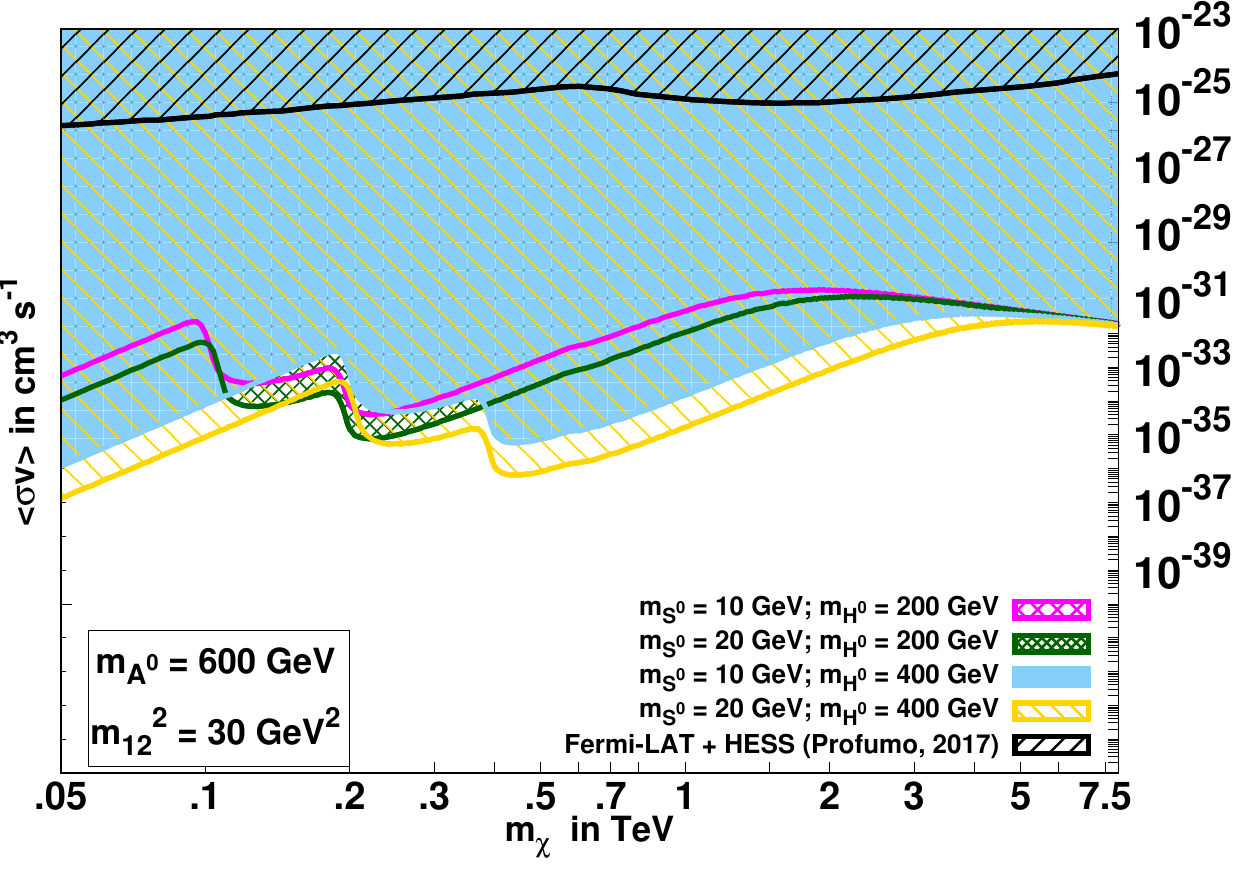}
  \caption{}
  \label{fig:Indirect_Detection_ss_p2e}
\end{subfigure}%
\begin{subfigure}{.48\textwidth}\centering
  \includegraphics[width=\columnwidth]{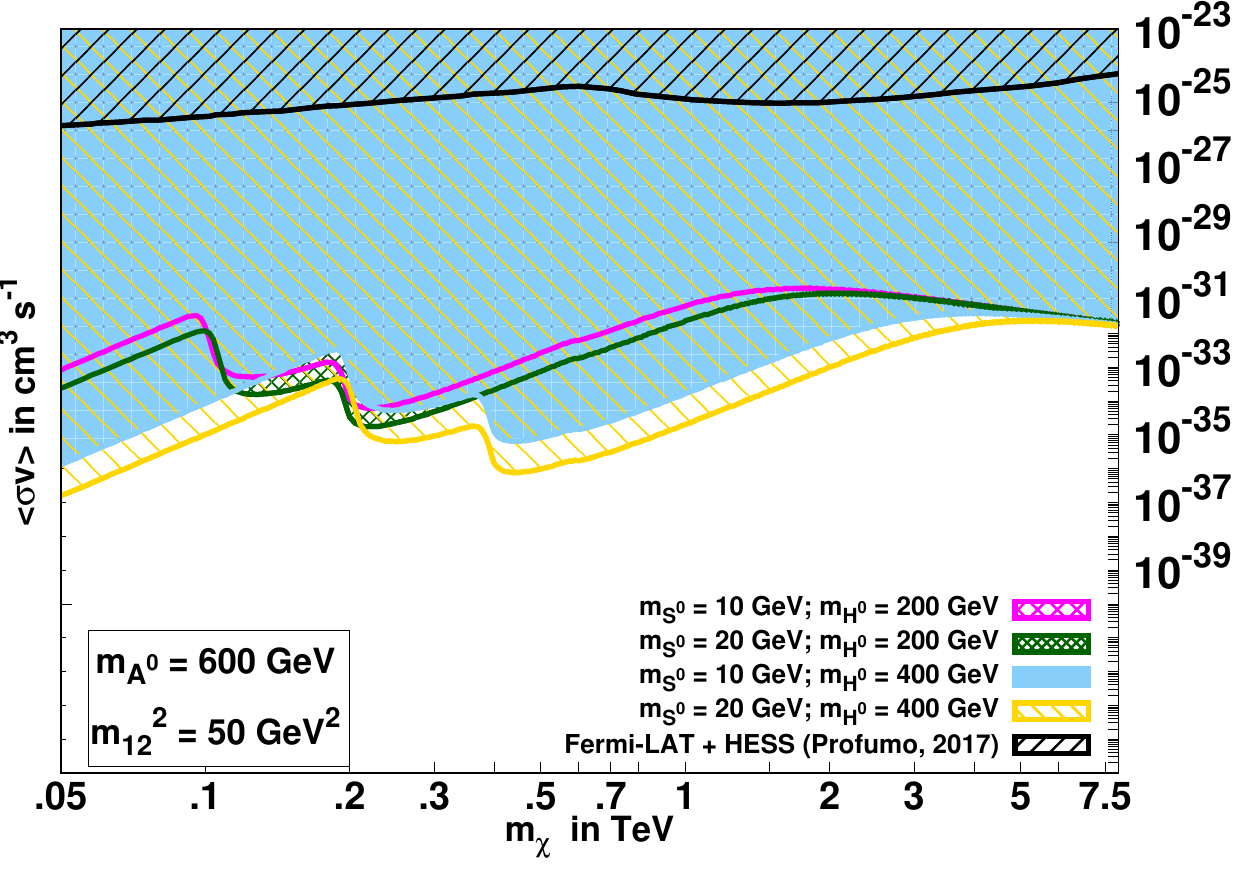}
  \caption{}
  \label{fig:Indirect_Detection_ss_p2f}
\end{subfigure}%
\caption{\small \em{Figures \ref{fig:Indirect_Detection_ss_p2a} to  \ref{fig:Indirect_Detection_ss_p2f} show the 
velocity-averaged scattering cross-section $<\sigma v>_{S^0 S^0 }$ variation with the $m_{\chi}$ for fixed $m_{H^\pm}$ = 600 GeV, $\delta_{13}$ = 0.2 and and  different choices of $m_{12}^2$ and $m_{A^0}$. All points on the contours satisfy the relic density 0.119 and also explain the discrepancy  $\Delta a_{\mu}=\,268(63)\,\times10^{-11}$. In the left and right panels, we plot the variation curves (bold lines) and  allowed (shaded) regions for five combinations of $m_{S^0}$ and $m_{H^0}$. The  upper limit  on  velocity-averaged annihilation cross-section for the process $\chi \chi\rightarrow S^0 S^0$ computed from 4$\tau$ final states from Fermi-LAT data\cite{Profumo:2017obk} is shown. }}   
\label{fig:Indirect_Detection_ssp2}
\end{figure}

 \begin{figure}[h!]
\centering 
\begin{subfigure}{.48\textwidth}\centering
  \includegraphics[width=\columnwidth]{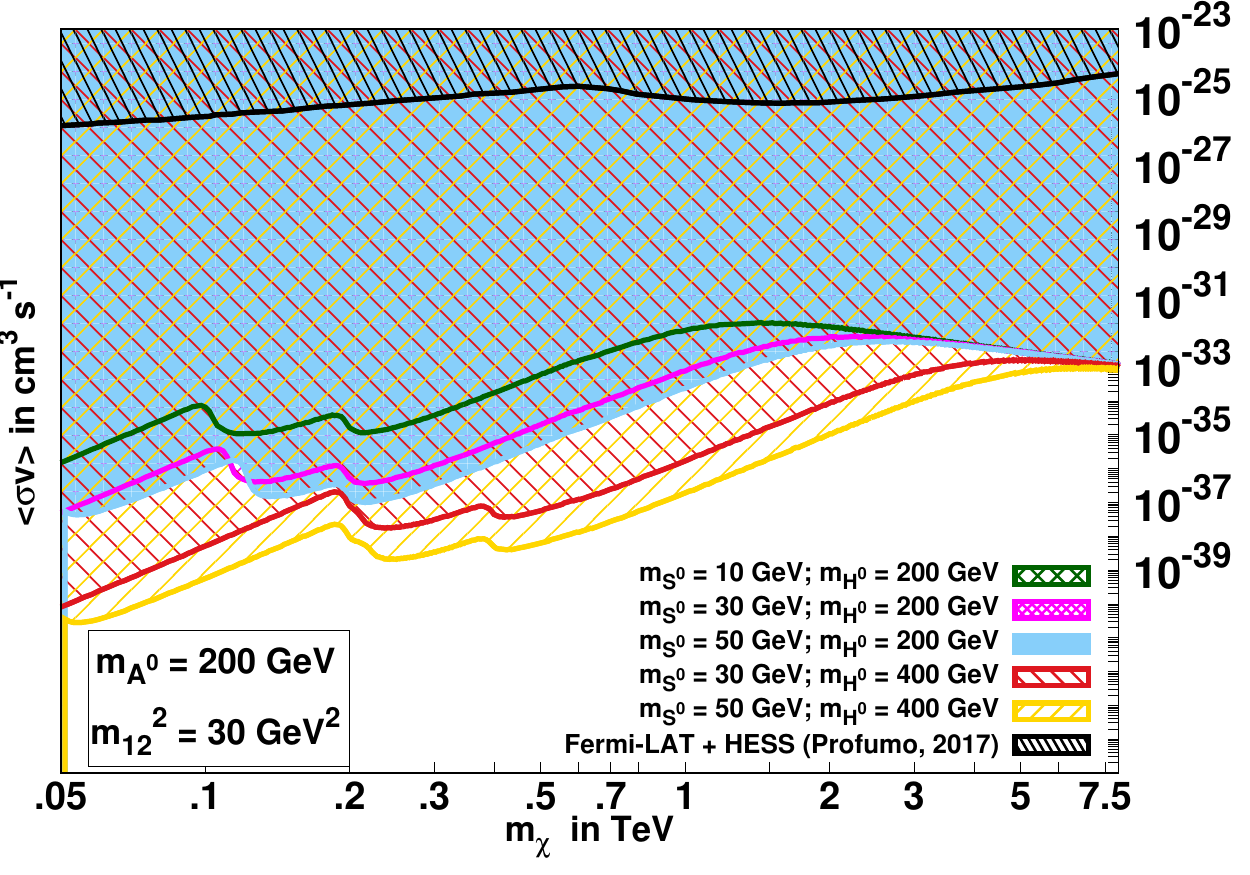}
  \caption{}
  \label{fig:Indirect_Detection_ss_p4a}
\end{subfigure}%
\begin{subfigure}{.48\textwidth}\centering
  \includegraphics[width=\columnwidth]{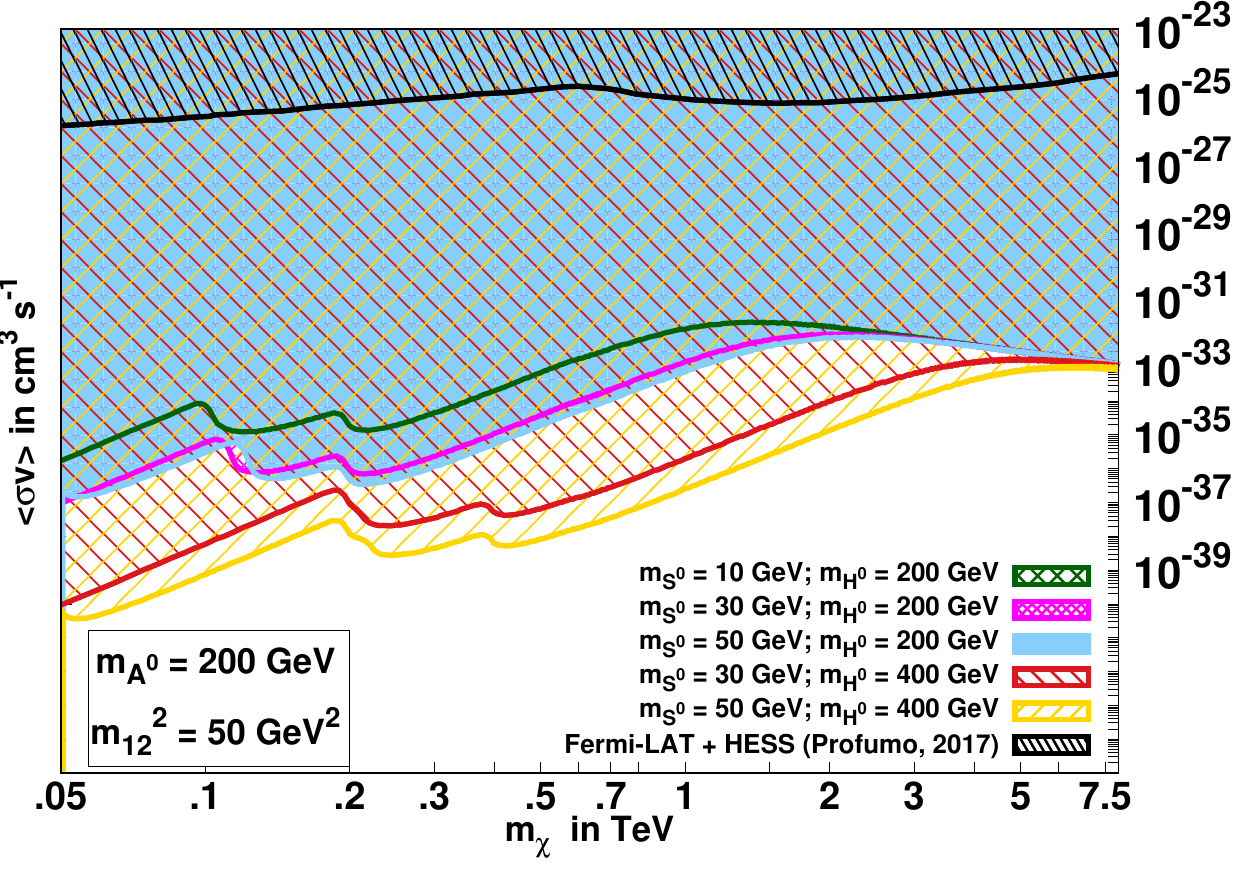}
  \caption{}
  \label{fig:Indirect_Detection_ss_p4b}
\end{subfigure}%

\begin{subfigure}{.48\textwidth}\centering
  \includegraphics[width=\columnwidth]{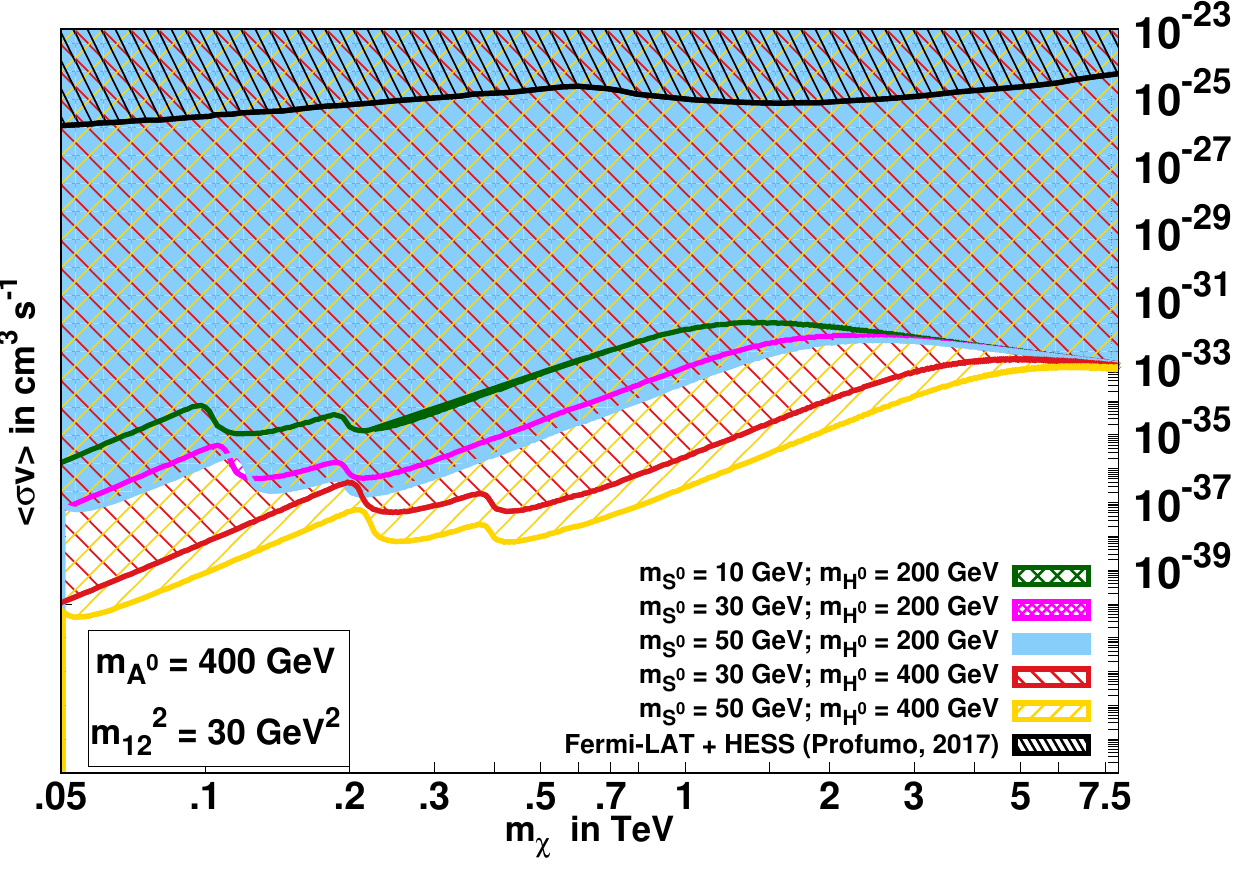}
  \caption{}
  \label{fig:Indirect_Detection_ss_p4c}
\end{subfigure}%
\begin{subfigure}{.48\textwidth}\centering
  \includegraphics[width=\columnwidth]{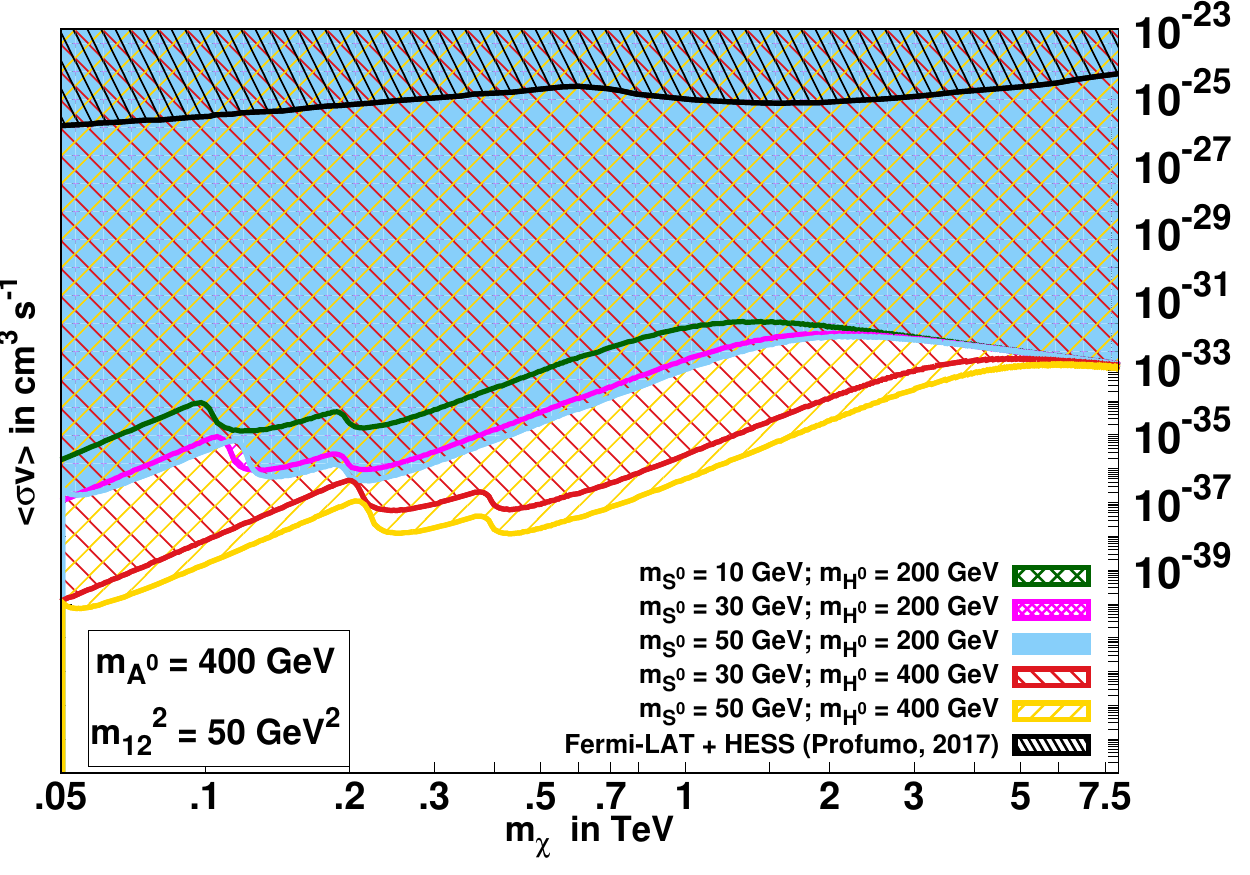}
  \caption{}
  \label{fig:Indirect_Detection_ss_p4d}
\end{subfigure}%

\begin{subfigure}{.48\textwidth}\centering
  \includegraphics[width=\columnwidth]{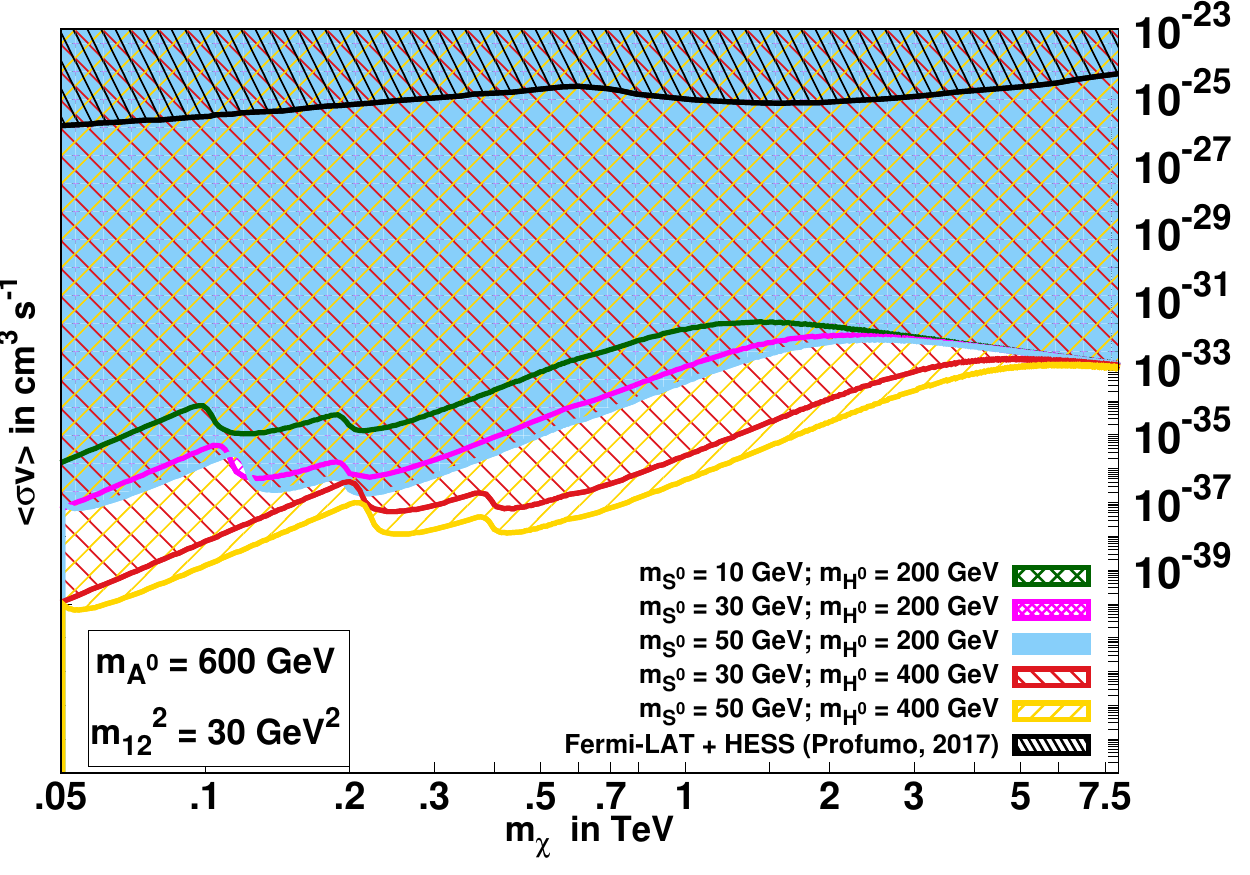}
  \caption{}
  \label{fig:Indirect_Detection_ss_p4e}
\end{subfigure}%
\begin{subfigure}{.48\textwidth}\centering
  \includegraphics[width=\columnwidth]{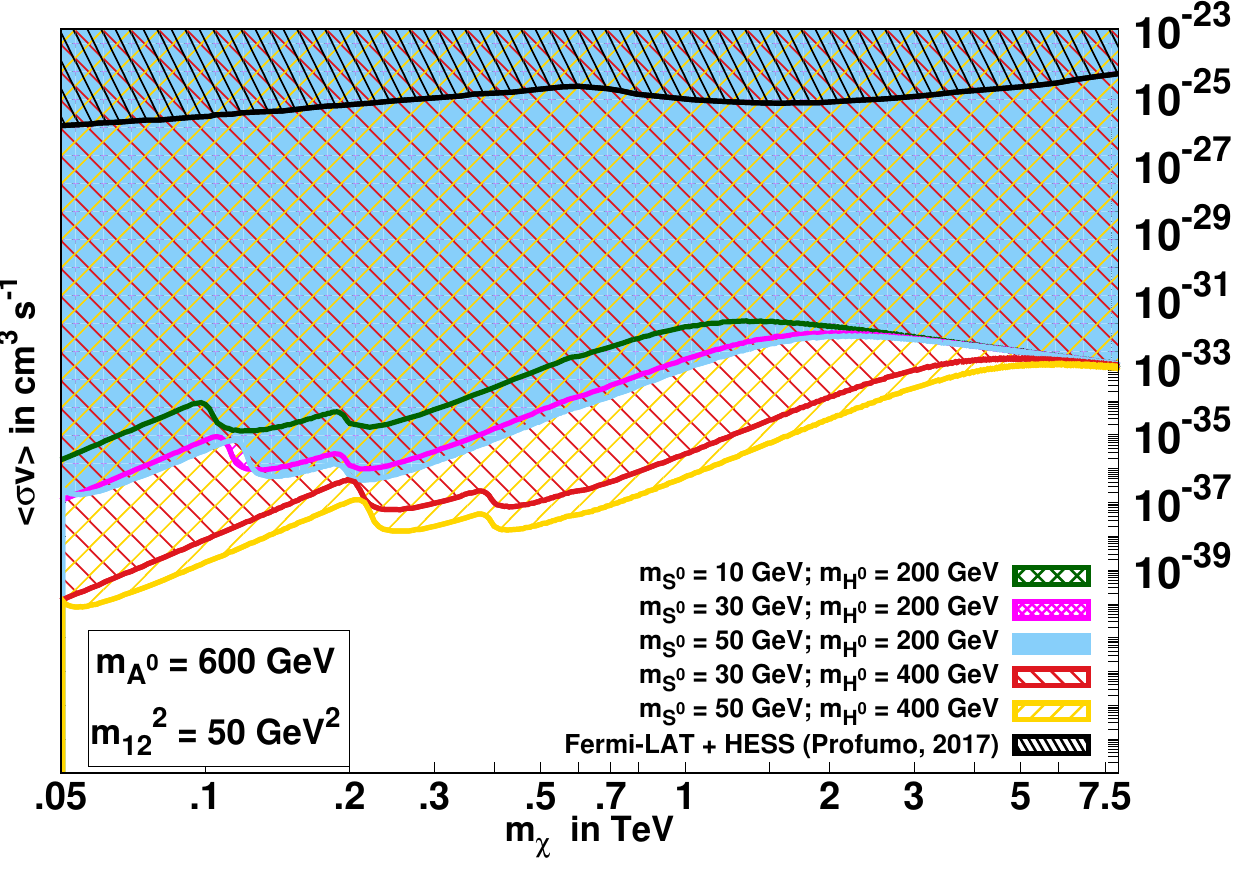}
  \caption{}
  \label{fig:Indirect_Detection_ss_p4f}
\end{subfigure}%
\caption{\small \em{Figures \ref{fig:Indirect_Detection_ss_p4a} to  \ref{fig:Indirect_Detection_ss_p4f} show the 
velocity-averaged scattering cross-section $<\sigma v>_{S^0 S^0 }$ variation with the $m_{\chi}$ for fixed $m_{H^\pm}$ = 600 GeV, $\delta_{13}$ = 0.4  and and  different choices of $m_{12}^2$ and $m_{A^0}$. All points on the contours satisfy the relic density 0.119 and also explain the discrepancy  $\Delta a_{\mu}=\,268(63)\,\times10^{-11}$. In the left and right panels, we plot the variation curves (bold lines) and  allowed (shaded) regions for five combinations of $m_{S^0}$ and $m_{H^0}$. The  upper limit  on  velocity-averaged annihilation cross-section for the process $\chi \chi\rightarrow S^0 S^0$ computed from 4$\tau$ final states from Fermi-LAT data \cite{Profumo:2017obk} is shown. }}   
\label{fig:Indirect_Detection_ssp4}
\end{figure}

\subsection{Indirect detection}
\label{subsection_Indirect_Detection}
Observations of diffused gamma rays  from the regions of our Galaxy, such as Galactic Center (GC) and dwarf spheroidal galaxies (dsphs), where DM density appears to be high, impose bounds on DM annihilation to SM particles. Experiments like Fermi-LAT  \cite{TheFermi-LAT:2015kwa,Ackermann:2015zua} and H.E.S.S.\,\cite{Moulin:2007qw} have investigated DM annihilation as a possible source of the incoming photon-flux. These experiments provide us with an upper-limit to velocity-averaged scattering cross-section for various channels, which can attribute to the observed photon-flux.

\par DM annihilations contribute to the photon-flux  through Final State Radiation (FSR) and radiative decays \cite{Mazziotta:2014ada,Ackermann:2013yva} from leptonic channels in  lepto-philic models. FSR contributions are important in understanding the photon-spectra from DM annihilations to charged final states and therefore are instrumental in calculation of the observed bounds by experiments like Fermi-LAT \cite{Ackermann:2015zua,Ackermann:2013yva,Essig:2009jx,Bringmann:2007nk}. 
The radiation emitted by the charged relativistic    final state fermions $f$  in the annihilation process  $\bar{\chi} + \chi \rightarrow f + \bar{f} + \gamma$  are approximately collinear with the charged  fermions. In this regime, the differential cross-section for the real emission process can be factorized into the   a collinear factor and cross-section $\sigma (\chi\chi\to f\bar{f})$ as discussed in the reference \cite{Birkedal:2005ep}. 
\beq\label{FSR}
\frac{d\sigma (\chi\chi\,\to\,f\bar{f}\gamma)}{dx} \approx  
\frac{\alpha_{em} Q_f^2}{\pi}\,{\cal F}_f(x)\,\log\left(\frac{s(1-x)}{m_f^2}\right) 
\sigma (\chi\chi\to f\bar{f}),
\eeq
where  $Q_f$ and $m_f$ are the 
electric charge and the mass of the $f$ particle, $s$ is the center-of-mass 
energy, and 
$x=2E_\gamma/\sqrt{s}$.
For fermion final states, the splitting function ${\cal F}$ is given by 
\beq
{\cal F}_f(x) = \frac{1 + \left(1-x\right)^2}{x}
\eeq
The suppression factor of $p$-wave suppressed thermal averaged cross-section  $ \left\langle \sigma (\chi\chi\to f\bar{f})\, v \right\rangle$   is mitigated in the thermal averaged cross-section of the real emission process  $\left\langle\sigma (\chi\chi\to f\bar{f} \gamma) \, v\right\rangle$ by the virtue of collinear factor given in equation \eqref{FSR}.

In the present model, the fermionic DM can annihilate to SM particles {\it via} $s-channnel$ through  the {\it scalar portal} as well as to a pair of singlet scalars through $t-channel$ diagrams. Recently authors of the reference \cite{Siqueira:2019wdg,Queiroz:2019acr} explored the discovery potential of the pair production of such lepto-philic scalars which pre-dominantly decay   into pairs of charged leptons at Cherenkov Telescope Array (CTA). Given the spectrum of pair produced SM particles through single scalar mediator and into two pairs of charged leptons through scalar pair production, we  should be able to simulate  the expected DM fluxes which will enable us to get the upper limits on the annihilation cross section for a given  mediator mass in the model.
\par We calculate the velocity averaged cross-sections analytically for the annihilation processes 
$\chi\,\bar\chi\to f \bar f$, $\chi\,\bar\chi\to Z^0 Z^0$, $\chi\,\bar\chi\to W^+ W^-$, $\chi\,\bar\chi\to \gamma \gamma$ and $\chi\,\bar\chi\to H_i\,  H_j$ and are given in equations \eqref{thermalfermion}, \eqref{thermalZ0}, \eqref{thermalW}, \eqref{thermalphoton} and \eqref{thermalscalar} respectively  where $H_i\equiv h^0,\, H^0, \, S^0$ and $A^0$ are  the scalars of the model.  In addition,  the   velocity averaged  annihilation cross-section for $\chi\,\bar\chi\to S^0 S^0$ through  the $t$ and $u$ channel diagrams are given  in \eqref{thermalS0S0}.  We observe that the velocity averaged scattering cross-sections for all these processes are $p$-wave suppressed and are, therefore, sensitive to the choice of velocity distribution of the DM in the galaxy.

\par The annihilation channels to fermions are proportional to the Yukawa coupling of the fermions with $S^0$.  We present the analysis for the most dominant $s-channel$ annihilation process  $\chi\,\bar\chi\to \tau^+\tau^-$, which is enhanced due to its coupling strength being proportional to $m_\tau \tan\beta$ and plot the variation of the velocity averaged scattering cross-section  $ \left\langle\sigma v\right\rangle(\chi\bar \chi \rightarrow \tau^+ \tau^-)$ as a function of the DM mass in figures \ref{fig:Indirect_Detectionp2} and \ref{fig:Indirect_Detectionp4} for mixing angle $\delta_{13}$ = 0.2 and 0.4 respectively.  The coupling  $g_{\chi S^0}$ for a given DM mass and all other parameters are chosen to satisfy the observed relic density and electro-weak constraints as shown in the figures \ref{fig:Relic_Densityp2} and \ref{fig:Relic_Densityp4}. Annihilation of DM pairs to gauge Bosons are proportional to the square of their masses and therefore it  is the second dominant process followed by the annihilation to $\tau\pm$ pairs. Similarly, we show   the variation of the velocity averaged scattering cross-section  $ \left\langle\sigma v\right\rangle(\chi\bar \chi \rightarrow W^+ W^-)$ as a function of the DM mass  in 
figures  \ref{fig:Indirect_Detection_wwp2} and \ref{fig:Indirect_Detection_wwp4}  for $\delta_{13}$ =  0.2 and 0.4 respectively.   The DM pair annihilation to  photons is loop suppressed and is not discussed further. The $s$ channel mediated DM pair annihilation to pair of scalars in the theory involve the triple scalar couplings, which are experimentally constrained and are therefore suppressed.

\par As mentioned above,  the t-channel pair production of singlet scalars  dominates over the other channels. The $S^0$ pair production through its decay to dominant $\tau$ pairs will modify the $\gamma$ ray spectrum that one would have expected from the two body decay processes.  We plot the  velocity averaged scattering cross-section  $ \left\langle\sigma v\right\rangle(\chi\bar \chi \rightarrow S^0 S^0)$ as a function of the DM mass  which satisfies  the relic density constraint in figures \ref{fig:Indirect_Detection_ssp2} and \ref{fig:Indirect_Detection_ssp4}  for $\delta_{13}$ =  0.2 and 0.4 respectively with all the other parameters fixed from the observed relic density and electro-weak constraints. The   experimental upper limit  on  velocity-averaged annihilation cross-section for the process $\chi \chi\rightarrow S^0 S^0$ for the varying DM mass are derived  from the  upper limits  on the events contributed to 4$\tau$ final states  at  Fermi-LAT \cite{Profumo:2017obk} and shown in figures  \ref{fig:Indirect_Detection_ssp2} and \ref{fig:Indirect_Detection_ssp4}.

\par  We find that the annihilation cross-sections for all these processes are three or more orders of magnitude smaller than the current upper-bounds from Fermi-Lat data \cite{Ackermann:2015zua,Profumo:2017obk}.

\section{Summary}
\label{section_summary}
In this article we have made an attempt to  address the observed discrepancy in anomalous magnetic moment of muon by considering a lepto-philic type {\bf X} 2-HDM and a singlet {\it scalar portal} for fermionic DM. We have presented the model in such a manner where most of it's scalar sector parameters can be constrained in terms of the lower bound on the physical neutral and charged scalar's masses derived from the direct and indirect searches at LEP and LHC. 
\par The model is  analysed in the alignment limit, where one of its scalar is identified with the Higgs Boson of SM and the Yukawa couplings of fermions with the singlet scalar are found to be proportional to mass of the fermions {\it i.e. non-universal}.  It is  then validated with low energy constraints. We have illustrated the constraints from anomalous magnetic moment in figures \ref{fig:d13_Vs_d23} and \ref{fig:tanbeta_Vs_ mS} and fixed the parameters $\tan\beta$ and $\delta_{23}$ for  a given $\delta_{13}$. We have considered two choices 0.2 and 0.4 respectively for the mixing angle $\delta_{13}$. Contrary to the results obtained in reference \cite{Agrawal:2014ufa} for the singlet scalar with mass lying between 10 - 300 MeV with {\it universal couplings} to leptons,  this study establishes the acceptability of the model to explain the discrepancy $\Delta a_\mu$ for singlet scalar mass lying between 10 GeV $\leq$ $m_{S^0}$ $\leq$ 80 GeV  with couplings to leptons being {\it non-universal}. The  requirement of the Yukawa   coupling $H^0 \,\tau^+\tau^-$ to remain perturbative further imposes an upper limit $\tan\beta\leq 485$ which in turn provides the upper bound on the allowed mass range of singlet scalars to be $\sim 80$ GeV. 
\par Exclusion limits on the couplings of SM gauge Bosons with the singlet scalars  are obtained from the process $e^+e^-\to Z^0 S^0$ at LEP-II experiment and  have been displayed in  table \ref{table2} for some chosen singlet scalar masses.
\par Validation of the model is further subjected to the observed total Higgs decay width at LHC \cite{pdg2018Higgs, Khachatryan:2016ctc}. It is shown that the parameter $m_{12}^2$, which has no bearing on the $\Delta a_\mu$, can now be constrained from the from upper bound on the triple scalar coupling involved in the decay of SM like Higgs to a pair of singlet scalars $h^0\to S^0S^0$. The observed total decay width of SM like Higgs $h^0$ restricts this additional channel and put a upper limit on the partial decay width, which has been shown in figures  \ref{fig:Higgs_Decay_Widthp2}  for $\delta_{13}$ = 0.2 and  in figures \ref{fig:Higgs_Decay_Width} and  \ref{fig:contmh2vsmh3} $\delta_{13}$ = 0.4 respectively. We have found that in the probed region of interest for singlet scalar mass,   $m_{12}^2$ greater than 100 GeV$^2$ and less than 0 GeV$^2$ are forbidden. 
\par We have addressed reasons for which there can be  a deviation from SM predicted universality in lepton-gauge Boson couplings. The precision constraints are also discussed for our model and found that corrections are  suppressed due to the smallness of mixing angle.

\par We augment our analysis by including a fermionic DM candidate $\chi$ and compute the relic density which are depicted in figures \ref{fig:Relic_Densityp2} \& \ref{fig:Relic_Densityp4}  for $\delta_{13}$ = 0.2 and 0.4 respectively. The parameter sets chosen corresponding to points lying on  contours satisfying relic density of 0.119 also fulfill the $\Delta a_\mu$ discrepancy and are consistent with the total Higgs decay width observed at LHC and LEP data.
\par The scalar portal induced DM interactions are now probed in the Direct-detection experiment by the DM-nucleon scattering propelled through the gluons. The variation of spin-independent scattering cross-sections with the DM mass are shown in figures \ref{fig:Direct_Detectionp2} and \ref{fig:Direct_Detectionp4} for $\delta_{13}$ = 0.2 and 0.4 respectively. It can be seen that most of the parameter space for $m_{S^0}$ lighter than 10 GeV is excluded by current Direct-detection constraints from PANDA 2X-II and XENON-1T experiments.
\par The velocity averaged cross-sections for dominant DM pair annihilation channels  like $\tau^+ \tau^-$, $W^+ W^-$, $S^0S^0$ and $\gamma\gamma$ are analytically derived, analysed and compared with the available  space borne indirect-detection experiments.  The velocity averaged cross-sections variation {w.r.t}  DM mass are shown for $\delta_{13}$ = 0.2 and 0.4  in figures  \ref{fig:Indirect_Detectionp2} and  \ref{fig:Indirect_Detectionp4} respectively for $\chi\bar\chi\to\tau^+\tau^-$,  in figures \ref{fig:Indirect_Detection_wwp2} and  \ref{fig:Indirect_Detection_wwp4}  respectively for $\chi\bar\chi\to W^+W^-$,  in figures \ref{fig:Indirect_Detection_ssp2} and  \ref{fig:Indirect_Detection_ssp4}   respectively  for $\chi\bar\chi\to S^0S^0$. We  find that the contribution to the gamma ray spectrum from the most dominant annihilation channel to $\tau^\pm$ pairs is  at least three orders of magnitude lower than  the current reach for the DM mass varying between 5 GeV - 8 TeV. 
\par In conclusion the lepton-specific type {\bf X} 2-HDM model with a singlet scalar portal for fermionic dark matter is capable of explaining both the observed discrepancy in the anomalous magnetic moment of the muon and the observed relic density. This model with the shrunk parameter space after being constrained by low energy experiments, LEP Data,  observed total decay width of Higgs at LHC and constrained by dark matter detection experiments  can now be tested  at the ongoing and upcoming collider searches.
\acknowledgments
Authors acknowledge the fruitful discussions with Mamta.
 SD and MPS acknowledge the partial financial
 support from the CSIR grant No. 03(1340)/15/EMR-II. MPS acknowledges the CSIR JRF fellowship for the partial financial support. SD and MPS thank IUCAA, Pune for providing the hospitality and facilities where this work was initiated.  
\newpage
\appendix
\begin{center}
{\bf \Large Appendix}
\end{center}
\section{Model Parameters}
\label{model_pram}
\par The parameters used in the Lagrangian for lepto-philic 2-HDM and  dark matter  portal singlet scalar given in equation \ref{notation1} are expressed in terms of the physical scalar masses, mixing angles $\alpha$  and $\beta$ and the model parameter $m_{12}^2$.
\begin{eqnarray}
  m^2_{11} &=& -\frac{1}{2}\left[m_{H^0}^2\cos^2\alpha
  +m_{h^0}^2\sin^2\alpha
  +\left\{\sin\alpha\cos\alpha\left(m_{H^0}^2-m_{h^0}^2\right)
 - 2\,m^2_{12}\right\}\tan\beta \right]
 \\
m^2_{22}&=&-\frac{1}{2}\left[m_{h^0}^2\cos^2\alpha
  +m_{H^0}^2\sin^2\alpha
+\left\{\sin\alpha\cos\alpha\left(m_{H^0}^2-m_{h^0}^2\right)
 - 2\,m^2_{12}\right\}\cot\beta \right]\\
\lambda_1&=& \frac{1}{v_o^2\cos^2\beta}\left[m_{H^0}^2\cos^2\alpha
  +m_{h^0}^2\sin^2\alpha- m^2_{12} \tan\beta \right]\\
\lambda_2&=&  \frac{1}{v_o^2\sin^2\beta}\left[m_{h^0}^2\cos^2\alpha
+m_{H^0}^2\sin^2\alpha- m^2_{12} \cot\beta  \right]\\
\lambda_3 &=& \frac{2}{v_o^2\,\sin\left(2\,\beta\right)}\left[
\sin\alpha\cos\alpha\left(m_{H^0}^2-m_{h^0}^2\right) -  m^2_{12} + \,m_{H^+}^2\sin\left(2\,\beta\right)\right] \\
\lambda_4&=& \frac{1}{v_o^2  \sin\left(2\,\beta\right)} \left[ 2 m^2_{12}+ \left(m_{A^0}^2 - 2\,m_{H^+}^2\right) \sin\left(2\,\beta\right)\right]\\
\lambda_5&=&   \frac{1}{v_o^2  \sin\left(2\,\beta\right)}\left[ 2\, m^2_{12}
- m_{A^0}^2 \,\sin\left(2\,\beta\right)\right]
\end{eqnarray}
\section{Decay widths of the singlet scalar $S^0$}
\label{s0_dky_wdth}
\par The tree level partial decay widths of the scalar mediator are  computed and are given by:
\begin{equation}
\Gamma (S^0\rightarrow f\,\bar{f}) = \frac{N_c }{8\,\pi }\left(\frac{\xi^{S}_f m_f}{v_o}\right)^2\,\,\, m_{S^0}\,\,\,\left(1-\frac{4m_f^2}{m_{S^0} ^2}\right)^{3/2}\,\,\, \theta \left(m_{S^0}-2\, mf\right)
\end{equation}
where $N_c$ = 1 for leptons and 3 for quarks
\begin{eqnarray}
\Gamma (S^0\rightarrow W^+\, W^-) &=& \frac{ (\xi^{S^0}_V)^2}{16\,\pi\, v_o^2 } \,m_{S^0}^3\,\left(1-\frac{4m_W ^2}{m_{S^0} ^2}\right)^{1/2}\,\left[12\,\left(\frac{m_W }{m_{S^0}}\right)^4-4\,\left(\frac{m_W}{m_{S^0}}\right)^2 +1\right]\nonumber
\\
&&\times \theta \left(m_{S^0}-2\,m_W\right)
\end{eqnarray}
\begin{eqnarray}
\Gamma (S^0\rightarrow Z^0\, Z^0 ) &=& \frac{1}{2}\,\,\Gamma (S^0\rightarrow W^+ W^-) \,\,\,{\rm with}\,\, m_W\rightarrow m_Z\\
\Gamma (S^0\rightarrow \chi\,\bar{\chi}) &=& \frac{ g_{\chi S^0} ^2}{8\pi } \,\,\,m_{S^0}\,\,\,\left(1-\frac{4m_{\chi}^2}{m_{S^0} ^2}\right)^{3/2}\,\,\, \theta \left(m_{S^0}-2\,m_{\chi}\right)
\end{eqnarray}
\par The one loop induced partial decay width of the scalar to gluons in this model arises mainly from relatively heavy quarks and is given by
\begin{equation}
\Gamma (S^0\rightarrow g g) = \left(\frac{m_t \xi^{S^0}_q}{v_o}\right)^2\,\, \frac{\alpha_s ^2}{72\, \pi^3}\,\,\, \frac{m_{S^0} ^3}{m_t ^2}\,\,\,\,\left\vert\sum_{q=c,b,t} I_q\right\vert^2  
\end{equation}
For the case of photons it is given by
\begin{equation}
\Gamma (S^0\rightarrow \gamma \gamma) = \frac{m_{S^0} ^3}{16 \pi v^2_0}\,\,\left( \frac{\alpha_{em}}{\pi} \right)^2 \left\vert\sum_q \xi^{S^0}_q Q^2_q I_q  +\sum_l\xi^{S^0}_l Q^2_q I_q -  \xi^{S^0}_{W^{\pm}} I_{W^\pm}  + C_{S H^+H^-} \frac{v_o}{2m^2_{H^{\pm}}} I_{H^\pm}\right\vert^2
\end{equation}

\par The integrals are given as
\begin{eqnarray}
\label{loopIntegral}
I_q = 3[2\lambda_q + \lambda_q(4\lambda_q-1)f(\lambda_q)]; && I_l = 2\lambda_q + \lambda_q(4\lambda_q-1)f(\lambda_q); \nonumber\\
I_W = 3\lambda_W(1-2\lambda_W)f(\lambda_W) - \lambda_W -\frac{1}{2}; && 
I_{H^\pm} = -\lambda_{H^\pm}[1+2\lambda_{H^\pm}f({H^\pm})].
\end{eqnarray}
The integrals are defined in terms of dimensionless parameter $\lambda_i=m^2_i/m^2_{S^0}$ and  its function $f(\lambda) $ as
 \begin{eqnarray}
 f(\lambda) &=& -2 \left(sin^{-1} \frac{1}{2\sqrt{\lambda}} \right)^2, for\, \lambda > \frac{1}{4}
\nonumber\\ 
 &=& \frac{1}{2} \left(\ln\frac{\eta^+}{\eta^-} \right)^2 -\frac{\pi^2}{2} - i \pi\frac{\eta^+}{\eta^-}, \,for\, \lambda < \frac{1}{4}
 \end{eqnarray}
 with $\eta^\pm=\frac{1}{2} \pm \sqrt{\frac{1}{4} - \lambda}$.

\section{Thermally averaged scattering cross-sections}
\label{thermalAveragedCrosssection}
We compute the thermal averaged annihilation cross-section of the fermionic DM {\it via} the singlet scalar portal $S^0$ to the SM final states. These processes contributes to the relic density of the universe and are directly used in computing the annihilation cross-section for indirect detection of the DM.
\begin{eqnarray} 
 \left\langle\sigma (\chi\bar \chi \rightarrow f \bar{f}) \, v\right\rangle&=&\left(\frac{m_f \xi^{S^0}_f}{v_o}\right)^2  g_{\chi S^0}^2\,\frac{1}{4\pi }\, \left(1-\frac{m_l ^2}{m_\chi ^2}\right)^{\frac{3}{2}} \frac{m_\chi ^2}{(4m_{\chi} ^2-m_{\phi}^2)^2} \left( \frac{3}{x_f}  \right) \theta\left(m_\chi-m_f\right)\nonumber\\&&\label{thermalfermion}
\\
\left\langle\sigma \left(\chi\bar \chi \rightarrow Z^0 Z^0\right)\, v\right\rangle&=& \left(\frac{\xi_V^{S^0}}{v_o}\right)^2 \frac{g_{\chi S^0} ^2}{8 \pi }\sqrt{1-\frac{m_Z^2}{m _\chi ^2}} \frac{(16m_\chi ^4 + 12m_Z ^4 -16 m_\chi ^2 m_Z ^2)}{(4m_{\chi} ^2-m_{S^0}^2)^2} \left(\ \frac{3}{x_f} \right) \nonumber \\&&\times\,\,\theta\left(m_\chi-m_{Z^0}\right) \label{thermalZ0}
\\
\left\langle\sigma \left(\chi\bar \chi \rightarrow W^+ W^-\right)\, v\right\rangle &=& \left(\frac{\xi_V^{S^0}}{v_o}\right)^2 \frac{g_{\chi S^0} ^2}{16 \pi }\sqrt{1-\frac{m_W^2}{m _\chi ^2}} \frac{(16m_\chi ^4 + 12m_W ^4 -16 m_\chi ^2 m_W ^2)}{(4m_{\chi} ^2-m_{S^0}^2)^2} \left(\ \frac{3}{x_f} \right) \nonumber \\&&\times\,\,\theta\left(m_\chi-m_{W^\pm}\right)\label{thermalW}
\\
\left\langle\sigma (\chi\bar \chi \rightarrow g g) \,v\right\rangle&=&  \left(\frac{\xi^{S^0}_q \alpha_s\, g_{\chi S^0} }{3 \pi^{3/2} v_o} \right)^2 \frac{ m_\chi ^4 }{(4m^2_\chi - m_{S^0} ^2)^2 + m_{S^0} ^2 \Gamma}\left(\frac{3}{2 x_f}\right)\left\vert\sum_{q=c,b,t} I_q\right\vert^2\label{thermalgluon}
\end{eqnarray}

 \begin{eqnarray}
 \left\langle \sigma \left(\chi\bar\chi\to\gamma\gamma\right)\, v\right\rangle =&& \frac{g_{\chi S^0}^2  \alpha^2_{em}}{2 \pi^3 v^2_o } \frac{m^4_\chi \left\vert\sum_q \xi^{S^0}_q Q^2_q I_q  +\sum_l\xi^{S^0}_l Q^2_q I_q -  \xi^{S^0}_{W^{\pm}} I_{W^\pm}  + C_{S H^+H^-} \frac{v_o}{2m^2_{H^{\pm}}} I_{H^\pm}\right\vert^2}{(4m_\chi-m^2_{S^0})^2 + m^2_{S^0} \Gamma^2}\nonumber
\\
 &&\times\left(\frac{3}{ x_f}\right)\label{thermalphoton}
 \end{eqnarray}

\begin{eqnarray}
\left\langle\sigma (\chi\bar \chi \rightarrow H^i\,H^j) \,v\right\rangle&=& c_0 \sum_{H^i;H^j\equiv H^0,\,A^0,\,H^\pm} C_{S^0H^iH^j}^2 \frac{g_{\chi S^0}^2 }{16 \pi} \lambda^{\frac{1}{2}}\left(1,\frac{m^2_{H^i}}{4m^2_\chi},\frac{m^2_{H^j}}{4m^2_\chi}\right) \nonumber
\\
&&\times\frac{1}{(4 m^2_\chi - m^2_\phi)^2}\left(\frac{3}{x_f}\right)\label{thermalscalar}
\end{eqnarray}
 \noindent where $C_{S^0H^iH^j}$ are the tri-linear scalar couplings given in the Appendix \ref{TripleScalarCoupling}; $c_0=\frac{1}{2}$ for i=j and $c_0=1$ for i$\neq$j ; $ \lambda(X,a,b)= X^2 + a^2 + b^2 -2ab - 2aX -2bX$.

In addition to {\it s-channel} processes considered above, we  also have contributions to the relic density from {\it t-channel} process $\chi \bar\chi\rightarrow S^0 S^0$, given by
\begin{eqnarray}
\left\langle\sigma \left(\chi \bar\chi \rightarrow S^0 S^0\right) \,v\right\rangle&=& \frac{3 g_{\chi S^0} ^4}{64 \pi m^2_\chi}\left(\frac{3}{x_f}\right)\,\theta\left(m_\chi-m_{S^0}\right) \label{thermalS0S0} 
\end{eqnarray}
\section{ Triple scalar coupling}
\label{TripleScalarCoupling}
Here, we  extract  the triple scalar coupling from the 2-HDM + singlet scalar Lagrangian in the alignment limit. Some of these scalars can be directly constrained from the ongoing experiments at the Colliders.  We define dimensionless ratios $r_0= \frac{m^2_{H^0}}{m^2_{h^0}}$ and $s_0 =\frac{m^2_{12}}{m^2_{h^0}}$. All  triple scalar couplings are defined in terms of $r_0$, $s_0$ and $\tan\beta$.
{\small
\begin{eqnarray}
C_{h^0 S^0 S^0} &=&  \frac{\delta^2_{13} m_{h^0}^2}{v_o}\left(-\frac{4 \delta_{13}^2 r_0^2}{\tan\beta}+\frac{2\delta_{13}^2 r_0}{\tan\beta }-\frac{8 r_0^3}{\tan^4\beta}
  +\frac{8 r_0^2 s_0}{\tan^3\beta }+\frac{4 r_0^2}{\tan^4\beta }-\frac{8 r_0 s_0}{\tan^3\beta}-\frac{8 r_0 s_0}{\tan\beta}\right.\nonumber
  \\
  &&
  +\left.\frac{2 r_0}{\tan^4\beta }-\frac{2 r_0}{\tan^2\beta}+\frac{2 s_0}{\tan^3\beta}+2 s_0 \tan\beta
  +\frac{4 \delta_{13}^2  s_0}{\tan\beta}-\frac{1}{ \tan^4\beta }-\frac{2}{\tan^2\beta }-1\right)\nonumber\\ \label{Higgsdecay} \end{eqnarray}
\begin{eqnarray}
C_{S^0 h^0 h^0} &=&  \frac{\delta_{13} m_{h^0}^2}{v_o} \left( -\frac{4 \delta_{13}^2 r_0^2}{\tan^2\beta}+\frac{2\delta_{13}^2 r_0}{\tan^2\beta }-2\delta_{13}^2 r_0-\frac{6 r_0}{\tan^3\beta}-\frac{4 r_0}{\tan\beta}\right) 
\end{eqnarray}
\begin{eqnarray}
C_{S^0 H^0 H^0} &=& \frac{\delta_{13} m_{h^0}^2}{v_o}\left( -\frac{8 \delta_{13}^2 r_0^3}{\tan^2\beta }+\frac{12 \delta_{13}^2 r_0^2}{\tan^2\beta }-\frac{4 \delta_{13}^2 r_0}{\tan^2\beta}-\frac{6  r_0^2}{\tan^5\beta}\nonumber
-\frac{4  r_0^2}{\tan^3\beta}+\frac{2 r_0^2}{\tan\beta}+\frac{6 r_0 s_0}{\tan^4\beta }+\frac{4  r_0 s_0}{\tan^2\beta}\right.\nonumber
\\
&&\left.-2 r_0 s_0 + \frac{3  r_0}{ \tan^5\beta}+\frac{ r_0}{ \tan^3\beta }-3  r_0 \tan\beta -\frac{7 r_0}{ \tan\beta}-\frac{3  s_0}{ \tan^4\beta}+3 s_0 \tan^2\beta 
-\frac{3 s_0}{ \tan^2\beta}+3  s_0 \right)\nonumber\\
\end{eqnarray}
\begin{eqnarray}
C_{S^0 S^0 H^0} &=& \frac{\delta^2_{13} m_{h^0}^2}{v_o} \left( -\frac{8 \delta_{13}^2 r_0^3}{\tan^2\beta}+\frac{8 \delta_{13}^2 r_0^2}{\tan^2\beta}-\frac{2\delta_{13}^2 r_0}{\tan^2\beta}+\frac{12 r_0^3}{\tan^5\beta}+\frac{4 r_0^3}{\tan^3\beta }-\frac{12 r_0^2 s_0}{\tan^4\beta }-\frac{4 r_0^2 s_0}{\tan^2\beta}-\frac{12 r_0^2}{\tan^5\beta}\right.\nonumber
 \\
 &&\left. +\frac{8 r_0^2}{\tan\beta}+\frac{12 r_0 s_0}{\tan^4\beta}+\frac{8 r_0 s_0}{\tan^2\beta}-4  r_0 s_0+\frac{3  r_0}{ \tan^5\beta}-\frac{ r_0}{ \tan^3\beta}-3  r_0 \tan\beta-\frac{11 r_0}{ \tan\beta}\right.\nonumber
 \\
 &&\left.-\frac{3 s_0}{ \tan^4\beta }+3 s_0 \tan^2\beta-\frac{3  s_0}{ \tan^2\beta}+3  s_0 \right)
\end{eqnarray}
\begin{eqnarray}
C_{S^0 H^0 h^0} && = \frac{\delta_{13} m_{h^0}^2}{v_o}\left(-\frac{4 \delta_{13}^2 r_0^3}{\tan^3\beta}+\frac{4 \delta_{13}^2 r_0^2}{\tan^3\beta}-\frac{4 \delta_{13}^2 r_0^2}{\tan\beta}-\frac{\delta_{13}^2 r_0}{\tan^3\beta}+\frac{3 \delta_{13}^2 r_0}{\tan\beta}+\frac{4 r_0^2}{\tan^4\beta}+\frac{4  r_0^2}{\tan^2\beta }-\frac{4  r_0 s_0}{\tan^3\beta}\right.\nonumber
 \\
 &&\left.-\frac{4 r_0 s_0}{\tan\beta}-\frac{3 r_0}{\tan^2\beta}- r_0+\frac{2  s_0}{\tan^3\beta }+2 s_0 \tan\beta+\frac{4 s_0}{\tan\beta}-\frac{1}{\tan^4\beta }-\frac{2}{\tan^2\beta}-1 \right)
\end{eqnarray}
\begin{eqnarray}
C_{S^0 H^+ H^-} &&=  \frac{\delta_{13} m_{h^0}^2}{v_o}\left(-\frac{2  r_0^2}{\tan^5\beta}+\frac{2  r_0^2}{\tan\beta}+\frac{2  r_0 s_0}{\tan^4\beta}+\frac{4  r_0 s_0}{\tan^2\beta }+2 r_0 s_0+\frac{ r_0}{\tan^5\beta }-\frac{ r_0}{\tan^3\beta}- r_0 \tan\beta\right.\nonumber
\\ 
 &&\left.-\frac{5 r_0}{\tan\beta}-\frac{ s_0}{\tan^4\beta}+  s_0 \tan^2\beta-\frac{ s_0}{\tan^2\beta}+ s_0-\frac{4  m_{H^\pm}^2 r_0}{\tan^3\beta m^2_{h^0}}-\frac{4 m_{H^\pm}^2 r_0}{\tan\beta m^2_{h^0}} \right)
\end{eqnarray}
\begin{eqnarray}
C_{S^0 A^0 A^0} &&=  \frac{\delta_{13} m_{h^0}^2}{v_o}\left( -\frac{4 m_{A^0}^2 r_0}{\tan^3\beta m_{h^0}^2}-\frac{4  m_{A^0}^2 r_0}{\tan\beta m_{h^0}^2}-\frac{2  r_0^2}{\tan^5\beta}+\frac{2 r_0^2}{\tan\beta}+\frac{2 r_0 s_0}{\tan^4\beta }+\frac{4  r_0 s_0}{\tan^2\beta}+2 r_0 s_0\right.\nonumber
 \\
 &&\left. +\frac{ r_0}{ \tan^5\beta}-\frac{ r_0}{ \tan^3\beta}- r_0 \tan\beta-\frac{5 r_0}{ \tan\beta}-\frac{ s_0}{ \tan^4\beta}+ s_0 \tan^2\beta-\frac{ s_0}{ \tan^2\beta }+ s_0 \right)
\end{eqnarray}
\begin{eqnarray}
C_{S^0 S^0 S^0} &&=  \frac{\delta^3_{13} m_{h^0}^2}{v_o} \left(-\frac{24 r_0^4}{\tan^5\beta}+\frac{24 r_0^3 s_0}{\tan^4\beta }+\frac{36 r_0^3}{\tan^5\beta}-\frac{12 r_0^3}{\tan^3\beta}-\frac{36 r_0^2 s_0}{\tan^4\beta }-\frac{12 r_0^2 s_0}{\tan^2\beta}-\frac{18 r_0^2}{\tan^5\beta}+\frac{12 r_0^2}{\tan^3\beta}\right.\nonumber
 \\
 &&+\frac{18 r_0^2}{\tan\beta}+\frac{18  r_0 s_0}{\tan^4\beta}+\frac{12  r_0 s_0}{\tan^2\beta}-6 r_0 s_0+\frac{3 r_0}{ \tan^5\beta}-\frac{3 r_0}{ \tan^3\beta }-3 r_0 \tan\beta-\frac{15 r_0}{ \tan\beta}\nonumber\\
 &&\left. -\frac{3 s_0}{\tan^4\beta}+3 s_0 \tan^2\beta-\frac{3 s_0}{ \tan^2\beta}+3 s_0 \right)
\end{eqnarray}
\begin{eqnarray} 
 C_{h^0 H^+ H^-} &=&  \frac{m^2_{h^0}}{v_o}\left(\frac{2 \delta^2_{13} r_0}{\tan\beta }+\frac{2 s_0}{\tan^3\beta }+2  s_0 \tan\beta+\frac{4  s_0}{\tan\beta }-\frac{1}{\tan^4\beta }-\frac{2 }{\tan^2\beta }-1-\frac{2 m_{H^\pm}^2}{\tan^4\beta m^2_{h^0} }\right.\nonumber
 \\
 &&\left.-\frac{4 m_{H^\pm}^2}{\tan^2\beta m^2_{h^0} }-\frac{2 m_{H^\pm}^2}{m^2_{h^0}}\right).
\end{eqnarray}
}

\end{document}